\documentclass[12pt]{article}
\usepackage[utf8]{inputenc}
\usepackage{amssymb}
\usepackage{amsfonts}
\usepackage{amsmath}
\usepackage{amsthm}
\usepackage{mathrsfs} 
\usepackage{graphicx}
\usepackage{mathtools}
\numberwithin{equation}{section}
\usepackage{xcolor}

\usepackage{yfonts}
\usepackage{amssymb}
\usepackage{amsmath}
\theoremstyle{theorem}
\newtheorem{theorem}{Theorem}[section]
\newtheorem{lemma}[theorem]{Lemma}
\newtheorem{proposition}[theorem]{Proposition}

\newtheorem{definition}[theorem]{Definition}
\newtheorem{remark}[theorem]{Remark}
\newtheorem{notation}[theorem]{Notation}
\newtheorem{assumption}[theorem]{Assumption}
\newcommand{\half}{\mbox{\footnotesize $\frac{1}{2}$}}
\newcommand{\raw}{\rightarrow}
\newcommand{\R}{\mathbb{R}}
\newcommand{\C}{\mathbb{C}}
\newcommand{\N}{\mathbb{N}}
\newcommand{\Z}{\mathbb{Z}}
\newcommand{\om}{\omega}
\newcommand{\cin}{C^{\infty}}
\newcommand{\ca}{$C^*$-algebra}
\newcommand{\er}{\eqref}
\topmargin = - 1 cm \textheight = 23 cm \textwidth = 15 cm
\begin{document}
\par
\bigskip
\large
\noindent
{\bf Strict deformation quantization of the state space of $M_k(\mathbb{C})$ with applications to the Curie--Weiss model}
\bigskip
\par
\rm
\normalsize

\noindent  {\bf Klaas Landsman$^{a}$$^{1}$}, {\bf Valter Moretti$^{b}$$^2$}, {\bf Christiaan J.F.  van de Ven$^{b}$$^3$}\\
\par

\noindent 
 $^a$  Institute for Mathematics, Astrophysics, and Particle Physics (IMAPP), \\
Radboud University,
Heyendaalseweg 135
6525 AJ Nijmegen,
The Netherlands\\
$^1$ Email: landsman@math.ru.nl\\

\noindent 
 $^b$ Department of  Mathematics, University of Trento, and INFN-TIFPA \\
 via Sommarive 14, I-38123  Povo (Trento), Italy\\
$^2$ Email: valter.moretti@unitn.it,  $^3$ Marie Sk\l odowska-Curie Fellow of the Istituto Nazionale di Alta Matematica. Email:
christiaan.vandeven@unitn.it\\
 \normalsize

\par

\rm\normalsize

\date{today}

\rm\normalsize


\par
\bigskip

\noindent
\small
{\bf Abstract}. Increasing tensor powers of the $k\times k$ matrices $M_k({\mathbb{C}})$ are known to give rise to a continuous bundle of $C^*$-algebras over 
$I=\{0\}\cup 1/\mathbb{N}\subset[0,1]$ with fibers $A_{1/N}=M_k({\mathbb{C}})^{\otimes N}$ and $A_0=C(X_k)$, where
 $X_k=S(M_k({\mathbb{C}}))$, the state space of $M_k({\mathbb{C}})$, which 
 is canonically a compact Poisson manifold (with stratified boundary). Our first result is the 
existence of a strict deformation quantization of $X_k$ \`{a} la Rieffel, defined by perfectly natural quantization maps $Q_{1/N}: \tilde{A}_0\rightarrow A_{1/N}$ (where $\tilde{A}_0$ is an equally natural dense Poisson subalgebra of $A_0$). 

We apply this quantization formalism to the Curie--Weiss model (an exemplary quantum spin with long-range forces) in the parameter domain where its  $\mathbb{Z}_2$ symmetry is spontaneously broken in the thermodynamic limit $N\raw\infty$. If this limit is taken with respect to the macroscopic observables of the model (as opposed to the quasi-local observables), it yields a classical theory with phase space $X_2\cong B^3$ (i.e\ the unit three-ball in $\R^3$).
Our quantization map then enables us to take the classical limit of the sequence of (unique) algebraic vector states induced by the ground state eigenvectors $\Psi_N^{(0)}$ of this model as $N\raw\infty$, in which the sequence converges to a probability measure $\mu$ on the associated classical phase space $X_2$.  This measure is a symmetric convex sum of two Dirac measures related by the underlying $\Z_2$-symmetry of the model, and as such the classical limit exhibits spontaneous symmetry breaking, too. Our proof of convergence is  heavily based on Perelomov-style coherent spin states and at some stage it relies on (quite strong) numerical evidence. Hence the proof is not completely analytic, but somewhat hybrid. 
\normalsize
\newpage
\tableofcontents
\newpage
\section{Introduction}
This paper addresses two important and seemingly unrelated topics in mathematical physics, which we both attempt to bring forward by relating them to each other: 
\begin{enumerate}
\item Strict (i.e.\ $C^*$-algebraic) deformation quantization \`{a} la Rieffel \cite{Rie89,Rie94,Lan98};
\item Spontaneous Symmetry Breaking (SSB) in quantum spin systems \cite{BR2,Lan17}.
\end{enumerate}
The relationship will  be that the second topic suggests an interesting phase space to quantize, namely the state space  $X_2=S(M_2({\mathbb{C}}))$ of the $C^*$-algebra of  $2 \times 2$ complex matrices (which is affinely homeomorphic to the unit three-ball in $\mathbb{R}^3$).  Though initially defined as the state space of a generic two-level \emph{quantum system}, $X_2$
 also plays the role of a \emph{classical} phase space underlying the Curie--Weiss model, which is an exemplary \emph{quantum} mean-field spin model exhibiting SSB (see for example \cite{ABN,CCIL,IL,Lan17,VGRL18} and references therein, as well as \S\ref{CWM} below). In particular, $X_2$  is a  compact convex manifold (with smooth boundary $\partial X_2\cong S^2$) canonically equipped with a Poisson structure, like its 
 its generalizations $X_k=S(M_k({\mathbb{C}}))$ for any $k\in\N$ (for $k>2$ the boundary is a {\em stratified} space though, see \S\ref{kgeq2}).  In that role, $X_k$ will be subjected to strict deformation quantization, which is our first achievement.
 
Once this has been accomplished, we will be able to define and compute a suitable limit of the sequence  $(\Psi_N^{(0)})_{N\in\N}$ of ground states of the Curie--Weiss model (which are unique up to a phase), where $N$ is the number of sites of the lattice on which the model is defined, as $N\raw\infty$. This limit is not so much taken of the vectors $\Psi_N^{(0)}$, but of the
 associated algebraic vector states; it exists in a suitable sense detailed below and yields a \emph{classical} state in the sense of a probability measure on the phase space $X_2$. This exhibits SSB, too, like the  thermodynamic limit of the quantum model.
 
 Let us now explain what this means, starting from the familiar phase space  $\R^{2n}$ (with its usual Poisson structure). To make the essential points clear we take the simplest functional-analytic situation, in which only smooth compactly supported functions $f\in C_c^{\infty}(\R^{2n})$ are quantized. Weyl \cite{Weyl} proposed the quantization maps
\begin{align}
Q_{\hbar}:  C_c^{\infty}(\R^{2n}) &\raw B_0(L^2(\R^n));\\
Q_{\hbar}(f)&=\int_{\R^{2n}} \frac{d^npd^nq}{(2\pi\hbar)^n}\,  f(p,q) \Omega_{\hbar}(p,q), \label{QW2tris}
\end{align}
 where  $\hbar\in (0,1]$ for simplicity; in modern parlance $B_0(H)$ is the $C^*$-algebra of compact operators on a Hilbert space $H$ (here $H=L^2(\R^n)$),   and for each point $(p,q)\in \R^{2n}$ the (bounded) operator $\Omega_{\hbar}(p,q):L^2(\R^n)\raw L^2(\R^n)$ is given by
\begin{equation}
\Omega_{\hbar}(p,q)\Psi(x)=2^n e^{2ip(x-q)/\hbar}\Psi(2q-x), \:\:\: \Psi\in L^2(\R^n). \label{omega}
\end{equation}
Eqs.\ \er{QW2tris} and \er{omega} are equivalent to Weyl's own (slightly rewritten) formula
\begin{equation}
Q_{\hbar}(f)\Psi(x) =\int_{\R^{2n}}
 \frac{d^npd^ny}{(2\pi\hbar)^n}\, e^{ip(x-y)/\hbar}f\left(p,\half (x+y)
 \right)\Psi(y). \label{QW2bis}
\end{equation}
In a $C^*$-algebraic framework (which was not available to Weyl), Rieffel \cite{Rie89,Rie94}, relying on Dixmier's concept of a continuous bundle (= field) of $C^*$-algebras (see \cite{Dix} for the original definition and \cite{Lan98,Lan17} for various useful reformulations), 
noted that:
\begin{enumerate}
\item The fibers $A_0= C_0(\R^{2n})$ and  $A_{\hbar}= B_0(L^2(\R^n))$, $h\in(0,1]$, can be combined into a (locally non-trivial) continuous bundle $A$ of $C^*$-algebras over $I=[0,1]$;
\item  $\tilde{A}_0=C^{\infty}_c(\R^{2n})$ is a  dense Poisson subalgebra of $A_0$;
 \item  Each quantization map  $Q_{\hbar}:\tilde{A}_0\to A_{\hbar}$ is linear, and if we also define $Q_0: \tilde{A}_0 \hookrightarrow A_0$ as the inclusion  map, then the ensuing family $Q = (Q_\hbar)_{\hbar \in I}$
 satisfies:
  \begin{enumerate}
\item  Each map $Q_{\hbar}$ is self-adjoint, i.e.\ $Q_{\hbar}(\overline{f})=Q_{\hbar}(f)^*$ (where $f^*(x)=\overline{f(x)}$). 
\item  For each $f\in \tilde{A}_0$ the following cross-section of the bundle is continuous:
\begin{align}
    &0\to f;\\
    &\hbar\to Q_{\hbar}(f) \ \ \  (\hbar\in I\backslash\{0\})).
\end{align}
\item  Each pair $f,g\in\tilde{A}_0$ satisfies the {\bf Dirac-Groenewold-Rieffel condition}:
\begin{align}
    \lim_{\hbar\to 0}\left|\left|\frac{i}{\hbar}[Q_{\hbar}(f),Q_{\hbar}(g)]-Q_{\hbar}(\{f,g\})\right|\right|_{\hbar}=0.\label{Diracgroenewold}
\end{align} 
\end{enumerate}
\end{enumerate}
This suggested the general concept of a strict deformation of a Poisson manifold $X$ \cite{Rie89,Lan98}, which we here state in the case of interest to us in which $X$ is compact (as already noted, our $X$'s will be manifolds with stratified boundary, see \S\ref{kgeq2}). In that case, examples from geometric quantization rewritten into the above language (e.g. \cite{BMS94}) suggest that the space $I$ in which $\hbar$ takes values cannot be all of $[0,1]$, but should be a subspace $I\subset [0,1]$ thereof that at least contains 0 as an accumulation point (and is typically discrete away from 0). This is assumed in what follows. Furthermore, the   Poisson bracket on $X$ is denoted, as usual, by $\{\cdot, \cdot\}: C^\infty(X)\times C^\infty(X) \to \mathbb{C}$.
 \begin{definition}\label{def:deformationq}
A {\bf strict deformation quantization} of a compact Poisson manifold $X$  consists of an index space  $I\subset [0,1]$ for $\hbar$ as detailed above, as well as:
 \begin{itemize}
\item  A continuous
 bundle of unital $C^*$-algebras $(A_{\hbar})_{\hbar\in I}$ over $I$  with $A_0=C(X)$;
 \item A dense Poisson suabalgebra  $\tilde{A}_0 \subseteq C^\infty(X) \subset A_0$ (on which $\{\cdot, \cdot\}$ is defined);
 \item  A  family $Q = (Q_\hbar)_{\hbar \in I}$ of  linear maps $Q_{\hbar}:\tilde{A}_0\to A_{\hbar}$ indexed by $\hbar\in I$ (called 
  {\bf quantization maps})  such that $Q_0$ is the inclusion map $\tilde{A}_0 \hookrightarrow A_0$, and the above 
  conditions (a) - (c) hold, as well as  $Q_{\hbar}(\mathrm{1}_X)=\mathrm{1}_{A_{\hbar}}$ (the unit of $A_{\hbar}$).\footnote{It follows from the definition of a continuous bundle of $C^*$-algebras that continuity properties like
$ \lim_{\hbar\to 0}\|Q_{\hbar}(f)\|_\hbar=\|f\|_{\infty}$ and 
     $\lim_{\hbar\to 0}\|Q_{\hbar}(f)Q_{\hbar}(g)-Q_{\hbar}(fg)\|_\hbar=0$
hold automatically (they may be \emph{imposed} in alternative definitions of strict quantization). See \cite{Lan98,Lan17}. }
\end{itemize}
\end{definition}
\noindent Perhaps surprisingly, mean-field quantum spin systems (such as the Curie--Weiss model) fit into this framework, with the index set $I$ given by ($0\notin\N=\{1,2,3,\ldots\}$)
\begin{equation}
I= \{1/N \:|\: N \in  \mathbb{N}\} \cup\{0 \}\equiv (1/\N) \cup\{0 \}, \label{defI}\
\end{equation}
with topology inherited from $[0,1]$. 
That is, we  put $\hbar=1/N$, where $N\in\mathbb{N}$ is interpreted as   the number of sites of the model; our interest is the limit $N\raw\infty$. 

In the framework of $C^*$-algebraic quantization theory, the analogy between the ``classical" limit $\hbar\raw 0$ in typical examples from mechanics
and the ``thermodynamic" limit $N\raw\infty$ in typical quantum spin systems 
is developed in detail in \cite{Lan17} and forms the basis of the present work.  The key point here is that for any unital $C^*$-algebra $B$ (where for applications to quantum spin systems one may  take $B=M_k(\C)$ for some $k\in\N$, with $k=2$ in the Curie---Weiss model), the following fibers may be turned into a continuous bundle of $C^*$-algebras over the base space $I=\{0\}\cup 1/\mathbb{N}\subset[0,1]$ (with relative topology, so that $(1/N)\raw 0$ as $N\raw\infty$):
 \begin{align}
 A_0&=C(S(B)); \label{A0}\\
 A_{1/N}&= {B}^{\otimes N}.\label{AN}
   \end{align}
Here $S(B)$ is the (algebraic) state space of $B$ equipped with the weak$\mbox{}^*$-topology (in which it is a compact convex set,
e.g.\ the three-ball $S(M_2(\C))\cong B^3\subset\R^3$),
 and ${B}^{\otimes N}$ is the $N$th tensor power of $B$ (often called $B^N$ in what follows).\footnote{Although this is irrelevant for our main application $B=M_k(\C)$, for general $C^*$-algebras $B$ one should equip $B^N$ with the {\em minimal} $C^*$-norm $\|\:\:\:\|_N$ \cite{Takesaki,Lan17}.} As in the case of vector bundles, the continuity structure of a bundle of $C^*$-algebras may be defined (indirectly) by specifying what the continuous cross-sections are. To do so for \er{A0} - \er{AN}, we need the {\em symmetrization operator} $S_N : {B}^N \to {B}^N$, defined as the unique linear continuous extension of the following map on elementary tensors:
\begin{align}S_N (a_1 \otimes \cdots \otimes a_N) = \frac{1}{N!} \sum_{\sigma \in {\cal P}(N)} a_{\sigma(1)} \otimes \cdots \otimes a_{\sigma(N)}. \label{defSN}\end{align}
Furthermore, for $N\geq M$ we need to generalize the definition of $S_N$ to give a bounded  operator  $S_{M,N}: {B}^M 	\to {B}^N$, defined by linear and continuous extension of  
\begin{align}S_{M,N}(b) = S_N(b \otimes \underbrace{I \otimes \cdots \otimes I }_{N-M \mbox{\scriptsize times}}),\quad b \in {B}^M. \label{defSMN}\end{align}
 We write cross-sections $a$ of \er{A0} - \er{AN} as sequences 
$(a_0,a_{1/N})_{N\in\N}$, where $a(0)=a_0$ etc. Following \cite{RW}, the part of the cross-section
 $(a_{1/N})_{N\in\N}$ away from zero (i.e.\ with $a_0$ omitted) is called {\bf symmetric}
if there exist $M \in \mathbb{N}$ and $a_{1/M} \in {B}^{\otimes M}$ such that 
\begin{align}\label{one}
    a_{1/N} = S_{M,N}(a_{1/M})\:\mbox{for all }  N\geq M,
\end{align}
and {\bf quasi-symmetric} if  $a_{1/N} = S_{N}(a_{1/N})$ if $N\in \mathbb{N}$,
and for every $\epsilon > 0$, there is a symmetric sequence $(b_{1/N})_{N\in\mathbb{N}}$
as well as  $M \in\mathbb{N}$ (both depending on $\epsilon$) such that 
\begin{align} \| a_{1/N}-b_{1/N}\| < \epsilon\: \mbox{ for all } N > M.
\end{align} 
The continuous cross-sections of the bundle  \er{A0} - \er{AN}, then, are the  sequences 
$(a_0,a_{1/N})_{N\in\N}$ for which the part  $(a_{1/N})_{N\in\N}$ away from zero is quasi-symmetric and \begin{align}
a_0(\omega)=\lim_{N\to\infty}\omega^N(a_{1/N})\label{8.46},
\end{align}
where $\omega\in S({B})$, and $\omega^N=  \underbrace{\omega\otimes \cdots \otimes \omega}_{N\: \mbox{\scriptsize times}} \in S({B}^{\otimes N})$,
is  the unique (norm) continuous linear extension of the following map that is defined on elementary tensors:
\begin{align}
\omega^N(b_1\otimes\cdot\cdot\cdot\otimes b_N)=\omega(b_1)\cdot\cdot\cdot\omega(b_N).\label{linearstateextension}
\end{align}
The limit in \er{8.46} exists provided  $(a_{1/N})_{N\in\N}$ is quasi-symmetric (as we assume), and by \cite[Theorem 8.4]{Lan17}, this choice of continuous cross-sections uniquely defines (or identifies) 
 a continuous bundle of $C^*$-algebras over  $I$ in \er{defI} with fibers \er{A0} - \er{AN}. 
 In particular, for $B=M_k(\C)$, the fibers of this continuous bundle are
  \begin{align}
 A_0&=C(S(M_k(\C))\equiv C(X_k); \label{B0}\\
 A_{1/N}&= M_k(\C)^{\otimes N}\cong M_{k^N}(\C).\label{BN}
   \end{align}
As already mentioned $X_k=S(M_k(\C))$ is canonically a compact Poisson manifold (see \S\ref{PoissonMk}), so that one may start looking for suitable Poisson subalgebras $\tilde{A}_0\subset C^{\infty}(X_k)$ on which, hopefully in accordance with Definition \ref{def:deformationq},
quantization maps 
\begin{equation}
Q_{1/N}: \tilde{A}_0\raw M_k(\C)^{\otimes N} \label{QCW}
\end{equation}
may be constructed. This can indeed be done. First, in suitable coordinates $\tilde{A}_0$ consists essentially of polynomial functions on $X_k$ (see \S\ref{thespace}). Second,  the construction of the maps $Q_{1/N}$ is given in section \ref{Sdf}, see especially \er{deformationqunatizaion}. This definition is natural and straightforward (perhaps even more so than Weyl quantization on $\R^{2n}$), but the proof that the choice \er{deformationqunatizaion} satisfies the conditions in Definition \ref{def:deformationq} is  nontrivial. 

Compared to a case like $X=\R^{2n}$, our quantization maps for $X_k=S(M_k(\C))$ \emph{as such} are of less value; typically (unlike $\R^{2n}$), physical observables for quantum spin systems are not constructed or interpreted through (deformation) quantization. The real point in having \er{QCW} lies in the possibility of defining classical limits of quantum states of the Curie--Weiss model (i.e.\ $k=2$). Namely, suppose we have a family of states $(\om_{1/N})_{N\in\N}$, where $\om_{1/N}$ is a state on $A_{1/N}$, as in \er{BN}, such that
 \begin{equation}
\om_0(f)=\lim_{N\raw\infty}  \om_{1/N}(Q_{1/N}(f))\label{om0limit}
\end{equation}
exists for all $f\in\tilde{A}_0$ and defines a state $\om_0$ on $A_0$ as in \er{B0}, that is, a probability measure $\mu_0$ on $X_k$ (so that
$\om_0(f)=\int_{X_k} d\mu_0\, f$).
Then, in complete analogy with the case  $X=\R^{2n}$, the state $\om_0$ may be regarded as the classical limit of the  family 
$(\om_{1/N})$. Of course, the given family should have a very specific $N$-dependence for the limits \er{om0limit} to exist, and we will see that this happens when $\om_{1/N}$ is the vector state defined by the ground state of the Curie--Weiss Hamiltonian for given $N$, i.e.,
\begin{equation}
\om_{1/N}(b)=\langle \Psi_N^{(0)}, b\Psi_N^{(0)}\rangle, \label{herdef}
\end{equation}
where $b\in A_{1/N}=M_2(\C)^N\cong B((\C^2)^{\otimes N})$ and $\Psi_N^{(0)}\in (\C^2)^{\otimes N}$ is the ground state 
of the Hamiltonian $h^{CW}_{1/N}$ of the model (see \S\ref{CWM}), seen in the usual (physics) way as a unit vector in the Hilbert space $(\C^2)^{\otimes N}$ on which the model on $N$ sites is defined. Even so, it is remarkable that the $N$-dependence of $h^{CW}_{1/N}$ precisely makes this work. 

\noindent To draw the analogy with the phase space $X=\R^{2n}$ in this respect, we apply the same procedure  to Weyl quantization. For a fixed unit vector $\Psi\in\ L^2(\R^n)$ this gives
\begin{equation}
\langle\Psi ,Q_{\hbar}(f)\Psi\rangle= \int_{\R^{2n}} \frac{d^npd^nq}{(2\pi\hbar)^n}\,  f(p,q) W^{\hbar}_{\Psi}(p,q),
\end{equation}
where $W^{\hbar}_{\Psi}: \R^{2n}\raw\R$
 is the famous \emph{Wigner function} associated to $\Psi$,  given by
\begin{equation}
W^{\hbar}_{\Psi}(p,q)=\hbar^{-n}\langle \Psi, \Omega_{\hbar}(p,q)\Psi\rangle \\
= \int_{\R^n} d^n ve^{ipv} \overline{\Psi(q+\half \hbar v)} \Psi(q-\half\hbar v).
\end{equation}
See \cite{BJ84,Wigner,Lan98}.
If $\|\Psi\|=1$, then $W^{\hbar}_{\Psi}$ gives a ``phase space portrait" of $\Psi$. 
However, this portrait is not a probability density  on $\R^{2n}$, since  Wigner functions are not necessarily positive.\footnote{This reflects the fact that Weyl's quantization map $Q_{\hbar}$ fails to be positive (in that $f\geq 0$ pointwise implies
$Q_{\hbar}(f)\geq 0$ as an operator). For $\R^{2n}$ this can be remedied by finding a quantization map that \emph{is} positive, i.e.\  Berezin quantization \cite{Lan98,Lan17}, whose associated ``Wigner" function is  the \emph{Husimi function}
(alas, for the bundle \er{B0} - \er{BN}  we were unable find a positive quantization map). 
Wigner and Husimi functions associated to the same family of states have the same limit, cf.\  \cite[Proposition II.2.6.3]{Lan98}; this is  how the claim about the double well in the main text is proved. Note that we only invoke Wigner functions in order to show that our limit \er{om0limit} is familiar; in our approach they 
 are secondary to quantization maps $Q_{\hbar}$. Indeed, for the quantum spin systems we study we do not (need to) construct the analogues of Wigner functions at all. 
 }  
Nonetheless, they are useful for studying the classical limit:  although the vectors $\Psi_{\hbar}$ by themselves have no limit, the associated Wigner functions $W^{\hbar}_{\Psi_{\hbar}}$ \emph{may} have a limit if they converge weakly to some probability measure $\mu_0$ on the classical phase space $\R^{2n}$, in the sense that for all $f\in C^{\infty}_c(\R^{2n})$ one has
\begin{equation}
\lim_{\hbar\raw 0} \int_{\R^{2n}} \frac{d^npd^nq}{(2\pi\hbar)^n}\,  f(p,q) W^{\hbar}_{\Psi_{\hbar}}(p,q)=\int_{\R^{2n}} d\mu_0 f.
\end{equation}
\emph{And this is precisely the limit} \er{om0limit}, provided $Q_{1/N}\equiv Q_{\hbar}$ is given by \er{QW2tris} or \er{QW2bis}.

As a case in point that is quite relevant for SSB we may take $\Psi_{\hbar}$ to be the (unique) ground state of the symmetric double well potential in $n=1$, for which it can be shown that $\mu_0=(\delta_++\delta_-)/2$, where $\delta_{\pm}$ are the Dirac measures localized at $(p=0,q=a_{\pm})\in\R^2$, where $a_-$ and $a_+$ are the left- and right minima of the double well, respectively \cite{Lan17,VGRL18} (for general families of states $W^{\hbar}_{\Psi_{\hbar}}$ may not have a limit!). 

As explained in detail in \cite{Lan17}, despite the above analogies there is one point in which the limit $N\raw\infty$ 
for quantum spin systems is richer than its counterpart $\hbar\raw 0$ for mechanical systems, in that the former 
may be taken in two entirely different ways, at least for mean-field models like the Curie--Weiss model. Which of the two limits applies depends on the class of observables one wants to take the limit of, namely either \emph{quasi-local}  or \emph{macroscopic} observables. The former are the ones traditionally studied for quantum spin systems  \cite{BR2,Sim}, but the 
latter relate these systems to strict deformation quantization, since macroscopic observables are precisely defined by (quasi-) symmetric sequences (see \cite{Lan17} for quasi-local observables). 

The plan of this paper is obvious from the table of contents. Apart from the general conceptual framework of applying strict deformation quantization to quantum spin systems, our main technical results are Theorem \ref{mainfirst} in \S\ref{3.2}, stating that
 the quantization maps \er{deformationqunatizaion} satisfy Definition \ref{def:deformationq}, and Theorem \ref{mainsecond}, in \S\ref{QZ}, establishing the classical limit of the ground state of the Curie--Weiss model, including its SSB. 
\section{Structures on  ${B} = M_k(\mathbb{C})$}\label{SecPoisson}
Unless stated otherwise, ${B}= M_k(\mathbb{C})$ is the unital $C^*$-algebra of $k\times k$ complex matrices equipped with  the natural $C^*$-norm, whose unit element is denoted by $I_k$ and whose $*$ operation is the standard hermitian conjugation. Furthermore, 
$M^h_k(\mathbb{C})$ is  the real linear subspace $M_k(\mathbb{C})$ containing all hermitian $k\times k$ matrices. We assume  familiarity with the basic theory of \ca s, as given in e.g.\ \cite{Dix} or \cite{Lan17,moretti}. 
\subsection{The state space of $M_k(\mathbb{C})$ as a set} 
The state space $S(B)$ of a general unital \ca\ $B$ with unit $I_B$ is defined as the set of linear functionals $\om:B\raw\C$ that satisfy $\om(I_B)=1$ and $\om(a^*a)\geq 0$ for any $a\in B$. It follows that $S(B)\subset B^*$ (the Banach dual of $B$), but $S(B)$ will always be 
equipped with the topology of pointwise convergence, i.e., the weak$\mbox{}^*$-topology (rather than the norm-topology inherited from $B^*$; for finite-dimensional $B$ this difference does not matter, though). In this topology $S(B)$ is a compact convex set. For ${B}= M_k(\mathbb{C})$, regarded as $B=B(\C^k)$, the algebra of (automatically) bounded linear operators on the  Hilbert space $\mathbb{C}^k$,
all  states are normal and hence bijectively correspond with  {\em  density matrices} (i.e.\ positive matrices $\rho$ with unit trace) via
 \begin{align}\omega_\rho(a) = \mathrm{tr}(\rho a) \quad \mbox{for every $a \in {B}$.}\label{id}\end{align}
The set of density matrices on $\C^k$ is denoted by $\mathcal{D}_k$. These form a convex set in their own right,  and hence $\mathcal{D}_k\cong S(M_k(\mathbb{C}))$ via \er{id} as an affine bijection (i.e.\ isomorphism) of  convex sets. We also give $\mathcal{D}_k$ the unique topology making this bijection a homeomorphism and in practice we often identify 
$\mathcal{D}_k$ and $S(M_k(\mathbb{C}))$.

We
 proceed by introducing some  useful  coordinate systems on $\mathcal{D}_k$ \cite{BZ06,BMMP}. 
\begin{definition}
{\em A {\bf parametrization} $(\mathcal{Q}_k,F_k)$ of   $\mathcal{D}_k$ consists of:
\begin{itemize}
\item[(a)]  a parameter set $\mathcal{Q}_k \subset \mathbb{R}^m$, where $m$ depends on $k$, i.e., $m = m(k)$;
\item[(b)] a bijective map $F_k : \mathcal{Q}_k \to \mathcal{D}_k$. 
\end{itemize} 
The parametrization is said to be {\bf affine} if it is (the restriction to $\mathcal{Q}_k$ of) an affine map  with respect to the natural real linear space structures
of $\mathbb{R}^m$ and $M^h_k(\mathbb{C})$.
\hfill $\blacksquare$}
\end{definition}
\begin{remark}\label{remcoord1} {\em The inverse map $F_k^{-1} :  \mathcal{D}_k \to \mathcal{Q}_k \subset \mathbb{R}^{m(k)}$ 
defines a (global) coordinate system on $\mathcal{D}_k$, though in a somewhat extended sense compared to the standard definition  for smooth 
manifolds (with boundary) when $k>2$. This is because, as we shall see shortly, $\mathcal{Q}_k$ has a more complicated structure than an
 open set possibly bounded by an embedded submanifold of $\mathbb{R}^{m(k)}$. \hfill $\blacksquare$}
\end{remark}
\noindent Clearly, the case $k=1$ is trivial, because $D_1 = \{1\}$. Therefore, in what follows we assume $k \geq 2$. We start with the simplest and simultaneously physically most relevant case $k=2$, which will later be applied to the Curie--Weiss model. 
\subsection{Smooth structure of the state space of $M_k(\mathbb{C})$}\label{kgeq2} 
We start with $k=2$. The {\em Pauli matrices} $\sigma_1,\sigma_2,\sigma_3$ together with the identity $I_2$ form a complex basis of the 
complex vector space $M_2(\mathbb{C})$, and a real basis of $M_2^h(\C)$, i.e.
\begin{align}
    a=\half(x_0I+x_1\sigma_1+x_2\sigma_2+x_3\sigma_3),
\end{align}
for  any $a\in M_{2}^{h}(\mathbb{C})$, where
 $x_j\in\R$ ($j=0,1,2,3$). Then $a$ is a density matrix, i.e.\ $a\in \mathcal{D}_2\subset  M_{2}^{h}(\mathbb{C})$, iff $x_0=1$ and $\mathbf{x}= (x_1,x_2,x_3)$ lies in the parameter set 
\begin{align}
    \mathcal{Q}_2=\{\mathbf{x}=\in\mathbb{R}^3 \mid  |\mathbf{x}|\leq 1 \}=B^3,
\end{align}
 the  
closed unit ball in $\mathbb{R}^3$. The corresponding map $F_2:\mathcal{Q}_2\to \mathcal{D}_2$ is given by
\begin{align}
    F_2(\mathbf{x})=\half I_2+\half \sum_{j=1}^3x_j\sigma_j.\label{F2sigma}
\end{align}
By construction, this map is onto $\mathcal{D}_2$, and is affine. An elementary argument based on the identity
$\mathrm{tr}(\sigma_k\sigma_l)= 2\delta_{kl}$ shows that  $F_2$ is also injective. Hence $(\mathcal{Q}_2,F_2)$ is an affine  parametrization of $\mathcal{D}_2$ with $m=3=k^2-1$, for $k=2$.
The key to generalizing this construction to $k>2$ lies in the fact that the anti-hermitian traceless matrices 
$(i\sigma_1, i\sigma_2, i\sigma_3)$ form a  basis of the Lie algebra  $\mathfrak{s\mathfrak{u(2)}}$ of the Lie group $SU(2)$; adding $iI_2$ gives a basis  $(iI_2,i\sigma_1, i\sigma_2, i\sigma_3)$ of  the Lie algebra  $\mathfrak{u(2)}$ of the Lie group $U(2)$. 

Similarly,  for $k\geq 2$ every $\rho\in\mathcal{D}_k$ is hermitian and hence it can be written as
\begin{align}
    \rho=\frac{1}{k}I_k+\sum_{j=1}^{k^2-1}x_j{b}_j,\label{generaldensitymatrix}
\end{align}
where $x_j\in\R$ and $b_j=iT_j$, for some basis
  $(T_j)_{j=1,\ldots, k^2-1}$    of
the Lie algebra  $\mathfrak{s\mathfrak{\mathfrak{\mathfrak{u(k)}}}}$ of  $SU(k)$, consisting of all traceless anti-hermitian $k\times k$ complex matrices, so that 
\begin{align}
b_{j}^{*}=b_j, \ \ \ \ \mathrm{tr}(b_j)=0, \quad (j = 1,\ldots, k^2-1). \label{proplamda}
\end{align}
Since the $T_j$ are a basis of $\mathfrak{s\mathfrak{\mathfrak{\mathfrak{u(k)}}}}$ as a vector space, as usual we  also have
\begin{align}\label{commrel}
[T_r, T_s]  =  \sum_{l=1}^{k^2-1}C_{rs}^l T _l; & &
[b_r, b_s]  = i \sum_{l=1}^{k^2-1}C_{rs}^l b_l,
\end{align}
for some real constants $C_{rs}^l$ antisymmetric in the lower indices and satisfying the  Jacobi identity.
The second part of \er{proplamda} guarantees $\mathrm{tr}(\rho)=1$ in \er{generaldensitymatrix}, but to turn $\rho$ 
into a density matrix the real  numbers $x_1,\ldots, x_{k^2-1}$ must also be  constrained in order that $\rho \geq 0$.  As for $k=2$, this  defines a set  $\mathcal{Q}_k \subset \mathbb{R}^{k^2-1}$ which we use to construct an affine parametrization of $\mathcal{D}_k$ based on (\ref{generaldensitymatrix}). For the moment we assume that $\mathcal{Q}_k$ has been defined  that way, so that the map $F_k$ (\ref{FK}) below  is surjective.

\noindent Compactness of the Lie group  $SU(k)$ implies that the matrices $({b}_j)$, which so far merely satisfy \er{proplamda}, can be 
chosen so as to also satisfy\footnote{With this choice of the normalization, for $k=2$, we find $b_j= 2^{-1/2}\sigma_j$ and also the coordinates $x_j$ in (\ref{FK}) below  correspond to $2^{-1/2}x_j$ in (\ref{F2sigma}).}
\begin{align}
\mathrm{tr}(b_i b_j)=\delta_{ij}. \label{proplambda2}
\end{align}
From (\ref{proplambda2}) and the same argument as for $k=2$, it follows that the surjective map 
 \begin{align} F_k : \mathcal{Q}_k \ni (x_1,\ldots x_{k^2-1})
 \mapsto \frac{1}{k}I_k+\sum_{j=1}^{k^2-1}x_j{b}_j \in \mathcal{D}_k\label{FK},
\end{align}
is also  injective. Indeed, multiplying both sides of (\ref{generaldensitymatrix}) with $b_i$, taking the trace, and using (\ref{proplambda2}) and the second identity in (\ref{proplamda}),
the inverse of $F_k$ reads
\begin{align}
    F_k^{-1}(\rho)=\left(\mathrm{tr}(\rho{b}_1), \ldots, \mathrm{tr}(\rho b_{k^2-1})\right), \quad \rho\in\mathcal{D}_k.
\end{align}
In terms of the state $\omega\in S(M_k(\mathbb{C}))$ related to the density matrix $\rho$, 
this gives an explicit coordinatization $\om\mapsto \left(x_1(\omega), \ldots, x_{k^2-1}(\omega)\right)$
of the former, given by
\begin{align}\label{Fkinv}
x_j(\om) =\omega(b_j)= \mathrm{tr}(\rho b_j) \:\:\: (j=1, \ldots, k^2-1).
\end{align}
To find $\mathcal{Q}_k$ more explicitly, we note that the  eigenvalues of $\rho \in \mathcal{D}_k$ are the roots $\lambda \in \mathbb{R}$ of the characteristic
polynomial $\text{det}(\lambda I_k-\rho)$, which has a unique representation 
\begin{align}
    \text{det}(\lambda I_k-\rho)=\sum_{j=1}^k(-1)^ja_j\lambda^{k-j}, \ \ a_0=1.\label{determinant}
\end{align}
Here the coefficients $a_j$ are uniquely determined by the choice of the  generators $b_j$ and are polynomials in the parameters $\mathbf{x} = (x_1,\ldots, x_{k^2-1})$, and hence they define continuous functions $a_j=a_j (\mathbf{x})$ for  $\mathbf{x}\in\mathbb{R}^{k^2-1}$. If $\lambda_1,\ldots,\lambda_k$ denote the roots of  $\text{det}(\lambda I_k-\rho)$, we obviously have
\begin{align}
    \sum_{j=1}^k(-1)^ja_j\lambda ^{k-j}=\prod_{j=1}^k(\lambda-\lambda_j).
\end{align}
From this, the characterization of the non-negativity of the eigenvalues follows:
\begin{align}
\lambda_j\geq 0 \ \ (j=1,...,k) \ \ \text{if and only if}\ \  a_j\geq 0 \ \ (j=1,...,k).
\end{align}
By definition,  $\mathcal{Q}_k$ is then the following subset in $\mathbb{R}^{k^2-1}$:
\begin{align}
   \mathcal{Q}_k =\{\mathbf{x}\in\mathbb{R}^{k^2-1}\ |\ \ a_j(\mathbf{x})\geq 0, \ j=1,...,k\}. \label{thesetQk}
\end{align}
As the  intersection of  closed sets (note that the maps $a_j$ are continuous), $\mathcal{Q}_k$ is closed. Also note
 that  $\mathcal{Q}_k$ has non-empty interior, because the set
\begin{align}
  \{\mathbf{x}\in\mathbb{R}^{k^2-1}\ |\ \ a_j(\mathbf{x})> 0, \ j=1,...,k\} \subset \mathcal{Q}_k
\end{align}
is open as a finite intersection of open sets,  and is not empty since it contains the  density matrix $\rho=I_k$, whose coordinates are $\mathbf{x}(I) = (0,0,\ldots, 0)$,
so that  $a_j(0,0,\ldots, 0) = \frac{1}{k^j}\binom{k}{j} >0$ for all $j$.  We now also show that $\mathcal{Q}_k$ is bounded in $\mathbb{R}^{k^2-1}$. Since $\rho \in \mathcal{D}_k$ is hermitian, $\rho \geq 0$, and $\mathrm{tr}(\rho)=1$,  we have 
\begin{equation}
\mathrm{tr}(\rho^2) \leq \mathrm{tr}(\rho)=1,
\end{equation}
as can be seen e.g.\ by diagonalizing $\rho$.
Representing $\rho$ as in  (\ref{generaldensitymatrix}) and 
taking advantage of  (\ref{proplambda2}) and the second identity in (\ref{proplamda}), the condition $\mathrm{tr}(\rho^2) \leq 1$ can be rephrased in a way that makes  boundedness of $\mathcal{Q}_k$ obvious, viz.\footnote{
The pure states are exactly those points in $\mathcal{Q}_k$  that saturate this inequality, since their density matrices satisfy $tr\rho^2 = tr \rho =1$. The pure states form $\partial_e\mathcal{Q}_k\cong \mathbb{CP}^{k-1}$ with  canonical (Fubini--Study)  Poisson structure, cf.\ \S\ref{PoissonMk} below. 
Strict deformation quantization of complex projective spaces is well known, for example as a special case of the constructions in \cite{BMS94} or \cite[\S8.1]{Lan17}. \label{fnS2}}
\begin{align}
 \frac{1}{k} + \sum_{j=1}^{k^2-1}|x_j|^2 \leq  1\quad \mbox{if $\mathbf{x} \in \mathcal{Q}_k$.}
\end{align}
Therefore, with $\mathcal{Q}_k$ defined as in (\ref{thesetQk}) and $F_k$ defined in (\ref{FK}), the pair $(\mathcal{Q}_k,F_k)$
 is an affine  parametrization for density matrices $\rho\in\mathcal{D}_k$ with $m=k^2-1$, and  (\ref{Fkinv}) defines a global coordinate system over $\mathcal{D}_k\equiv S(M_k(\mathbb{C}))$ in the sense of Remark \ref{remcoord1}.  Coming from an affine map, this coordinate system preserves the convex structure of  $S({B})$, so that  
  $\mathcal{Q}_k$ is a compact convex subset of $\mathbb{R}^{k^2-1}$ with  non-empty interior.
To conclude this section, few remarks about the differentiable structure of $\mathcal{Q}_k$ are in order. 
We have seen that  $\mathcal{Q}_2\cong B^3$ is a $3$-dimensional manifold with boundary $\partial B^3\cong S^2$ (the two-sphere), where the topological boundary also coincides with the  extreme boundary  $\partial_e \mathcal{Q}_2$  as defined in convexity theory (which defines the pure states).  However, this simple picture is misleading, since for  $k>2$ the set $\mathcal{Q}_k$ is no longer  a (smooth) manifold with boundary \cite{GKM05}, as the boundary is not a manifold but a stratified space \cite{Pflaum}.
Indeed, for $k>2$, we have the following situation:
\begin{itemize} \item[(1)]  Under the isomorphism $\mathcal{Q}_k\cong \mathcal{D}_k$ 
the interior $\mathrm{int}(\mathcal{Q}_k)$  of $\mathcal{Q}_k$ corresponds to the rank-$k$ density matrices and  is  a connected $k^2-1$ dimensional smooth manifold. Points in the interior precisely correspond to {\em faithful states} on $M_k(\mathbb{C})$.\footnote{A state $\omega$ on a $C^*$algebra  ${B}$  is called {\em faithful} if
$\omega(a^*a)=0$ implies $a=0$, for $a\in {B}$.}
 \item[(2)] The topological boundary $\partial \mathcal{Q}_k$ now differs from the extreme boundary  $\partial_e \mathcal{Q}_k$:
 \begin{itemize}
\item  $\partial \mathcal{Q}_k$ is the disjoint union of $k-1$ smooth embedded submanifolds $\mathcal{Q}_k^{(l)}$ of $\mathbb{R}^{k^2-1}$, 
 where $l=1, \ldots, k-1$, and $\mathcal{Q}_k^{(l)}$ contains all points corresponding to  density matrices with rank $l$ (rank $l=k$ corresponding to the interior).
 \item $\partial_e \mathcal{Q}_k=\mathcal{Q}_k^{(1)}\subset \partial \mathcal{Q}_k\subset \mathcal{Q}_k$ corresponds to the pure state space on $M_k(\C)$.
 \end{itemize}
   \item[(3)]  
Every point of $\partial \mathcal{Q}_k$ is a limit point of $\mathrm{int}(\mathcal{Q}_k)$ and clearly 
$\mathcal{Q}_k=\mathrm{int}(\mathcal{Q}_k)\cup \partial \mathcal{Q}_k$.
\end{itemize}
Finally, all properties of  $\mathcal{Q}_k$ we established are independent of the choice of the basis $\{ib_j\}_{j=1,\ldots, k^2-1}$ used to define $(\mathcal{Q}_k,F_k)$, as one easily proves: each different choice of basis just defines a different global coordinate system compatible with the linear  structure, the topology, and  the  differentiable structures involved.
In that sense, these properties are  {\em intrinsic}, and eventually come from $\mathcal{D}_k\cong  S(M_k(\mathbb{C}))$.
\subsection{Poisson structure of state space of $M_k(\mathbb{C})$}\label{PoissonMk}
We now show that the state space $X_k$, so far realized in three different ways as 
\begin{equation}
X_k=S(M_k(\mathbb{C}))\cong \mathcal{D}_k\cong  \mathcal{Q}_k, \label{four}
\end{equation}
 carries a canonical Poisson structure \cite{Bona,DW,Lan17}.\footnote{A Poisson bracket $\{\cdot, \cdot\}$ on a commutative algebra $A$ is a Lie bracket satisfying the Leibniz rule $\{a,bc\}=\{a,b\}c+\{a,c\}b$, or: for each $a\in A$ the (linear) map $\delta_a:A\raw A$ defined by $\delta_a(b)=\{a,b\}$ is a derivation, i.e.\ $\delta_a(bc)=\delta_a(b)c+\delta_a(c)b$. We take $A=C^{\infty}(X_k)$ with pointwise mutliplication.}
 If $X_k$ were a manifold, this structure would be defined as a  Poisson bracket on $C^{\infty}(X_k)$, but we have just seen that $X_k$ is not even a manifold with boundary. We circumvent this problem by recalling 
 \begin{equation}
 \mathcal{Q}_k\subset \R^{k^2-1}, \label{QRn}
\end{equation}
with $\dim(\mathrm{int}(\mathcal{Q}_k))=k^2-1$, as shown in the previous section,  and hence we simply define $f\in \cin(\mathcal{Q}_k)$ iff $f$ is the restriction of some $\tilde{f}\in\cin(\R^{k^2-1})$.
 
We also recall that  if $\mathfrak{g}$ is any (finite-dimensional) Lie algebra, then the dual space  $\mathfrak{g}^*$ has a canonical Poisson structure coming from the Lie bracket on $\mathfrak{g}$ \cite{MR94}. The Poisson bracket is completely defined 
 by its value on linear functions on $\mathfrak{g}^*$; each $X\in \mathfrak{g}$ defines such a function $\hat{X}$ through
 $\hat{X}(\theta)=\theta(X)$, where $\theta\in\mathfrak{g}^*$, and
 \begin{equation}
\{\hat{X},\hat{Y}\}=\widehat{[X,Y]}. \label{LPB}
\end{equation}
If $(T_1, \ldots, T_n)$ is a basis of  $\mathfrak{g}$ ($n=\dim(\mathfrak{g})$) with structure constants $C_{ab}^c$ given by 
\begin{equation}
[T_a,T_b]=\sum_c C_{ab}^cT_c, \label{TC}
\end{equation}
then one has an identification $\mathfrak{g}^*\cong\R^n$ in that $\mathbf{x}=(x_1, \ldots, x_n)\in\R^n$ corresponds to
$\theta=\sum_a x_a\omega^a$, where $(\omega^a)$ is the dual basis to $(T_a)$ (i.e., $\omega^a(T_b)=\delta^a_b$), so that
\begin{equation}\label{pfg}
    \{f,g\}(\mathbf{x})=\sum_{a,b,c=1}^n C_{ab}^cx_c
    \frac{\partial f(\mathbf{x})}{\partial x_a}\frac{\partial g(\mathbf{x})}{\partial x_b}.\end{equation}
    In particular,  the coordinate functions $f(\mathbf{x})=x_a$ reproduce the Lie bracket, i.e.,
    \begin{align}\label{xx}
\{x_a,x_b\} = \sum_{c=1}^{k^2-1} C_{ab}^c x_c. 
\end{align}
Applying this to $\mathfrak{g}=\mathfrak{s\mathfrak{\mathfrak{\mathfrak{u(k)}}}}$, so  $n=k^2-1$, see  \er{commrel}, then gives a Poisson structure on $\R^{k^2-1}$ and hence, by restriction, on $\mathcal{Q}_k$:
\begin{definition} \label{PBSB}{\em  The {\bf Poisson bracket} of $f,g \in C^\infty(\mathcal{Q}_k)$  is given by
\begin{equation}
\{f,g\} = \{\tilde{f},\tilde{g}\}_{|_{\mathcal{Q}_k}},\label{PBQk}
\end{equation}
where $\tilde{f},\tilde{g} \in C^\infty(\R^{k^2-1})$ are arbitrary extensions of $f$ and $g$ respectively, cf.\ \er{QRn}, and 
the Poisson bracket $\{\tilde{f},\tilde{g}\}$ on $C^\infty(\R^{k^2-1})$ is defined by (\ref{pfg}) for $\mathfrak{g}=su(n)$.
   \hfill $\blacksquare$}
\end{definition}
This definition is meaningful because of the following facts:
\begin{enumerate}
\item The bracket $\{f,g\}$ does not depend on the choice of the extensions $\tilde{f},\tilde{g} \in C^\infty(\R^{k^2-1})$, 
  because every point of $\partial \mathcal{Q}_k$ is a limit point of the interior of $\mathcal{Q}_k$.
 \item The function $\{f,g\}$ trivially lies in $C^\infty(\mathcal{Q}_k)$, which by definition means that it has a smooth extension to $\R^{k^2-1}$, since 
$\{\tilde{f},\tilde{g}\}$ is such an extension. 
\item  The bracket does not depend on the choice of the basis $\{T_j\}_{j=1,\ldots, k^2-1}$ of $\mathfrak{s\mathfrak{\mathfrak{\mathfrak{u(k)}}}}$ (with $b_j=iT_j$), since a linear change of basis induces  a change in the structure constants $C_{ab}^c$ in \er{TC} \emph{and} a linear change of the coordinates in $\R^{k^2-1}$ coming from identifying $\mathfrak{s\mathfrak{\mathfrak{\mathfrak{u(k)}}}}\cong \R^{k^2-1}$, which cancel out in \er{pfg} and hence in \er{PBQk}. 
\end{enumerate}
The last point can also be seen from the  more intrinsic form
 the  bracket takes in terms of the other two entries in \er{four}. First, for the density matrices $\mathcal{D}_k$ we have
\begin{equation}
 \mathcal{D}_k\subset M^h_k(\mathbb{C})_1\cong  M^h_k(\mathbb{C})_0=i \mathfrak{s\mathfrak{\mathfrak{\mathfrak{u(k)}}}}\cong i \mathfrak{s\mathfrak{\mathfrak{\mathfrak{u(k)}}}}^*,\label{five}
\end{equation}
where $M^h_k(\mathbb{C})_t$
is the  space of hermitian $k\times k$ matrices $\rho$ with  trace $t$. The first inclusion is given by \er{generaldensitymatrix}, the subsequent isomorphism is given by $(I_k/k)+b\mapsto b$, where $b\in  M^h_k(\mathbb{C})_0$, and the last isomorphism $\mathfrak{s\mathfrak{\mathfrak{\mathfrak{u(k)}}}}\cong \mathfrak{s\mathfrak{\mathfrak{\mathfrak{u(k)}}}}^*$ comes from (minus) the (negative definitie) Cartan--Killing inner product on $\mathfrak{s\mathfrak{\mathfrak{\mathfrak{u(k)}}}}$, which is given by
\begin{equation}
 -B(X,Y)=-2k\: \mathrm{tr}(XY)=2k\: \mathrm{tr}(X^*Y)= 2k\langle X,Y\rangle_{HS},
\end{equation}
where the right-hand side is the Hilbert--Schmidt inner product on $M_k(\C)$. If we now equip $\mathcal{D}_k$ with a differentiable structure through the last isomorphism in \er{four}, as detailed in the previous section, and 
define  $f\in \cin( \mathcal{D}_k)$ iff $f$ is the restriction of some $\tilde{f}\in\cin(M^h_k(\mathbb{C})_1)$,  or, by \er{five}, iff it is the restriction of some  $\tilde{f}\in\cin(i\mathfrak{s\mathfrak{\mathfrak{\mathfrak{u(k)}}}})$, and transfer the Poisson structure on $\mathfrak{s\mathfrak{\mathfrak{\mathfrak{u(k)}}}}^*$ to $\mathfrak{s\mathfrak{\mathfrak{\mathfrak{u(k)}}}}$ through \er{five}, then we clearly obtain an intrinsic Poisson structure on $ \mathcal{D}_k$, essentially given also by \er{LPB}.

Finally, perhaps as the mother of all of the above, for any unital finite-dimensional \ca\ $B$ (and with due modifications,  even for infinite-dimensional ones), the state space $S(B)$ has a natural structure as a Poisson manifold (with stratified boundary, as above). The Poisson bracket is most easily written down through the isomorphism $B^h\cong A(S(B))$ of real Banach spaces, where $B^h$ is the set of hermitian (= self-adjoint) elements of $B$ and for any compact convex set $K$, $A(K)$ is the space of continuous real-valued affine functions on $K$, equipped with the supremum-norm. This isomorphism is given by $b\mapsto\hat{b}$, where $b\in B^h$ and $\hat{b}\in A(S(B))$ is given by  $\hat{b}(\omega) = \omega(b)$, and, as in \er{LPB}, the Poisson bracket is fully defined by
\begin{equation}
    \{\Hat{a},\Hat{b}\} = i\widehat{[a,b]}.\label{Poissonlin}
\end{equation}
The relationship with the previous constructions may be inferred from the inclusion
\begin{equation}
  S(M_k(\mathbb{C}))\subset M^h_k(\mathbb{C})^*_1\cong  (M^h_k(\mathbb{C})_0)^*=(i\mathfrak{s\mathfrak{\mathfrak{\mathfrak{u(k)}}}})^*,
\end{equation}
where $M^h_k(\mathbb{C})^*_1$ is the set of linear functionals $\varphi: M_k(\mathbb{C})\raw \C$ that are hermitian ($\varphi(a^*)=\overline{\varphi(a)}$) and normalized ($\varphi(I_k)=1$); the remainder is obvious from \er{five}.
\section{Strict deformation quantization of $S(M_k(\mathbb{C}))$}\label{Sdf}
In this section we state and prove our first main theorem.\footnote{Some ideas in the proof were inspired by techniques in \cite{Bona,DW,RW}, as rewritten in terms of continuous bundles of \ca s in \cite[Ch.\ 8]{Lan17}.
The relationship between the strict deformation quantization of $X_k$ (constructed below) and of its extreme boundary $\mathbb{CP}^{k-1}$ (cf.\  footnote \ref{fnS2} and \cite{BMS94,Lan17}) is unclear; even for $k=2$ the fiber algebras $A_{1/N}$ are  different, namely $A_{1/N}=M_2(\C)^N$ for $X_2\cong B^3$ and $A_{1/N}=M_{N+1}(\C)$ for $\mathbb{CP}^{1}\cong S^2$. On the other hand, the ground state of the quantum Curie--Weiss model lies in $\C^{N+1}$, see \er{SYmN1}, whereas its classical counterpart(s) lie in $S^2$, so perhaps one should look for embeddings $M_{N+1}(\C)\hookrightarrow M_2(\C)^N$ ``quantizing" $S^2\hookrightarrow B^3$, etc. 
}
We wish to construct a strict deformation quantization of $X_k= S(M_k(\mathbb{C}))$
 according to Definition \ref{def:deformationq}, where the continuous bundle of \ca s is given by the fibers \er{B0} - \er{BN}, with continuity structure as explained before these equations, and  Poisson structure on $X_k$ as defined in
 the previous subsection. We therefore still need to construct:
\begin{enumerate}
\item[(i)]A suitable dense Poisson subalgebra  $\tilde{A}_0$ of $C^\infty({X}_k)$;
\item[(ii)] Quantization maps $Q_{1/N}: \tilde{A}_0  \to M_k(\mathbb{C})^N$, cf.\ \er{QCW}.
\end{enumerate}
  Although the space we quantize is $X_k$, we will (often without comment) use both identifications $X_k\cong 
\mathcal{D}_k$ and $X_k\cong \mathcal{Q}_k$ explained in the previous chapter, the latter equipped with
 the Poisson structure of Definition \ref{PBSB}. As usual, $B=M_k(\C)$.
\subsection{Choice of the Poisson subalgebra   $\tilde{A}_0$}   \label{thespace}
As before, we choose  a basis $\{b_1,...b_{k^2-1}\}$ of $i\mathfrak{s\mathfrak{\mathfrak{\mathfrak{u(k)}}}}$
satisfying (\ref{proplamda}) and \er{commrel}, where $\mathfrak{s\mathfrak{\mathfrak{\mathfrak{u(k)}}}}$ is a \emph{real} vector space. Using {\em complex} coefficients, the hermitian  matrices $(I_k,b_1,...,b_{k^2-1})$ then form 
 a basis of the complex vector space $M_k(\mathbb{C})$. 
We  introduce a  subspace of  $\bigoplus_{M=0}^{\infty} {B}^{M}$ making use of 
the   {\bf symmetrized tensor product} \begin{align} 
a_1\otimes_s \cdots \otimes_s a_N = S_N(a_1\otimes \cdots \otimes a_N),\label{defPS}\end{align} where $S_N$ is defined in \er{defSN} and  we  adopt
 the Einstein summation convention. We define
$Z\subset\bigoplus_{M=0}^{\infty} {B}^{M}$ as the subspace consisting of
all  elements of the form
\begin{align}
z=  c_0I_k\oplus c_1^{j_1}b_{j_1}\oplus c_2^{j_1j_2}b_{j_1}\otimes_sb_{j_2}\oplus...\oplus c_M^{j_1\cdot\cdot\cdot j_M}b_{j_1}
\otimes_s\cdot\cdot\cdot\otimes_sb_{j_M}, \:\:\: (M=0,1,\ldots), \label{zN}
\end{align}
where the coefficients $c_L^{j_1\cdots j_L} \in \mathbb{C}$ are symmetric,
$j_i\in\{1,...,k^2-1\}$, and  $i=1,...,N$.
\begin{remark}  
{\em \begin{itemize}
\item[(1)] The matrices $I_k$ and all of the $b_{j_1}\otimes_s\cdot\cdot\cdot\otimes_sb_{j_N}$, where $j_i\in\{1,...,k^2-1\}$ and  $i=1,...,N$, are linearly independent and form a  basis of $Z$. 
\item[(2)] $Z$ does not depend on the initial choice of the  basis    $\{b_1,...,b_{k^2-1}\}$ of $i\mathfrak{s\mathfrak{\mathfrak{\mathfrak{u(k)}}}}$. \hfill $\blacksquare$
\end{itemize} }\end{remark}
\noindent We  now introduce an important auxiliary  linear map $\chi:Z\to C(S({B}))$,
through which we will construct $\tilde{A}_0$. By linearity, $\chi$ is 
completely defined if, for $\omega \in S({B})$,
\begin{align} \chi(I_k)(\omega) &=1, \:\:\ \mbox{i.e. }\: \chi(I_k)=1_{S(B)};\label{definitionf1} \\
    \chi(b_{j_1}\otimes_s\cdots\otimes_sb_{j_N})(\omega)&=\omega^{N}(b_{j_1}\otimes_s\cdots \otimes_sb_{j_N}) =
 \omega(b_{j_1})\cdots \omega(b_{j_N}).\label{definitionf2}
\end{align}
By definition of weak $*$-topology we have $\chi(z) \in C(X_k)$, since $z \in Z$ 
is a finite sum.
\begin{lemma}\label{lemmainj}
The map $\chi: Z\to C(S({B}))$, is injective, so that in particular all  functionals $1_{S(B)}$ and  $\chi(b_{j_1}\otimes_s\cdots\otimes_sb_{j_N})$ are linearly independent.
\end{lemma}
\noindent For the proof see Appendix \ref{proofs}.
We can now define our Poisson subalgebra as
\begin{align}\Tilde{A}_0 = \chi(Z). \label{tA0}\end{align} 
Then $\Tilde{A}_0$ is a $\|\cdot\|_\infty$ dense subspace of  $C(S({B}))$ by injectivity of $\chi$  and the Stone--Weierstrass theorem (if necessary using the homomorphism $S(B)\cong  \mathcal{Q}_k$, or directly in its \ca ic version). 
Indeed, it follows from (\ref{Fkinv}) and (\ref{definitionf1}) -  (\ref{definitionf2})  that (using the  Einstein summation convention) generic elements of $\Tilde{A}_0$ take the form 
\begin{align} &\chi( c_0I_k\oplus c_1^{j_1}b_{j_1}\oplus c_2^{j_1j_2}b_{j_1}\otimes_sb_{j_2}\oplus\cdots \oplus
 c_M^{j_1\cdots  j_M}b_{j_1}\otimes_s\cdots \otimes_sb_{j_M}) (\omega)\nonumber \\ &=
c_0+ c_1^{j_1}x_{j_1}+ c_2^{j_1j_2}x_{j_1}x_{j_2}+\cdots + c_M^{j_1\cdot\cdot\cdot j_M}x_{j_1}\cdots x_{j_M}. \end{align}
Since (under $X_k\cong  \mathcal{Q}_k$) elements of
$\Tilde{A}_0$ are polynomials, we also have $\Tilde{A}_0 \subset C^\infty(X_k)$, and  using 
Definition \ref{PBSB},  it is also clear that $\Tilde{A}_0$ is a Poisson subalgebra of $\cin(X_k)$.
\subsection{Quantization maps}\label{3.2}  
 We define
$Q_{1/N}: \Tilde{A}_0 \to M_k(\C)^N$ by linear extension of its values on the basis vectors $\chi(I_k)$ and $\chi(b_{j_1}\otimes_s\cdot\cdot\cdot\otimes_sb_{j_L})$ of $\Tilde{A}_0$ ($L \in \mathbb{N}$), (\ref{definitionf1}) -  (\ref{definitionf2}). On those, we define
\begin{align}\label{deformationqunatizaion}
 Q_{1/N}(\chi(b_{j_1}\otimes_s\cdot\cdot\cdot\otimes_sb_{j_L})) &=
\begin{cases}
    S_{L,N}(b_{j_1}\otimes_s\cdot\cdot\cdot\otimes_sb_{j_L}), &\ \text{if} \ N\geq L \\
    0, & \ \text{if} \ N < L,
\end{cases}\\
Q_{1/N}(\chi(I_k)) &= \underbrace{I_k \otimes \cdots \otimes I_k}_{\scriptsize N \: times}. \label{deformationqunatizaion2}
\end{align}
\noindent
\begin{remark}\label{remzz1} {\em Suppose that $z\in Z$ takes the form (\ref{zN}) with not all coefficients $c_M^{j_1\cdots j_M}$ vanishing.
 Then there exists $z_1 \in {B}^M$,  such that 
\begin{align}\label{QQ}
Q_{1/N}(\chi(z))  = S_{M,N}(z_1)\quad \mbox{if $N\geq M$}.
\end{align}
To construct $z_1$ from $z$, it is sufficient to replace every 
summand 
\begin{equation}
c_L^{j_1\cdot\cdot\cdot j_L}b_{j_1}\otimes_s\cdot\cdot\cdot\otimes_sb_{j_L} \in {B}^L 
\end{equation}
in (\ref{zN}) by a corresponding term
\begin{equation}
c_L^{j_1\cdot\cdot\cdot j_L}b_{j_1}\otimes_s\cdot\cdot\cdot\otimes_sb_{j_L} \otimes_s I_k\otimes _s \cdots \otimes_s I_k\in {B}^M, 
\end{equation}
 where the factor $I_k$ occurs $M-L$ times, so that
\begin{align}
z_1= & \: (c_0\underbrace{I_k \otimes_s \cdots \otimes_s I_k}_{\scriptsize M \: times})\oplus (c_1^{j_1}b_{j_1}\otimes_s 
\underbrace{I_k \otimes_s \cdots \otimes_s I_k}_{\scriptsize M-1 \:times})\nonumber \\
&\oplus \cdots\oplus
( c_M^{j_1\cdot\cdot\cdot j_M}b_{j_1}\otimes_s\cdot\cdot\cdot\otimes_sb_{j_M}),\label{zN33}
\end{align}
With $z$ as in (\ref{zN}), where not all $C_M^{j_1\cdots j_M}$ vanish, and  $z_1\in {B}^M$ as in (\ref{zN33}), it  immediately follows from the definition of $Q_{1/N}$
that (\ref{QQ}) holds. \hfill $\blacksquare$}
\end{remark}
\begin{theorem}\label{mainfirst} 
Let $S(M_k(\C))$ be the state space of $M_k(\C))$.
The following data give a strict  deformation quantization of $S( M_k(\C))$ in the sense of Definition \ref{def:deformationq}:
  \begin{enumerate}
\item The continuous bundle of \ca s over the base space \er{defI} with fibers \er{B0} - \er{BN}, with continuity structure as explained before these equations;
\item  The  (canonical) Poisson structure on  $S( M_k(\C))$ defined in \S\ref{PoissonMk}:
\item The dense Poisson subalgebra $\Tilde{A}_0\subset \cin(S( M_k(\C)))\subset A_0$  defined by (\ref{tA0});
\item  The maps $Q_{1/N}: \Tilde{A}_0 \to M_k(\C)^N$ defined by linear extension of \er{deformationqunatizaion} - \er{deformationqunatizaion2}.
\end{enumerate}
\end{theorem}
\noindent
\begin{proof} 
For each $a_0\in \tilde{A}_0$, the following map is a continuous section of the bundle:
\begin{align}
    &0\to a_0\\
    &1/N\to Q_{1/N}(a_0) \ \ \  (N>0).
\end{align}
This is true because continuous sections are given by (quasi) symmetric sequences and the sequence 
of the $Q_{1/N}(a_0)$ defined in 
(\ref{deformationqunatizaion}) - \eqref{deformationqunatizaion} is even symmetric due to (\ref{QQ}). The only nontrivial part of the proof is the  Dirac-Groenewold-Rieffel condition
\begin{align}\label{DGR}
    \lim_{N\to \infty}\left|\left|iN[Q_{1/N}(f),Q_{1/N}(g)]-Q_{1/N}(\{f,g\})\right|\right|_{N}=0,
\end{align}
where $h,g\in\tilde{A}_0$.
Since both terms in the norm in (\ref{DGR}) are bilinear in $f$ and $g$, and the case where  $f$ or $g$ equals $1_{S( M_k(\C))}$ is trivially satisfied (since $Q_{1/N}(1_{S( M_k(\C))})$ is the unit operator in $B^N$),  
 it is is sufficient to prove this for basis elements
of $\tilde{A}_0$:
\begin{align}\label{hg}
    f= \chi(b_{i_1}\otimes_s\cdot\cdot\cdot\otimes_sb_{i_M}) ,\quad g= \chi(b_{j_1}\otimes_s\cdot\cdot\cdot\otimes_sb_{j_L}).
\end{align}
For these functions, we have by definition
\begin{align}\label{hgx}
    f(x_1,\ldots, x_{k^2-1})= x_{i_1}\cdots x_{i_M}; & &  g(x_1,\ldots, x_{k^2-1})= x_{j_1}\cdots x_{j_L}.
\end{align}
As a consequence of (\ref{pfg}) and (\ref{xx}), we obtain
\begin{align}
\{f, g\} =& \left(\sum_l C_{i_1j_1}^l x_l  x_{i_2} \cdots x_{i_M}  x_{j_{2}}  \cdots  x_{j_{L}}  \nonumber\right. 
\nonumber \\
& + \sum_l C_{i_1j_2}^l x_l x_{i_2} \cdots  x_{i_M}   x_{j_{1}} 
x_{j_3}  \cdots x_{j_{L}}  \nonumber \\
&+\cdots +  \left.\sum_l C_{i_Mj_L}^l  x_l x_{i_1} x_{j_1} \cdots x_{i_{M-1}} x_{j_{1}} 
x_{j_3}  \cdots 
x_{j_{L-1}}  \right)\nonumber,
\end{align}
where all possible Poisson brackets  $\{x_{i_l}, x_{j_m}\}= \sum_l C_{i_lj_m}^l x_l$ are considered for $l=1,\ldots, M$, $m=1,\ldots, L$.
From this  expression we compute $Q_{1/N}(\{f,g\})$ in (\ref{DGR}):
\newpage
\begin{align}
Q_{1/N}(\{f,g\}) &=
S_{M+L-1, N}\left( \sum_l C_{i_1j_1}^l b_l \otimes b_{i_2} \otimes \cdots \otimes b_{i_M} \otimes  b_{j_{2}} 
\otimes  \cdots \otimes b_{j_{L}}  \nonumber\right. \\
 &+  \sum_l C_{i_1j_2}^l b_l \otimes b_{i_2} \otimes \cdots \otimes b_{i_M} \otimes  b_{j_{1}} 
\otimes b_{j_3}\otimes   \cdots \otimes b_{j_{L}}  \nonumber \\
&+\cdots +\nonumber \\
&+  \left.\sum_l C_{i_Mj_L}^l  b_l \otimes b_{i_1} \otimes \cdots \otimes b_{i_{M-1}} \otimes  b_{j_{1}} 
\otimes b_{j_3}\otimes   \cdots \otimes b_{j_{L-1}}  \right)\label{Q1Nhg}.
\end{align}
Let us  pause to analyse the remaining term  in the norm in (\ref{DGR}), more precisely,  
\begin{equation}
[Q_{1/N}(f), Q_{1/N}(g)]=[S_{M,N}(f^{-1}(f)), 
S_{L,N}(f^{-1}(g))].\end{equation}
\begin{lemma}\label{lemmacommSN}
Consider elements $a_1 \otimes \cdots \otimes a_N$ and $b_1 \otimes \cdots \otimes b_N$ of ${B}^N$.
Then
\begin{align}\label{formula}
&\left[S_N(a_1 \otimes\cdots \otimes a_N), S_N(a'_1 \otimes \cdots \otimes a'_N)\right]\nonumber \\  &= \frac{1}{N!} \sum_{\pi \in {\cal P}(N)} 
\left(S_N\left( a_1 a'_{\pi(1)}\otimes 
\cdots   \otimes a_N a'_{\pi(N)}\right) -  S_N\left(a'_{\pi(1)}a_1\otimes 
\cdots  \otimes  a'_{\pi(N)}a_N\right) \right).
\end{align}
\end{lemma}
\begin{proof} See Appendix \ref{proofs}.
\end{proof}
\noindent Let us  to evaluate the commutator
\begin{equation}
[Q_{1/N}(f), Q_{1/N}(g)]=[S_{M,N}(f^{-1}(f)), S_{L,N}(f^{-1}(g))]
\end{equation}
 in the concrete case  from where $f$ and $g$ are given by
(\ref{hg}). Then the relevant sequences in ${B}^N$ are   
\begin{align}\label{aaaa} a_1 \otimes \cdots \otimes a_N =  
b_{i_1} \otimes\cdots \otimes b_{i_M}\otimes I_k \otimes
 \cdots \otimes I_k;\\
  a'_1 \otimes \cdots \otimes a'_N = b_{j_1} \otimes \cdots \otimes b_{j_L}\otimes I_k \otimes \cdots \otimes I_k,
  \label{aaaa2} \end{align}
since, from (\ref{hg}) and the  definition of $S_{P,N}$, i.e.,
\begin{align}S_{M,N}(f^{-1}(f))&= S_N(a_1 \otimes \cdots \otimes a_N );\\
S_{L,N}(f^{-1}(g))&= S_N(a'_1 \otimes \cdots \otimes a'_N ).\label{SSaabb}\end{align}
Keeping (\ref{aaaa}) and (\ref{aaaa2}), for $L \leq M$ fixed and large $N$ there are three types of  permutations $\pi \in {\cal P}(N)$ 
classified by  the following distinct properties of the  elements $$a_1 a'_{\pi(1)}\otimes 
\cdots   \otimes a_N a'_{\pi(N)} \hspace{20pt} \mbox{or }  \hspace{20pt} a'_{\pi(1)}a_1\otimes 
\cdots   \otimes a'_{\pi(N)}a_N)
$$ in the right-hand side of (\ref{formula}): 
\newpage
\begin{trivlist}
\item[I.] For every  factor $a_l a'_{\pi(l)}$ (resp. $a'_{\pi(l)}a_l$), either  $a_l=I_k$ or $a'_{\pi(l)}=I_k$ (or both);
\item[II.] There is exactly one  factor $a_l a'_{\pi(l)}$  (resp.\ $a'_{\pi(l)}a_l$)  with both  $a_l \neq I_k$ and $a'_{\pi(l)} \neq I_k$;
\item[III.]\hspace{-7pt} There is more than one factor $a_l a'_{\pi(l)}$  (resp.\ $a'_{\pi(l)}a_l$)  with both  $a_l \neq I_k$ and $a'_{\pi(l)} \neq I_k$.
\end{trivlist}
We accordingly decompose ${\cal P}(N)$  into three pairwise disjoint parts as 
 \begin{align}{\cal P}(N)= {\cal P}(N)_I \cup {\cal P}(N)_{II} \cup {\cal P}(N)_{III}.\label{PPP}\end{align}
This decomposition induces a corresponding  decomposition of $[Q_{1/N}(f), Q_{1/N}(g)]$
arising from the right-hand side of (\ref{formula}), taking (\ref{SSaabb}) into account, where a sum over $\pi \in {\cal P}(N)$ shows up. We symbolically write this decomposition as
\begin{align}
[Q_{1/N}(f), Q_{1/N}(g)] & = [Q_{1/N}(f), Q_{1/N}(g)]_I\nonumber \\ &+[Q_{1/N}(f), Q_{1/N}(g)]_{II}\nonumber \\ &+[Q_{1/N}(f), Q_{1/N}(g)]_{III}.
\end{align}
It should be clear that 
\begin{equation}
\sum_{\pi \in {\cal P}(N)_I} 
\left(S_N\left( a_1 a'_{\pi(1)}\otimes 
\cdots   \otimes a_N a'_{\pi(N)}\right) -  S_N\left(a'_{\pi(1)}a_1\otimes 
\cdots  \otimes  a'_{\pi(N)}a_N\right) \right)=0,
\end{equation}
so that $[Q_{1/N}(f), Q_{1/N}(g)]_I =0$. The term
 $[Q_{1/N}(f), Q_{1/N}(g)]_{II}$  is proportional to 
 \begin{align}&
\sum_{\pi \in {\cal P}(N)_{II}} 
\left(S_N\left( a_1 a'_{\pi(1)}\otimes 
\cdots   \otimes a_N a'_{\pi(N)}\right) -  S_N\left(a'_{\pi(1)}a_1\otimes 
\cdots  \otimes  a'_{\pi(N)}a_N\right) \right) \nonumber \\
= &\sum_{\pi \in {\cal P}(N)_{II}} S_N\left( [a_1, a'_{\pi(1)}]\otimes a_2a'_{\pi(2)} \otimes 
\cdots   \otimes a_N a'_{\pi(N)}\right)\nonumber \\ 
+ &\sum_{\pi \in {\cal P}(N)_{II}} S_N\left( a_1 a'_{\pi(1)}\otimes [a_2,a'_{\pi(2)}] \otimes 
\cdots   \otimes a_N a'_{\pi(N)}\right) + \cdots\nonumber \\
+ &\sum_{\pi \in {\cal P}(N)_{II}} S_N\left( a_1 a'_{\pi(1)}\otimes a_2a'_{\pi(2)} \otimes 
\cdots   \otimes [a_N, a'_{\pi(N)}]\right), \label{sum2}
\end{align}
where, for each fixed $\pi \in {\cal P}(N)_{II}$, there is exactly one pair $a_l, a'_{\pi(l)}$ with both $a_l \neq I_k$ and $a'_{\pi(l)} \neq I_k$
(so that  at most  the  commutator $[a_l, a'_{\pi(l)}]$ does not vanish and  the overall sum above contains at most  one non-vanishing summand depending on  
$\pi$).

 Let us focus on the first summand in the right-hand side of (\ref{sum2}) and consider the generic summand therein for some $\pi \in {\cal P}(N)_{II}$, namely
\begin{equation}
 S_N\left( [a_1, a'_{\pi(1)}]\otimes a_2a'_{\pi(2)} \otimes 
\cdots   \otimes a_N a'_{\pi(N)}\right),\end{equation}
 where we assume, to avoid a trivial case, that $a_1 \neq I_k$ and $a'_{\pi(1)} \neq I_k$. 
Recall that  
\begin{align}
a_1 \otimes \cdots \otimes a_N &= b_{i_1} \otimes \cdots \otimes b_{i_M}\otimes I_k \otimes \cdots \otimes I_k;\\
a'_1 \otimes \cdots \otimes a'_N &= b_{j_1} \otimes  \cdots \otimes b_{j_L}\otimes I_k \otimes \cdots \otimes I_k,
\end{align}
where we assume $M\geq L$.
Since,  in every pair $a_j, a'_{\pi(j)}$ with $j>2$ at least one of the elements must coincide with $I_k$, the following identity must hold:
\begin{align} \label{sdf} &S_N\left( [a_1, a'_{\pi(1)}]\otimes a_2a'_{\pi(2)} \otimes 
\cdots   \otimes a_N a'_{\pi(N)}\right)= \nonumber \\
&S_N\left( [b_{i_1}, b_{j_{\pi(1)}}] \otimes b_{i_2} \otimes \cdots \otimes b_{i_M} \otimes  b_{j_{\pi(2)}} 
\otimes  \cdots \otimes b_{j_{\pi(L)}}\otimes I_k \otimes \cdots \otimes I_k\right).
\end{align}
The number of all permutations $\pi$ of type II and with fixed value $\pi(1)$ can be easily evaluated as (see Appendix \ref{proofs} for a more general formula)
\begin{align} C_N =\frac{(N-L)! (N-M)!}{(N-L-M +1)!}.\label{CN}\end{align}
Each of these permutations makes the same contribution (\ref{sdf}) to the first summand in the right-hand side of (\ref{sum2}), because changing $\pi$ in this way just amounts to keeping the factor  $[b_{i_1}, b_{j_{\pi(1)}}]$ and permuting the remaining factors in  the argument of $S_N$ in the right-hand side of (\ref{sdf}). This cannot change the final value in view of the very presence of the symmetrizer $S_N$. An identical argument applies 
to the remaining terms in the right-hand side of (\ref{sum2}).
Summing up, we can now write 
\begin{align*} &
C_N^{-1}\sum_{\pi \in {\cal P}(N)_{II}} 
\left(S_N\left( a_1 a'_{\pi(1)}\otimes 
\cdots   \otimes a_N a'_{\pi(N)}\right) -  S_N\left(a'_{\pi(1)}a_1\otimes 
\cdots  \otimes  a'_{\pi(N)}a_N\right) \right) \\
&= S_N\left([b_{i_1}, b_{j_1}] \otimes b_{i_2} \otimes \cdots \otimes b_{i_M} \otimes  b_{j_{2}} 
\otimes  \cdots \otimes b_{j_{L}}\otimes I_k \otimes \cdots \otimes I_k  \right) \\
 &+  S_N\left([b_{i_1}, b_{j_2}] \otimes b_{i_2} \otimes \cdots \otimes b_{i_M} \otimes  b_{j_{1}} 
\otimes b_{j_3}\otimes   \cdots \otimes b_{j_{L}}\otimes I_k \otimes \cdots \otimes I_k  \right) \\
&+\cdots + \\
&+  S_N\left( b_{i_1} \otimes \cdots \otimes b_{i_{M-1}} \otimes  b_{j_{1}} 
\otimes b_{j_3}\otimes   \cdots \otimes b_{j_{L-1}}\otimes [b_{i_M}, b_{j_L}] \otimes I_k \otimes \cdots \otimes I_k  \right),
\end{align*}
where all possible commutators $[b_{i_l}, b_{j_m}]$ are considered for $l=1,\ldots, M$ and $m=1,\ldots, L$.
We finally have that the term $iN[Q_{1/N}(f),Q_{1/N}(g)]_{II}$ equals
\begin{align*}
& iN[Q_{1/N}(f),Q_{1/N}(g)]_{II}= \\
& i\frac{N}{N!} \sum_{\pi \in {\cal P}(N)_{II}} 
\left(S_N\left( a_1 a'_{\pi(1)}\otimes 
\cdots   \otimes a_N a'_{\pi(N)}\right) -  S_N\left(a'_{\pi(1)}a_1\otimes 
\cdots  \otimes  a'_{\pi(N)}a_N\right) \right)\\
= & \frac{-C_N}{(N-1)!} S_N\left(\sum_l C_{i_1j_1}^l b_l \otimes b_{i_2} \otimes \cdots \otimes b_{i_M} \otimes  b_{j_{2}} 
\otimes  \cdots \otimes b_{j_{L}}\otimes I_k \otimes \cdots \otimes I_k  \right. \\
 &+  \sum_l C_{i_1j_2}^l b_l \otimes b_{i_2} \otimes \cdots \otimes b_{i_M} \otimes  b_{j_{1}} 
\otimes b_{j_3}\otimes   \cdots \otimes b_{j_{L}}\otimes I_k \otimes \cdots \otimes I_k   \\
&+\cdots + \\
&+  \left.\sum_l C_{i_Mj_L}^l  b_l \otimes b_{i_1} \otimes \cdots \otimes b_{i_{M-1}} \otimes  b_{j_{1}} 
\otimes b_{j_3}\otimes   \cdots \otimes b_{j_{L-1}}\otimes I_k \otimes \cdots \otimes I_k  \right),
\end{align*}
which can be rearranged to
\begin{align}
&iN[Q_{1/N}(f),Q_{1/N}(g)]_{II} =\nonumber \\
& \frac{-C_N}{(N-1)!} S_{M+L-1, N}\left(\sum_l C_{i_1j_1}^l b_l \otimes b_{i_2} \otimes \cdots \otimes b_{i_M} \otimes  b_{j_{2}} 
\otimes  \cdots \otimes b_{j_{L}}  \nonumber\right. \\
 &+  \sum_l C_{i_1j_2}^l b_l \otimes b_{i_2} \otimes \cdots \otimes b_{i_M} \otimes  b_{j_{1}} 
\otimes b_{j_3}\otimes   \cdots \otimes b_{j_{L}}  \nonumber \\
&+\cdots +\nonumber \\
&+  \left.\sum_l C_{i_Mj_L}^l  b_l \otimes b_{i_1} \otimes \cdots \otimes b_{i_{M-1}} \otimes  b_{j_{1}} 
\otimes b_{j_3}\otimes   \cdots \otimes b_{j_{L-1}}  \right) \label{iNQQ},
\end{align}
where we have freely rearranged the order of some  factors (which we may do since this order is  irrelevant in view of the presence of the symmetizator $S_N$).

\noindent We now observe that the last expression for $iN[Q_{1/N}(f),Q_{1/N}(g)]_{II}$  is identical to the expression of $Q_{1/N}(\{f,g\})$
found in (\ref{Q1Nhg}), up to the factor $\frac{C_N}{(N-1)!}$. However,  a direct computation using Stirling's formula proves that, for fixed $M,L$,
 \begin{align}\frac{C_N}{(N-1)!}  \to  1\quad \mbox{for $N\to \infty$}\label{limCN}.\end{align}
To conclude the proof of (\ref{DGR}), exploiting  (\ref{limCN}) and the triangle inequality for  the norm in (\ref{DGR}),
it is therefore sufficient to prove that 
\begin{align}\left|\left|\left[\frac{C_N}{(N-1)!}- 1\right] Q_{1/N}(\{f,g\})\right|\right|_N &\to 0;\label{onelast}\\
\| iN[Q_{1/N}(f),Q_{1/N}(g)]_{III}\|_{N}& \to 0\label{twolast}\end{align}
are both valid as $N\to \infty$. 
The former  is  true as a consequence of (\ref{Q1Nhg}) and the following property of the maps $S_{M,L}$ for $M\leq L$, which is easy to prove:
\begin{equation}
\| S_{M,L}(a_1\otimes \cdots \otimes a_M)\|_M \leq \max\{ \|a_j\|^M\:|\: j=1,\ldots, M\}.
\end{equation}
This implies that $\|Q_{1/N}(\{f,g\})\|$
is a bounded function of $N$ (for $f$ and $g$ given as  above), so that (\ref{limCN}) implies (\ref{onelast}). Regarding the latter, we observe that the conjunction of
(\ref{SSaabb}), the property \begin{equation}
N[Q_{1/N}(f), Q_{1/N}(g)]=N[S_{M,N}(f^{-1}(f)), S_{L,N}(f^{-1}(g))],
\end{equation} and  Lemma \ref{lemmacommSN} imply
\begin{align} \|iN[Q_{1/N}(f), Q_{1/N}(g)]_{III}\|_N \leq \frac{2 C}{(N-1)!} \#{\cal P}(N)_{III}\label{est2}\end{align}
for the constant \begin{equation}
C= \max\left\{ \left. \|b_{i_m}\|^M \|b_{j_l}\|^L\:\right|\: m=1,\ldots, M, \: l=1,\ldots, L\right\}.
\end{equation}
 Referring to the discussion just before (\ref{PPP}), one can prove  (see Appendix \ref{proofs}) that the number $\# {\cal P}(N)_K $ of elements  $\pi \in {\cal P}(N)$  for which the string $$a_1 a'_{\pi(1)}\otimes 
\cdots   \otimes a_N a'_{\pi(N)},$$ or,  equivalently,  $$a'_{\pi(1)}a_1\otimes 
\cdots   \otimes a'_{\pi(N)}a_N,$$  in the right-hand side of (\ref{formula}) includes {\em exactly $K$} factors $a_l a'_{\pi(l)}$  (resp.\ $a'_{\pi(l)}a_l$)  with both  $a_l \neq I_k$ and $a'_{\pi(l)} \neq I_k$ is equal to
\begin{align}\# {\cal P}(N)_K = \frac{L!M!(N-L)!(N-M)!}{K!(L-K)!(M-K)!(N-L-M+K)!},\label{speriamo}\end{align}
where we assumed $0\leq K\leq L\leq M$ and $N$ large.
Hence
\begin{align}
\frac{\#{\cal P}(N)_{III}}{(N-1)!} = \frac{1}{(N-1)!}\sum_{K=2}^L \frac{L!M!(N-L)!(N-M)!}{K!(N-M-L+K)!(L-K)!(M-K)!}.
\end{align}
As a consequence, for some constant $A>0$ depending on $L,M$, we have
\begin{align}
\frac{\#{\cal P}(N)_{III}}{(N-1)!} &\leq  \frac{A(N-L)!(N-M)!}{(N-1)!(N-M-L+2)!}\nonumber \\
& = \frac{AC_N}{(N-1)!}\frac{1}{(N-M-L+2)},
\end{align}
where we used (\ref{CN}). Taking advantage of (\ref{limCN}), we obtain
\begin{equation}
\frac{\#{\cal P}(N)_{III}}{(N-1)!} \to 0 \quad \mbox{for $N\to \infty$}.\end{equation}
 This result  implies that (\ref{twolast}) holds because of (\ref{est2}), which concludes the proof.
\end{proof}
\begin{remark}
{\em Observe that we can rearrange (\ref{speriamo}) as
\begin{equation}
\#{\cal P}(N)_K = (N-L)!L!\binom{M}{K}\binom{N-M}{L-K}.\end{equation}
As a consequence, expoliting the well-known {\em Chu-Vandermonde identity}, we find 
\begin{align}
\sum_{K=0}^L \#{\cal P}(N)_K &=  (N-L)!L!\sum_{K=0}^L \binom{M}{K}\binom{N-M}{L-K}\nonumber \\ &= (N-L)!L!  \binom{N}{L} =N!,\end{align}
that is,
\begin{equation}
\sum_{K=0}^L \#{\cal P}(N)_K =  \#{\cal P}(N),\end{equation}
as it must be.\hfill $\blacksquare$}
\end{remark}
\newpage
\section{Application to the Curie--Weiss model} \label{CWM}
We now apply our quantization maps $Q_{1/N}$ of Theorem \ref{mainfirst} to the (quantum) Curie--Weiss model,\footnote{This model exists in both a classical and a quantum version and is a mean-field approximation to the  Ising model. See e.g.\ \cite{FV2017} for a mathematically rigorous treatment of the classical version, and \cite{CCIL,IL} for the quantum version. 
 Quantum mean field theories (starting with the BCS model of superconductivity) have been subjected to intense mathematical scrutiny since the 1960s, starting with work of Bogoliubov and Haag; see the notes to \S 10.8 on pages 432-433 of \cite{Lan17} for extensive references and history. As already mentioned, for our approach the papers \cite{Bona,DW,RW}  played an important role. See also \cite{ABN} for a very detailed discussion of the quantum Curie--Weiss model.}
 which corresponds to the case $k=2$. The {\bf quantum Curie--Weiss Hamiltonian}, defined on a lattice with $N$ sites (whose geometric configuration, including its dimension, is irrelevant, as is typical for mean-field models),
  is
  \begin{align}
h^{CW}_{1/N}: &  \underbrace{\mathbb{C}^2 \otimes \cdots  \otimes\mathbb{C}^2}_{N \: times}  \to 
\underbrace{\mathbb{C}^2 \otimes \cdots  \otimes\mathbb{C}^2}_{N \: times}; \\
h^{CW}_{1/N} &=\frac{1}{N} \left(-\frac{J}{2N} \sum_{i,j=1}^N \sigma_3(i)\sigma_3(j) -B \sum_{j=1}^N \sigma_1(j)\right).\label{CWham}
\end{align}
Here $\sigma_k(j)$ stands for $I_2 \otimes \cdots \otimes \sigma_k\otimes \cdots \otimes I_2$, where $\sigma_k$ occupies  the $j$-th slot, and 
 $J,B \in \mathbb{R}$ are given constants defining the strength of the spin-spin coupling and the (transverse) external magnetic field, respectively. Note that 
 \begin{equation}
 h^{CW}_{1/N} \in \mathrm{Sym}(M_2(\mathbb{C})^{\otimes N}),
\end{equation}
where $\mathrm{Sym}(M_2(\mathbb{C})^{\otimes N})$ is the range of the symmetrizer $S_N$ defined in \er{defSN}; in other words, as a sequence indexed by $N\in\N$ the operators \er{CWham} form a symmetric sequence. 
Our interest will lie in the limit $N\raw\infty$. As such, we rewrite \er{CWham} as
\begin{align}
h^{CW}_{1/N}&= -\frac{J}{2N(N-1)} \sum^N_{i \neq j, \:i,j=1} \sigma_3(i)\sigma_3(j) - \frac{B}{N} \sum_{j=1}^N \sigma_1(j) + O(1/N).\nonumber \\
&= Q_{1/N}(h^{CW}_0)  + O(1/N)\label{QNh}
\end{align}
where $O(1/N)$ is meant in norm, and
the {\bf  classical Curie--Weiss Hamiltonian} is 
\begin{align}\label{hcv0}
h^{CW}_0: B^3 &\mapsto\R;\\
h^{CW}_0(x,y,z)  &= -\left(\frac{J}{2}z^2 + Bx\right), \quad \mathbf{x}=(x,y,z) \in \mathcal{Q}_2. \label{hcwc}
\end{align}
Recall that $B^3= \{ \mathbf{x} \in \mathbb{R}^3 \:|\: \|\mathbf{x}\|\leq 1\}$ is the closed unit ball in $\R^3$, arising in this context as the parameter space $\mathcal{Q}_2$, as explained in \S\ref{kgeq2}. Clearly, recalling \er{tA0},
\begin{equation}
h^{CW}_0\in \tilde{A}_0 \subset C^\infty(B^3)\subset C(B^3)\cong C(S(M_2(\C)))=A_0.
\end{equation}
Therefore, up to a small error as $N\raw\infty$,
 the quantum  Curie--Weiss  Hamiltonian \er{CWham} is given by deformation quantization of its classical counterpart \er{hcwc}. 
\subsection{Classical limit and Spontaneous Symmetry Breaking}\label{QZ}
We henceforth assume  $B\in (0,1)$ and $J=1$, in which regime the quantum 
CW model exhibits Spontaneous Symmetry Breaking (SSB). 
One can prove (see \cite{IL}, \cite[\S5.3]{vandeVen},  \cite{VGRL18}) that  for each $N=1,2,3,\ldots$ the ground-state vector $\Psi^{(0)}_N$ of $h^{CW}_{1/N} $
 is {\em unique} (up to phase factors and normalization) and belongs to the symmetric space
\begin{equation}
\mathrm{Sym}^N(\mathbb{C}^2) = 
\underbrace{\mathbb{C}^2 \otimes_s \cdots \otimes_s \mathbb{C}^2}_{ N \:times}\cong \C^{N+1}. \label{SYmN1}
\end{equation}
Instead of looking for a possible limit of  $\Psi^{(0)}_N$ as a vector in some Hilbert space,  which would involve the messy infinite tensor products of von Neumann (see \cite[\S 8.4]{Lan17}), we redefine the notion of a state in the spirit of the algebraic formulation of quantum theory, and consider the so-called (algebraic) vector states \er{herdef}, i.e.,
\begin{align}
 \omega_{1/N}^{(0)}(\cdot) = \langle \Psi^{(0)}_N,  \cdot \Psi^{(0)}_N\rangle,\label{ALST}
 \end{align} 
which are associated to the unit vectors $\Psi^{(0)}_N$ (these are positive normalized functionals on the $C^*$-algebras  $M_2(\C)^{N}$ and hence are states in the $C^*$-algebraic sense).  Each state $\omega_{1/N}^{(0)}$ is defined on the fiber
$A_{1/N}=M_2(\C)^N$ of our continuous bundle of $C^*$-algebras \er{B0} - \er{BN}, and we hope that the sequence $( \omega_{1/N}^{(0)})_{N\in\N}$ converges to some state $\om_0^{(0)}$ on $A_0= C(S(M_2(\C)))$ in the sense of \er{om0limit}. If it does, by the Riesz representation theorem the limit state $\om_0$ corresponds to a probability measure $\mu^{(0)}$ on
$S(M_2(\C))\cong B^3$;  pure states then correspond to Dirac measures, which are  concentrated at single points of $B^3$.
As a hallmark of SSB,\footnote{See \cite[\S10.3]{Lan17} or \cite[\S1]{VGRL18} for the algebraic picture of SSB we use here.}
 we note that unlike the case where $N$ is finite, for $0<B<1$ and $J=1$ the ground state of  the classical CW hamiltonian  (\ref{hcv0}) is not unique:  first interpreting the notion of a ground state in the usual way, i.e.\ as a point $\mathbf{x}\in B^3$ where the function $h^{CW}_0$ assumes an absolute minimum, for  example for $B=1/2$, $J=1$ we find {\em two} 
 such minima $\mathbf{x}_\pm$, given by
 \begin{align}\label{xpm}
\mathbf{x}_\pm = \left(\half , 0,\pm \half\sqrt{3}\right).
\end{align}
Algebraically, these define Dirac measures $\mu^{(0)}_\pm$ localized at $\mathbf{x}_\pm$, or the corresponding functionals  $\om^{(0)}_\pm$ on $C(B^3)$, given by $\om^{(0)}_\pm(f)=f(\mathbf{x}_\pm)$, where $h\in C(B^3)$. If we now look at the $\Z_2$-symmetry of the  classical CW hamiltonian  (\ref{hcv0}), given by 
\begin{equation}
(x,y,z)\mapsto (x,-y,-z), \label{xyz}
\end{equation}
then clearly neither $\mathbf{x}_+$ nor $\mathbf{x}_-$ is invariant under this symmetry: instead,  $\mathbf{x}_\pm$ is mapped to $\mathbf{x}_\mp$. Thus no pure invariant ground state exists. However, the mixture
\begin{equation}
\om^{(0)}=\half(\om^{(0)}_++\om^{(0)}_-), \label{defom0}
\end{equation}
which also qualifies as a ground state in the algebraic sense, is invariant {\em but not pure}. At least in the language of algebraic quantum theory this is the essence of SSB:
\begin{center}
 \emph{Pure ground states are not invariant, whilst
invariant ground states are not pure}.
\end{center}
In contrast, for any $N<\infty$ the quantum CW model has no SSB, since it has a unique invariant pure ground state  \cite{IL}, \cite[\S5.3]{vandeVen},  \cite{VGRL18}.\footnote{Again, in the algebraic sense; the physicists's ground state vector $\Psi^{(0)}_N$ is unique up to a phase.}
The relevant $\Z_2$-symmetry of $h^{CW}_{1/N}$ is given by the $N$-fold tensor power of the automorphism of $M_2(\C)$ given by 
\begin{equation}
a\mapsto  \sigma_1 a \sigma_1, \label{asigma1}
\end{equation}
see  \S\ref{Dick} or \cite[\S10.8]{Lan17}.
If $\zeta$ is the nontrivial element (-1) of $\Z_2$, we denote the automorphism of $M_2(\C)^N$ induced by \er{asigma1} by
$\zeta^{(1/N)}$, and  write 
 the pullback of \er{xyz} to $C(B^3)$ as $\zeta^{(0)}$. Then $\zeta^{(0)}$ leaves $\tilde{A}_0$ invariant, and each map $Q_{1/N}$ is {\em equivariant}:
\begin{align}\label{symma}
Q_{1/N} \circ \zeta^{(0)} = \zeta^{(1/N)}\circ Q_{1/N}.\end{align}
Furthermore, since  the ground state $\Psi^{(0)}_N$ of $h^{CW}_{1/N}$ (seen as a unit vector) is unique up to a phase,  its associated algebraic state
 $\omega_{1/N}^{(0)}$ is strictly {\em invariant} under $\Z_2$, i.e.
\begin{align} \omega_{1/N}^{(0)}\circ \zeta^{(1/N)} = \omega_{1/N}^{(0)}.\label{INVomega}\end{align} 
Combining \er{symma} and \er{INVomega}, we obtain, for any $h\in\tilde{A}_0$,
\begin{equation}
 \omega_{1/N}^{(0)}(Q_{1/N}(\zeta^{(0)}(f)))= \omega_{1/N}^{(0)}(Q_{1/N}(f)),
\end{equation}
so that if the limit  \er{om0limit} exists, the limit state $\omega_0^{(0)}$ satisfies
\begin{equation}
\omega_0^{(0)}\circ  \zeta^{(0)} =\omega_0^{(0)}.
\end{equation}
One may therefore expect that
 the sequence $(\omega_{1/N}^{(0)})_N$ of (pure) ground states of the quantum CW Hamiltonian
converges to the invariant state $\om^{(0)}$ as $N\raw\infty$ in the sense of  \er{om0limit}, and this is indeed what we shall prove,
at least for $f\in\tilde{A}_0$. 
Part of the proof of Theorem \ref{mainsecond} relies on (convincing) numerical evidence about the large $N$ behavior of $\Psi^{(0)}_N$, summarized  in  
Assumption \ref{assumption} in \S\ref{Exis} below. Those who only accept strictly analytic proofs might prefer to state this evidence, i.e.\ Assumption \ref{assumption}, as an hypothesis for the theorem, but we consider it part of the proof.\footnote{The question why in nature one of the pure symmetry-breaking states $\om^{(0)}_\pm$ is found, rather than the mixture $\om^{(0)}$, is answered in \cite{VGRL18}, partly  based on the ``tower of states"  of P.W. Anderson.}
\begin{theorem}\label{mainsecond}
Let  $Q_{1/N}: \Tilde{A}_0 \to M_2(\C)^N$  be the quantization maps defined by linear extension of \er{deformationqunatizaion} - \er{deformationqunatizaion2}, cf.\ Theorem \ref{mainfirst}, and let $\Psi_N^{(0)}$ be the (unit) ground state vector  in \er{ALST} of the  Hamiltonian \er{CWham} of the quantum Curie--Weiss model. Then
\begin{equation}
\lim_{N\raw\infty}\omega_{1/N}^{(0)}( Q_{1/N}(f))=\om^{(0)}(f),\label{question}
\end{equation}
for all $h \in \tilde{A}_0$, where $\omega_{1/N}^{(0)}$ and $\om^{(0)}$ are defined in \er{ALST} and 
\er{defom0}, respectively.
\end{theorem}
\noindent Unfolding \er{question} on the basis of  \er{ALST} and 
\er{defom0}, the theorem therefore states that
\begin{equation}
\lim_{N\raw\infty}  \langle \Psi^{(0)}_N,  Q_{1/N}(f) \Psi^{(0)}_N\rangle=\half(f(\mathbf{x}_+)+f(\mathbf{x}_-)),
\end{equation}
for any polynomial function $f$ on $B^3$ (parametrizing the state space of $M_2(\C)$),
where the points $\mathbf{x}_\pm\in B^3$ are given by \er{xpm}. This is our second main result.
\subsection{Coherent spin  states and Dicke basis in $\mathrm{Sym}^N(\mathbb{C}^2)$}\label{secOmegaDicke}
Our proof relies on the  large-$N$ behaviour of the components of $\Psi_N^{(0)}$. By permutation symmetry of the Hamiltonian and uniqueness of the ground state we know that  $\Psi_N^{(0)}$ lies in the symmetric subspace
 $\mathrm{Sym}^N(\mathbb{C}^2)$ of $(\mathbb{C}^2)^{N\otimes}$. We will introduce a certain  
bases of  that subspace with respect to which the asymptotics of $\Psi_N^{(0)}$ will be studied. 

Let $|\!\uparrow\rangle, |\!\downarrow\rangle$ denote the eigenvectors of $\sigma_3$ in $\mathbb{C}^2$, so that
$\sigma_3|\!\uparrow\rangle=|\!\uparrow\rangle$ and $\sigma_3|\!\downarrow\rangle=- |\!\downarrow\rangle$. 
 If
$\Omega \in {S}^2$, with  polar angles  
$\theta_\Omega \in (0,\pi)$, $\phi_\Omega \in (-\pi, \pi)$, we define\footnote{In the literature there are some inequivalent definitions of the overall  non-constant phase affecting  $|\Omega\rangle_1$  \cite{Pe72, BZ06},  but all choices have the same important properties  listed here.}
\begin{align}\label{om1}
|\Omega\rangle_1  = \cos \frac{\theta_\Omega}{2} |\!\uparrow\rangle + e^{i\phi_\Omega}\sin   \frac{\theta_\Omega}{2} |\!\downarrow\rangle.
\end{align}
Writing $\underline{\sigma}=(\sigma_1,\sigma_2,\sigma_3)$, 
it is easy to prove that 
\begin{equation}
\Omega \cdot \underline{\sigma} |\Omega\rangle_1 =  |\Omega\rangle_1.
\end{equation}
 If $N \in \mathbb{N}$, the associated {\bf $N$-coherent spin state} $|\Omega\rangle_N\in \mathrm{Sym}^N(\mathbb{C}^2)$,
  equipped with the usual scalar product $\langle \cdot ,\cdot \rangle_N$ inherited from $(\C^2)^N$, 
   is defined as follows \cite{Pe72}:
\begin{align}\label{om2}
|\Omega\rangle_N =  \underbrace{|\Omega\rangle_1 \otimes \cdots \otimes |\Omega \rangle_1}_{N \: times}.
\end{align}
We occasionally also adopt the alternative  notation $|\Omega_{\theta, \phi}\rangle_N$, which emphasizes  the dependence of $\Omega$ of the polar angles $(\theta,\phi)$.
An explicit expression of $|\Omega\rangle_N$ can be presented through the so-called {\bf Dicke basis} of $\mathrm{Sym}^N(\mathbb{C}^2)$, given by 
\begin{equation}
\{|k, N-k\rangle\:|\: k=0,1 \ldots, N \}, \label{Dickebasis}
\end{equation}
where $|k,N-k \rangle$ is the normalized vector obtained   by symmetrization of a tensor product of $N$ vectors in $\mathbb{C}^2$ whose  $k$ factors are of type $|\!\uparrow\rangle$ and the remaining  $N-k$ factors are of type $|\!\downarrow\rangle$.
A simple computation relying upon  (\ref{om1}) and (\ref{om2}) yields
\begin{align}\label{important}
    |\Omega_{\theta,\phi}\rangle_N=\sum_{k=0}^N\sqrt{N\choose k}\cos{(\theta/2)}^k\sin{(\theta/2)}^{N-k} e^{i(N-k)\phi}|k,N-k\rangle.
\end{align}
Coherent spin states form an 
{\em overcomplete set of vectors} for 
$\mathrm{Sym}^N(\mathbb{C}^2)$, in that
\begin{align}\label{overcomplete}
\langle \Psi,\Phi \rangle_N = \frac{N+1}{4\pi}\int_{{S}^2}  \langle \Psi,\Omega \rangle_N \langle \Omega, \Phi \rangle_N d\Omega,\quad \mbox{for all $\Psi,\Phi \in \mathrm{Sym}^N(\mathbb{C}^2)$.}
\end{align}
Here  $d\Omega$ indicates the unique $SO(3)$-invariant Haar measure on ${S}^2$ with $\int_{{S}^2} d\Omega = 4\pi$, which, in turn,  coincides with the measure generated by  the metric induced  to the embedded submanifold ${S}^2$ from  $\mathbb{R}^3$.  
Another property relevant for our computations, which  straightforwardly follows from (\ref{om1}) - (\ref{om2}), is
\begin{align}\label{quasidelta}
|\langle \Omega,\Omega' \rangle_N|^2 = \left( \frac{1+ \cos \Phi(\Omega,\Omega')}{2}\right)^{N},
\end{align}
where \begin{align}\label{cos} \cos \Phi(\Omega_{\theta,\phi},\Omega_{\theta',\phi'}) = \cos \theta \cos \theta'  + \sin\theta \sin \theta' \cos(\phi-\phi')\end{align}
 is the cosine of the angle $\Phi$ between $\Omega_{\theta,\phi}$ and $\Omega_{\theta',\phi'}$.
\subsection{Proof of Theorem \ref{mainsecond}} \label{Exis}
\begin{figure}[!htb]
\centering
   \includegraphics[width=9cm]{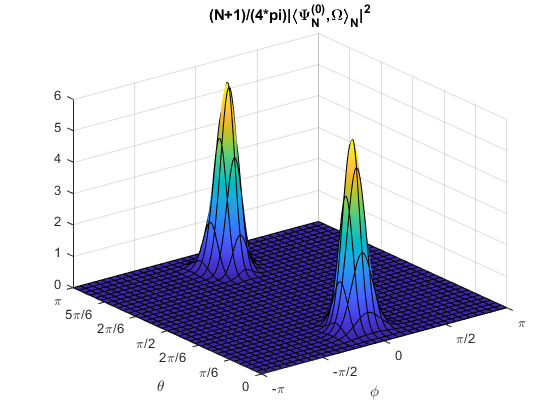}
 \caption{$\frac{N+1}{4\pi}|\langle \Psi^{(0)}_N, \Omega_{\theta,\phi}\rangle|^2$ as a function of $\theta$ and $\phi$, for $N=150, J=1$, $B=1/2$.}
    \label{plotN=150Leftpart}
\end{figure}

\begin{figure}[!htb]
\centering
   \includegraphics[width=9cm]{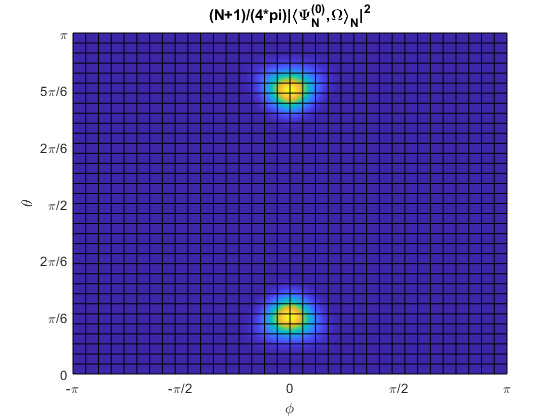}
 \caption{Top view of  the previous plot.}
    \label{plotN=150Leftparttopview}
\end{figure}

With the help of a good numerical evidence, we are now in a position to prove  (\ref{question}). 
We will take advantage of some preparatory results we are going  to  discuss. The first one is a  pivotal proposition whose proof is unfortunately a bit technical.\footnote{Here, and henceforth in similar statements, when dealing with differentiable functions defined on ${S}^2$ we always refer  to
 the differentiable structure induced on ${S}^2$ by $\mathbb{R}^3$. }
\begin{proposition}\label{lemmadelta}
Let  $h: {S}^2 \to \mathbb{C}$  be a  bounded measurable function that  is  $C^1(A)$ for some open set  $A\subset {S}^2$.
 Then the following properties hold for every $\Omega'\in A$:
\begin{itemize}
\item[{\bf (a)}] If $\ell>0$, then
\begin{align}\label{delta1} 
h(\Omega') = \lim_{N\to \infty} \frac{\ell (N+1)}{4\pi} \int_{{S}^2}  h(\Omega)|\langle \Omega',\Omega\rangle_N|^{2\ell} d \Omega.\end{align}
\item[{\bf (b)}]  In particular,
\begin{align}\label{delta2} 
 \left|h(\Omega') - \frac{\ell (N+1)}{4\pi} \int_{{S}^2}  h(\Omega)|\langle \Omega',\Omega\rangle_N|^{2\ell} d \Omega\right| \leq \frac{B_\ell \|h\|_\infty + C^{(A)}_\ell\|dh\|^{(A)}_\infty}{\sqrt{N}},\end{align}
where
\begin{align} \|dh\|^{(A)}_\infty = \sup_{\Omega \in A}\sqrt{{\bf g}_\Omega(d\overline{h},dh)},\label{defdf}\end{align}
 in which ${\bf g}_\Omega$ is the inner product on $T_\Omega^*{S}^2$ induced from $\mathbb{R}^3$, and 
 $B_\ell, C^{(A)}_\ell\geq 0$ are constants independent of $h$ and $\Omega'$ (but $C^{(A)}_\ell$ depends on $A$).
\end{itemize}
\end{proposition}
\begin{proof} See Appendix \ref{proofs}.
\end{proof}
\begin{remark}\label{remA} 
\begin{itemize}{\em 
\item[(1)] Here $\|dh\|^{(A)}_\infty $ could be infinite and, in that case, (\ref{delta2}) is trivially valid for every choice  of $C^{(A)}_\ell$.
It is, however, always possible to restrict $A$ to a smaller open  set with compact closure included in the initial set $A$ where $h$ is $C^1$.  In that case,  $\|dh\|^{(A)}_\infty $ is finite.  This observation applies to all similar statements we will establish in the rest of the work.
\item[(2)]  The apparently cumbersome formulation of Proposition \ref{lemmadelta}, where $A$ does not coincide with ${S}^2$, is really necessary, since we will use this and  similar results exactly where the  functions in question are not everywhere $C^1$. \hfill $\blacksquare$}
\end{itemize}
\end{remark}
\noindent Another crucial building block of the proof of Theorem \ref{mainsecond}
is good numerical evidence about the behaviour  of the coherent components of $\langle \Psi^{(0)}_N,\Omega\rangle$ for  large $N$ (see Appendix \ref{numerical}). 
Namely, for sufficiently large $N$, we  have for $\ell = 1$ and $\ell =1/2$, 
\begin{align}\label{ne2}
\frac{N+1}{4\pi}|\langle \Psi^{(0)}_N ,\Omega_{\theta, \phi}
\rangle_N|^{2\ell} \approx 
\frac{N+1}{4\pi 2^\ell} |\langle  \Omega_+ ,\Omega_{\theta,\phi}\rangle_N|^{2\ell}+\frac{N+1}{4\pi 2^\ell} |\langle \Omega_- , \Omega_{\theta,\phi}\rangle_N|^{2\ell}  ,
\end{align}
where $\Omega_\pm$ define  a pair of corresponding unit vectors $\mathbf{x}_\pm$ as in (\ref{xpm}), always assuming $J=1$ and $B=1/2$. In terms of polar  angles $\theta, \phi$, we have
\begin{align}
(\theta_+, \phi_+) = (\pi/6, 0), \quad (\theta_-, \phi_-) = (5\pi/6, 0).
\end{align}
\begin{remark}\label{remdeltA}
{\em The practical meaning of (\ref{ne2}) is that, as $N$ increases, the map 
$\Omega \mapsto \frac{N+1}{4\pi}|\langle \Psi^{(0)}_N ,\Omega
\rangle_N|^{2\ell}$ increasingly accurately approximates a linear combination of two functions, each of which, in turn, tends to  a Dirac delta-function  centered at $\Omega_+$  and $\Omega_-$ respectively, in accordance with part (a) in  
Proposition \ref{lemmadelta}. In particular,  the set of points $\Omega$  where  $\frac{N+1}{4\pi}|\langle \Psi^{(0)}_N ,\Omega
\rangle_N|^{2\ell}$ is apreciably different from zero tends to concentrate around $\Omega_+$ and $\Omega_-$.
\hfill $\blacksquare$}
\end{remark}
\noindent In figures \ref{plotN=150Leftpart} and \ref{plotN=150Leftparttopview} the function  $(\theta,\phi)\mapsto \frac{N+1}{4\pi}|\langle \Psi^{(0)}_N, \Omega_{\theta,\phi}\rangle|^2$ is computed for $N=150$; the  peaks at the values $(\theta,\phi)=(\pi/6,0)$ and $(\theta,\phi)=(5\pi/6,0)$ are clearly visible.

\noindent In figure 10 (see Appendix), the angle $\phi=0$ is fixed and a plot of the two functions
\begin{align}
\theta&\mapsto\frac{N+1}{4\pi}|\langle \Psi^{(0)}_N, \Omega_{\theta,0}\rangle|^2;\\
\theta&\mapsto\frac{N+1}{8\pi}|\langle \Omega_{+}, \Omega_{\theta,0}\rangle|^2+\frac{N+1}{8\pi}|\langle \Omega_{-},\Omega_{\theta,0}\rangle|^2
\end{align}
is given. It is evident that the two  graphs are almost indistinguishable and this fact becomes  more and more evident as  $N$ increases.
Similarly, in figure 11, the angle $\theta=\pi/6$ is fixed and a plot of the two functions
\begin{align}
\phi&\mapsto\frac{N+1}{4\pi}|\langle \Psi^{(0)}_N, \Omega_{\pi/6,\phi}\rangle|^2;\\
\phi&\mapsto\frac{N+1}{8\pi}|\langle \Omega_{+}, \Omega_{\pi/6,\phi}\rangle|^2+\frac{N+1}{8\pi}|\langle \Omega_{-},\Omega_{\pi/6,\phi}\rangle|^2
\end{align}
is displayed.
It is once again evident that the two  graphs are almost indistinguishable and this fact becomes the more evident the  $N$ increases. 
We repeated the same analysis for the point $5\pi/6$, but omitted this plot as its graph looks similar due to symmetry. Moreover, in the appendix we produce similar plots for  $\ell=1/2$. 

Concerning assumptions (a) and (b) below, we  will employ an $L^2$  intepretation  of  (\ref{ne2}) for $\ell=1$ partially  suggested by  Remark \ref{remdeltA}, and an even weaker interpretation for $\ell=1/2$.
As a matter of fact, the proof of Theorem \ref{mainsecond} directly uses the three requirements in Assumption \ref{assumption} below  which are  
 supported  by numerical evidence (Appendix \ref{numerical}), independently of  (\ref{ne2}). 
 
  To state item (c) in these assumptions,
we define,  for  $\Omega_0 \in {S}^2$ and $r>0$,
\begin{equation}
D_r(\Omega_0)=  \{ \Omega \in {S}^2 \:|\: \Phi(\Omega, \Omega_0) <r\}.
\end{equation}
It is clear that $D_r(\Omega_0)$ is a geodesical disk on ${S}^2$
centered at $\Omega_0$ with  radius $r$.
\begin{assumption}\label{assumption}
On numerical evidence, we assume the following properties:
\begin{itemize}
\item[{\bf (a)}]   $\lim_{N\raw\infty}$
\begin{align}
\label{firstlim}\int_{{S}^2}  \left( \frac{N+1}{4\pi}|\langle \Psi^{(0)}_N ,\Omega
\rangle_N|^{2} -
\frac{N+1}{8\pi} |\langle \Omega_+, \Omega \rangle_N|^{2}-\frac{N+1}{8\pi} |\langle \Omega_-, \Omega\rangle_N|^{2}\right)d\Omega =0.
\end{align}
\item[{\bf (b)}] There is a constant $G\geq 0$ such that for every $N \in \mathbb{N}$ and  $\ell=1/2, 1$,
\begin{align} \label{decreasingsequence} \int_{{S}^2}  \left| \frac{N+1}{4\pi}|\langle \Psi^{(0)}_N ,\Omega
\rangle_N|^{2\ell} -
\frac{N+1}{4\pi 2^\ell} |\langle \Omega_+, \Omega \rangle_N|^{2\ell}-\frac{N+1}{4\pi 2^\ell} |\langle \Omega_-, \Omega \rangle_N|^{2\ell}\right|d\Omega\leq G.\end{align}
\item[{\bf (c)}] For every  $n \in \mathbb{N}$ and $\ell=1/2,1$,
the sequence of maps \begin{equation}
{S}^2\setminus D_{1/n}(\Omega_+)\cup D_{1/n}(\Omega_-) \ni \Omega \mapsto \frac{N+1}{4\pi}|\langle \Psi^{(0)}_N ,\Omega
\rangle_N|^{2\ell} \label{firstlim2} \end{equation}
is bounded by some constant $K_n \geq 0$  and pointwise converges to $0$.
\end{itemize}
\end{assumption}

\begin{remark}\label{remCU} $\null$ \\
{\em {\bf (a)} Using Lebesgue's dominated
convergence theorem,  item (c) implies in particular that, if $A\subset {S}^2$ is a given open set containing $\Omega_+$ and $\Omega_-$, then 
\begin{equation}
\lim_{N\raw\infty} \int_{{S}^2 \setminus A}\frac{N+1}{4\pi}|\langle \Psi^{(0)}_N ,\Omega
\rangle_N|^{2\ell}= 0.
\end{equation}
{\bf(b)} For given $\ell=1/2$ or $1$,  the class of functions
\begin{equation}
{S}^2 \ni \Omega \mapsto \frac{N+1}{4\pi 2^\ell} |\langle \Omega_+, \Omega \rangle_N|^{2\ell}-\frac{N+1}{4\pi 2^\ell} |\langle \Omega_-, \Omega \rangle_N|^{2\ell}\end{equation}
also satisfies (c), as is clear from the Proof of Proposition \ref{lemmadelta}.
 \hfill $\blacksquare$}
\end{remark}
\noindent  Together with Proposition \ref{lemmadelta} and the elementary 
facts about the states $|\Omega\rangle$ presented in Section \ref{secOmegaDicke}, these properties of $\Psi^{(0)}_N$ (assumed valid on the basis of their numerical evidence) are the source of the following two lemmas:
\begin{lemma}\label{lemmaA} Let  $h: {S}^2 \to \mathbb{C}$ be
a bounded measurable  function that  is 
$C^1(A)$ for some open set $A\subset {S}^2$  containing both $\Omega_+$ and $\Omega_-$.
On  Assumption  \ref{assumption}, where (b) and (c) are required only for 
$\ell=1$, one has
\begin{align}\label{lemmaAid}
\lim_{N\to \infty} \frac{(N+1)}{4\pi}
\int_{{S}^2}  \hskip -6 pt h(\Omega)
 |\langle \Psi^{(0)}_N ,\Omega\rangle_N|^{2} d \Omega =  \half h(\Omega_+) + \half h(\Omega_-) .\end{align}
\end{lemma}

\begin{proof} See Appendix \ref{proofs}.
\end{proof}

\begin{notation} {\em From now on, $S$ denotes the {\bf south pole} of ${S}^2$ determined by $\theta=\pi$ in standard spherical polar coordinates.\hfill $\blacksquare$\\}
\end{notation} 
\begin{lemma}\label{lemma23} Let  $h: {S}^2 \to \mathbb{C}$ be
a bounded measurable  function that  is 
$C^1(A)$ for some open set $A\subset {S}^2$ that does not contain  $S$. 
On  Assumption  \ref{assumption}, where (b) and (c) are required only for 
$\ell=1$,  for any $\Omega' \in A$,  $M\in\N$, and $N>M$ one has
\begin{align}\label{tpb}
\left|\int_{{S}^2}   \frac{N+1}{4\pi}\langle \Psi^{(0)}_N, \Omega\rangle_N h(\Omega) \langle \Omega,\Omega'\rangle_{N-M}
d\Omega - 
\langle \Psi^{(0)}_N, \Omega'\rangle_N h(\Omega') \right|\nonumber \\
\leq \frac{K^{(A)}\|h\|_\infty+\sqrt{C\|f\|^2_\infty + 
D^{(A)}\|dF\|^{(A)}_\infty}}{(N-M)^{1/4}},
\end{align}
where the  constants $C, K^{(A)}, D^{(A)} \geq 0$ may depend on $M$, and $K^{(A)}$ and $D^{(A)}$ may also depend on $A$,  but $C, K^{(A)}, D^{(A)}$ are independent of 
$\Omega'$,  $h$, and $F$, where 
\begin{equation}
F(\Omega)= |h(\Omega) - h(\Omega')|^2.
\end{equation}
\end{lemma}

\begin{proof} See Appendix \ref{proofs}.
\end{proof}
\noindent After these preparations we are finally in a position to prove Theorem \ref{mainsecond}.

\begin{proof} Let us start the analysis of the large-$N$ behavior of the expectation value $\langle \Psi_N^{(0)}, Q_{1/N}(f) \Psi^{(0)}_N\rangle$ for some fixed polynomial $f= f(\mathbf{x})$ in the components $x_1,x_2,x_3$ of $\mathbf{x} \in B^3$ (always supposing $J=1$, $B=1/2$).
From   (\ref{overcomplete}) we have
\begin{align}\langle \Psi_N^{(0)}, Q_{1/N}(f) \Psi^{(0)}_N\rangle
=\frac{N+1}{4\pi} \hskip -3 pt\int_{{S}^2} \hskip -3 pt d\Omega  \langle \Psi_N^{(0)}, \Omega \rangle_N \langle \Omega, Q_{1/N}(f)\Psi^{(0)}_N\rangle_N.\label{start}\end{align}
We argue that the above  limit for $N\to \infty$  can be  computed by restricting the integration set to  ${S}^2 \setminus E$, where $E$ is the closure of an open neighborhood of $S$  such that $E$ does not include $\Omega_+$ and $\Omega_-$.
Indeed,
\begin{align}
& \left|\frac{N+1}{4\pi} \hskip -3 pt\int_{E} \hskip -3 pt d\Omega  \langle \Psi_N^{(0)}, \Omega \rangle_N \langle \Omega, Q_{1/N}(f)\Psi^{(0)}_N\rangle_N\right|\leq\nonumber \\ &
\frac{N+1}{4\pi} \hskip -3 pt\int_{E} \hskip -3 pt d\Omega  |\langle \Psi_N^{(0)}, \Omega \rangle_N | \:  \|\Omega\rangle_N\| \|\Psi^{(0)}_N\| \|Q_{1/N}(f)\|, \end{align}
where 
$\|\Omega\rangle_N\|^2 = \|\Psi^{(0)}_N\|^2 =1$, and $\|Q_{1/N}(f)\|_N\to \|f\|_\infty$ as  $N\to \infty$. 
Shrinking $E$ if necessary,  assumption (c) and Remark \ref{remCU} part (a) 
 therefore imply that 
\begin{equation}
\left|\frac{N+1}{4\pi} \hskip -3 pt\int_{E} \hskip -3 pt d\Omega  \langle \Psi_N^{(0)}, \Omega \rangle_N \langle \Omega, Q_{1/N}(f)\Psi^{(0)}_N\rangle_N\right|\to 0.\end{equation}
 In summary, decomposing the integration set in (\ref{start})
as ${S}^2 = E \cup ({S}^2 \setminus E)$, we conclude that
\begin{align}L&= \lim_{N\to \infty}\langle \Psi_N^{(0)}, Q_{1/N}(f) \Psi^{(0)}_N\rangle\nonumber \\ &= \lim_{N\to \infty}
\frac{N+1}{4\pi} \hskip -3 pt\int_{{S}_E^2} \hskip -3 pt d\Omega  \langle \Psi_N^{(0)},\Omega \rangle_N \langle \Omega,Q_{1/N}(f)\Psi^{(0)}_N\rangle_N, \label{middle}\end{align}
where we have defined ${S}_E^2  ={S}^2 \setminus E$.
Taking this result into account and exploiting   (\ref{overcomplete}) again, our final task just consists of  computing the limit
\begin{equation}
L= \lim_{N\to \infty}\frac{(N+1)^2}{(4\pi)^2} \hskip -3 pt\int_{{S}^2_E} \hskip -3 pt d\Omega' \hskip -3 pt \int_{{S}^2} \hskip -3 pt d\Omega \langle \Psi_N^{(0)}, \Omega \rangle_N \langle \Omega, Q_{1/N}(f)\Omega' \rangle_N \langle \Omega', \Psi^{(0)}_N\rangle.
\end{equation}
In view of the definitions  of $Q_{1/N}$ and $\tilde{A}_0$,  and taking advantage of linearity, it is sufficient to prove the claim for  polynomials of the form 
\begin{equation}
f(\mathbf{x}) = x_{j_1}\cdots x_{j_M},\quad j_r \in \{1,2,3\}, \quad r=1,\ldots, M.\end{equation}
In this case, if $N\geq M$, we have 
\begin{equation}
Q_{1/N}(f) = S_{M,N} (\sigma_{j_1}\otimes \cdots \otimes \sigma_{j_M} \otimes I_2 \otimes \cdots \otimes I_2).
\end{equation}
The decisive observation for applying the technical results we have accumulated is that, as the states $|\Omega\rangle_N$  are factorized as in (\ref{om2}), we must have
\begin{equation}
\langle \Omega,  Q_{1/N}(f) \Omega'\rangle_N = \langle \Omega ,\Omega'\rangle_{N-M}  \langle \Omega, Q_{1/M}(f) \Omega'\rangle_M,
\end{equation}
where 
\begin{align}\label{finew}\langle \Omega, Q_{1/M}(f)\Omega'\rangle_M = \langle \Omega, \sigma_{j_1}\otimes \cdots \otimes \sigma_{j_M}\Omega\rangle_M. \end{align}
This  entails 
\begin{align}\label{start3}
L= \lim_{N\to \infty}\frac{(N+1)^2}{(4\pi)^2} \hskip -3 pt\int_{{S}^2_E} \hskip -3 pt d\Omega' \hskip -3 pt \int_{{S}^2} \hskip -3 pt d\Omega \langle \Psi_N^{(0)}, \Omega \rangle_N \langle \Omega ,\Omega'\rangle_{N-M}  \langle \Omega, Q_{1/M}(f) \Omega'\rangle_M\langle \Omega', \Psi^{(0)}_N\rangle.\end{align}
The idea is now to  apply Lemma \ref{lemma23} to the inner integral
\begin{equation}
\frac{N+1}{4\pi}\int_{{S}^2} \hskip -3 pt d\Omega \langle \Psi_N^{(0)}, \Omega \rangle_N \langle \Omega, \Omega'\rangle_{N-M}  \langle \Omega, Q_{1/M}(f) \Omega'\rangle_M, \end{equation}
where the function $h\equiv f$ in the hypotheses of the lemma is now specialised to  
\begin{equation}
{S}^2  \ni \Omega \mapsto
k(\Omega,\Omega')=  \langle \Omega, Q_{1/M}(f)\Omega'\rangle,
\end{equation}
which depends also {\em parametrically} on $\Omega'$.
The map ${S}^2 \times {S}^2  \ni (\Omega, \Omega') \mapsto k(\Omega,\Omega')$ is trivially  bounded and measurable (also in each variable separately). Furthermore, for every fixed $\Omega'\in {S}^2_E$, the restriction
${S}^2  \ni \Omega \mapsto k(\Omega,\Omega')$  is $C^1(A)$ with $A= 
{S}^2_E = {S}^2 \setminus E$ and the $\Omega$-derivatives of $k(\Omega,\Omega')$ are jointly continuous on $A\times A$. If necessary we can redefine   $E$ as a smaller set,  in order that  the continuity of those derivatives  
remains  still valid on the compact set $\overline{A}$. In this way, we obtain
\begin{equation}
\|d_\Omega K(\cdot,\cdot) \|_\infty^{(A\times A)}=   \sup_{\Omega, \Omega' \in A}{\bf g}_\Omega(d_\Omega K(\Omega,\Omega'), d_\Omega K(\Omega,\Omega'))< \infty,\end{equation}
where 
$K(\Omega, \Omega') = |\langle \Omega, Q_{1/M}(f) \Omega'\rangle - \langle \Omega', Q_{1/M}(f) \Omega'\rangle|^2$.
For every fixed $\Omega'\in {S}^2_E $, we can apply  Lemma \ref{lemma23} with the open set $A= {S}^2_E $ in common for all $\Omega'$. Thus we obtain a  first $\Omega'$-dependent   bound 
\begin{align}&
\left|\frac{N+1}{4\pi}\int_{{S}^2} \hskip -3 pt d\Omega \langle \Psi_N^{(0)}, \Omega \rangle_N \langle \Omega ,\Omega'\rangle_{N-M}  \langle \Omega, Q_{1/M}(f) \Omega'\rangle_M -   \langle \Psi_N^{(0)}, \Omega' \rangle_N \langle \Omega', Q_{1/M}(f) \Omega'\rangle_M \right|\nonumber \\ & \mbox{} \hspace{50pt}
\leq \frac{K^{(A)}\|k(\cdot,\Omega')\|_\infty+\sqrt{C\|k(\cdot,\Omega')\|^2_\infty +  D^{(A)}\|d_\Omega K(\cdot,\Omega')\|^{(A)}_\infty}}{(N-M)^{1/4}}.\end{align}
where  according to Lemma \ref{lemma23}  the constants $K^{(A)},C, D^{(A)}$ do not depend on the  function $k(\cdot,\Omega')$, i.e., they do not depend on  $\Omega'$ (the constants $K^{(A)}, D^{(A)}$ do depend on the  set $A$ which, however, is the same  for all choices of $\Omega'$).  Finally,  since
\begin{equation}
\|k(\cdot,\Omega')\|_\infty \leq \|k(\cdot,\cdot)\|_\infty\quad\mbox{and}\quad\|d_\Omega K(\cdot,\Omega')\|^{(A)}_\infty \leq \|d_\Omega K(\cdot,\cdot)\|^{(A\times A)}_\infty,\end{equation}
for sufficiently large $N$ we also have a $\Omega'$-uniform bound:
\begin{align}
&\left|\frac{N+1}{4\pi}\int_{{S}^2} \hskip -3 pt d\Omega \langle \Psi_N^{(0)}, \Omega \rangle_N \langle \Omega ,\Omega'\rangle_{N-M}  \langle \Omega, Q_{1/M}(f)\Omega'\rangle_M -   \langle \Psi_N^{(0)}, \Omega' \rangle_N \langle \Omega', Q_{1/M}(f)\Omega'\rangle_M \right|\nonumber \\
&\leq \frac{K^{(A)}\|k(\cdot,\cdot)\|_\infty+\sqrt{C\|k(\cdot,\cdot)\|^2_\infty +  D^{(A)}\|d_\Omega K(\cdot,\cdot)\|^{(A\times A)}_\infty}}{(N-M)^{1/4}}=  \frac{C^{(A)}}{(N-M)^{1/4}}.\label{unifE}\end{align}
Plugging  this result  in the right-hand side of (\ref{start3}), we immediately have
\begin{align}
L &=  \lim_{N\to \infty}\frac{N+1}{4\pi}  \int_{{S}^2_E} d\Omega' \langle \Psi^{(0)}_N, \Omega'\rangle_N 
\langle \Omega', Q_{1/M}(f) \Omega'\rangle_M \langle \Omega', \Psi^{(0)}_N\rangle\nonumber \\
&+\lim_{N\to \infty}\frac{N+1}{4\pi}  \int_{{S}^2_E} d\Omega'  R_N(\Omega')\langle \Omega', \Psi^{(0)}_N\rangle,\label{secondL}\end{align}
where $R_N(\Omega')$ is given by the expression
\begin{equation}
\frac{N+1}{4\pi}\int_{{S}^2} \hskip -3 pt  \langle \Psi_N^{(0)}, \Omega \rangle_N \langle \Omega ,\Omega'\rangle_{N-M}  \langle \Omega, Q_{1/M}(f) \Omega'\rangle_M d\Omega -   \langle \Psi_N^{(0)}, \Omega' \rangle_N \langle \Omega', Q_{1/M}(f)\Omega'\rangle_M.\end{equation}
Let us focus on the second limit in (\ref{secondL}). First of all, observe that (b) in Assumption \ref{assumption}, together with (a) in Proposition \ref{lemmadelta} with $\ell=1/2$ and $f=1$ constant, imply that the integral
$\int_{{S}^2} (N+1) |\langle \Omega', \Psi^{(0)}_N\rangle| d\Omega'$ is bounded when $N$ increases, so that  the corresponding  integral over ${S}^2_E$
 must be bounded as well. Since 
 \begin{equation}
 |R_N(\Omega')| \leq C^{(A)}/(N-M)^{1/4},
\end{equation}
 where $C^{(A)}$ from (\ref{unifE}) does not depend on $\Omega'$,  we conclude that 
 the second limit in (\ref{secondL}) is $0$.
In summary,
\begin{equation}
L =  \lim_{N\to \infty}\frac{N+1}{4\pi}  \int_{{S}^2\setminus E} d\Omega' |\langle \Psi^{(0)}_N, \Omega'\rangle_N|^2
\langle \Omega', Q_{1/M}(f) \Omega'\rangle_M.\end{equation}
We can rearrange  the above integral  into
\begin{equation}
L =  \lim_{N\to \infty}\frac{N+1}{4\pi}  \int_{{S}^2} d\Omega' Z(\Omega')
|\langle \Psi^{(0)}_N, \Omega'\rangle_N|^2,\end{equation}
where $Z(\Omega') = \langle \Omega', Q_{1/M}(f)\Omega'\rangle_M$ if $\Omega' \in {S}^2\setminus E$ and 
$Z(\Omega') = 0$ otherwise. With this change,  we may apply Lemma \ref{lemmaA} to the function $Z$, because it satisfies all requirements, finding
\begin{equation}
\lim_{N\to \infty}\langle \Psi_N^{(0)}, Q_{1/N}(f) \Psi^{(0)}_N\rangle = L = \half (Z(\Omega_+) + Z(\Omega_-)).
\end{equation}
However,   since $\Omega_\pm \in {S}^2\setminus E$, the very definition of $Z$ yields
\begin{equation}
\lim_{N\to \infty}\langle \Psi_N^{(0)}, Q_{1/N}(f) \Psi^{(0)}_N\rangle  =  \frac{1}{2}\langle \Omega_+, Q_{1/M}(f) \Omega_+\rangle_M + \frac{1}{2}\langle \Omega_-, Q_{1/M}(f)\Omega_-\rangle_M.
\end{equation}
From (\ref{definitionf1}) -  (\ref{definitionf2}),
(\ref{deformationqunatizaion}),  (\ref{deformationqunatizaion2}), (\ref{om2}), and
(\ref{finew}) we have
\begin{equation}
\langle \Omega_\pm, Q_{1/M}(f) \Omega_\pm\rangle_M = \omega^{(0)}_\pm (f), \end{equation}
so that finally, 
\begin{equation}
\lim_{N\to \infty}\langle \Psi_N^{(0)}, Q_{1/N}(f) \Psi^{(0)}_N\rangle =  \half  \omega^{(0)}_+(f) + \half  \omega^{(0)}_-(f)=\omega^{(0)}(f), \end{equation}
and the proof is complete. 
\end{proof}

\section*{Acknowledgments} Christiaan van de Ven is Marie Sk\l odowska-Curie fellow of the Istituto Nazionale di Alta Matematica and is funded by the INdAM Doctoral Programme in Mathematics and/or Applications co-funded by Marie 
Sk\l odowska-Curie Actions, INdAM-DP-COFUND-2015, grant number 713485. The authors thank  C. Agostinelli for his help with  numerical simulation programs, as well as F. Serra Cassano for helpful technical discussions.
\appendix
\section{Proof of some technical propositions}\label{proofs}
\subsection{Quantization map}
{\bf Proof of Lemma \ref{lemmainj}}.  Since $\chi$ is linear, the claim is equivalent to the implication $\chi(z)=0 \raw z=0$, where $z\in Z$ has the generic form
\begin{align}
z=  c_0I_k\oplus c_1^{j_1}b_{j_1}\oplus c_2^{j_1j_2}b_{j_1}\otimes_sb_{j_2}\oplus...\oplus c_M^{j_1\cdot\cdot\cdot j_M}b_{j_1}\otimes_s\cdot\cdot\cdot\otimes_sb_{j_M},\label{zN3}
\end{align}
The requirement 
$\chi(z)=0$ means
$\chi(z)(\omega)=0$, for all $\omega\in C(S({B}))$. Thinking of the states $\omega$ as density matrices of $\mathcal{D}_k$ 
represented by the affine parametrization $(\mathcal{Q}_k,F_k)$ defined in  (\ref{FK}), the map 
 $S({B}) \ni \omega \mapsto \chi(z)(\omega)$ is clearly the restriction of a polynomial in $k^2-1$  
variables $(x_1,\ldots,x_{k^2-1})\in \mathbb{R}$, which determine $\omega$ through $F_k$ when 
restricted to $\mathcal{Q}_k$, that is,
\begin{equation}\chi(z)(\omega) = c_0 +  c_1^{j_1}  x_{j_1}  + c_2^{j_1j_2}  x_{j_1} x_{j_2} + \cdots + c_M^{j_1\ldots j_M}x_{j_1}\cdots x_{j_M},\end{equation}
where we have taken (\ref{Fkinv}) into account.
Since the interior of $\mathcal{Q}_k$ is not empty (and open by definition)
and the polynomial therefore vanishes on some open nonempty set, it vanishes everywhere, hence all coefficients $c_N^{j_1\ldots j_N}$ are zero. From  (\ref{zN3}), 
we have proven that, for $z \in Z$ the condition $\chi(z)=0$ implies that $z=0$, as wanted.  \hfill $\Box$\\

\noindent {\bf Proof of Lemma \ref{lemmacommSN}}.
The definition (\ref{defSN}) of $S_N$ implies 
\begin{align} & S_N(a_1 \otimes \cdots \otimes a_N)S_N(a'_1 \otimes \cdots \otimes a'_N) = \frac{1}{N!^2} \sum_{\sigma\in {\cal P}(N)}\sum_{\pi\in {\cal P}(N)}
a_{\sigma(1)}a'_{\pi(1)} \otimes \cdots \otimes a_{\sigma(N)} a'_{\pi(N)}\nonumber \\
& \mbox{} \hspace{50pt} =\frac{1}{N!^2} \sum_{\sigma}\sum_{\pi}
a_{\sigma(1)}a'_{\sigma \circ\pi(1)} \otimes \cdots \otimes a_{\sigma(N)} a'_{\sigma\circ \pi(N)}, \end{align}
since, for any given  $\sigma \in {\cal P}(N)$, the map $\pi\mapsto \sigma\circ \pi$ is a bijection of the permutation group ${\cal P}(N)$. Exploiting the definition of $S_N$ once again yields  
\begin{equation} \frac{1}{N!^2} \sum_{\sigma}\sum_{\pi}
a_{\sigma(1)}a'_{\sigma \circ\pi(1)} \otimes \cdots \otimes a_{\sigma(N)} a'_{\sigma\circ \pi(N)} 
= \frac{1}{N!} \sum_{\pi \in {\cal P}(N)} S_N\left(a_1a'_{\pi(1)} \otimes \cdots \otimes a_N a'_{\pi(N)}\right), \end{equation}
so that
\begin{equation}S_N(a_1 \otimes \cdots \otimes a_N)S_N(a'_1 \otimes \cdots \otimes a'_N) = \frac{1}{N!} \sum_{\pi \in {\cal P}(N)} S_N\left(a_1a'_{\pi(1)} \otimes \cdots \otimes a_N a'_{\pi(N)}\right).\end{equation}
A similar arguments gives
\begin{equation}S_N(a'_1 \otimes \cdots \otimes a'_N)S_N(a_1 \otimes \cdots \otimes a_N) = \frac{1}{N!} \sum_{\pi \in {\cal P}(N)} S_N\left(a'_{\pi(1)}a_1 \otimes \cdots \otimes  a'_{\pi(N)}a_N\right),\end{equation}
proving the claim.
 \hfill $\Box$\\

\noindent {\bf Proof of Equation (\ref{speriamo})}.
We have to compute the number of all possible bijective maps $f_\pi$  (corresponding to  permutations $\pi^{-1}$ when $\pi \in {\cal P}(N)_K$) whose domain consists of the following $N$ elements: $L$ elements  $\{b_{j_1}, \ldots, b_{j_L}\}$ 
together with $N-L$ identities $I_k$. 
All those elements are viewed as {\em distinct objects}. The codomain of $f_\pi$ consists of $N$ elements: $M$ elements  $\{b_{i_1}, \ldots, b_{i_M}\}$ 
together with $N-M$ identities $I_k$. Again, all those elements are viewed as distinct objects. We assume $L\leq M$ and the maps we want to count are those  that map  {\em exactly} $K$ elements among those in  $\{b_{j_1}, \ldots, b_{j_L}\}$ to distinct elements of the subset    $\{b_{i_1}, \ldots, b_{i_M}\}$   of the codomain. 

We start by choosing $K$ couples whose  first element is chosen from the set  $\{b_{j_1}, \ldots, b_{j_L}\}$
and the corresponding second element (the image of the former according to $f_\pi$) is from the set $\{b_{i_1}, \ldots, b_{i_M}\}$.
We can do this in 
\begin{equation}\frac{L(L-1) \cdots (L-K+1) M(M-1) \ldots (M-K+1)}{K!} \end{equation}
different  ways, where the factor $1/K!$ is needed because the order we use to select the said $K$ couples does not matter.
This number can be rewritten as
\begin{equation}\frac{L!}{(L-K)!}\frac{M!}{(M-K)!}\frac{1}{K!}.\label{KL}\end{equation}
We have now to assign the images via $f_\pi$ of the remaining $L-K$ elements  of the set $\{b_{j_1}, \ldots, b_{j_L}\}$ in the domain  (having removed the $K$ elements as above),
which must be injectively  mapped to the subset of the codomain consisting of $N-M$ unit elements $I_k$.
Keeping the initial order of those $L-K$ elements, the image of the first one can be taken in $(N-M)$ ways, the image of the second one
in $(N-M-1)$ ways, and so on.
This leads to a number of
 \begin{equation}(N-M)(N-M-1) \cdots (N-M-(L-K)+1)=
\frac{(N-M)!}{(N-L-M+K)!}\label{KL2} \end{equation} choices. The total number of choices is the product of \er{KL} and \er{KL2}.
To conclude,  we have to injectively assign the values of the reamaing $N-L$ elements $I_k$ of the domain of $f_\pi$ into the set of remaining $N-L$ values of the codomain: this gives $(N-L)!$ choices. The total amount of choices is then identical  to  (\ref{speriamo}):
\begin{equation}\frac{1}{K!}\frac{L!}{(L-K)!}\frac{M!}{(M-K)!}\frac{(N-L)!(N-M)!}{(N-L-M+K)!}.\end{equation}
\hfill $\Box$

\subsection{Classical limit}
\noindent {\bf Proof of Proposition \ref{lemmadelta}}.  From now on, ${S}^2$ is viewed as an embedded submanifold of $\mathbb{R}^3$
endowed with the differentiable structure, the  metric  and the associated measure (which coincides with $d\Omega$) 
induced by $\mathbb{R}^3$.  

\textbf{Proof of (a)}.  Since the   measure $d\Omega$ and $\cos \Phi(\Omega,\Omega')$ are both rotationally invariant,
 we assume without loss of generality that   $\Omega'$ coincides with ${\bf e}_z$ and we only demonstrate  the claim for this choice. Writing $N'= N+1$, for $\ell >0$ we have 
\begin{equation}I_N= \frac{\ell N'}{4\pi} \int_{{S}^2}  h(\Omega)|\langle \Omega,\Omega'\rangle_N|^{2\ell} d \Omega =
\frac{\ell N'}{2^{\ell N} 4\pi} \int_{[0,\pi) \times (-\pi,\pi]}  \hskip -30 pt h(\theta, \phi) (1+ \cos \theta)^{\ell N} \sin \theta d\theta d\phi   .\end{equation}
Notice that the integral is well defined because $|\langle \Omega, \Omega'\rangle_N|^2$ is smooth and bounded by some constant when
$\Omega$ ranges in  ${S}^2$, $h$ is $L^1$ with respect to $d\Omega$ because it is measurable and bounded, and  ${S}^2$ has finite measure. The same argument applies 
to the integrals appearing in the rest of the proof.
To go on,  we decompose 
\begin{equation}h(\Omega) = h(\Omega') + h(\Omega) -h(\Omega')\end{equation}
so that
\begin{align}I_N =  h(\Omega')\frac{\ell N'}{2^{\ell N} 4\pi} \int_{{S}^2} \hskip -5 pt (1+ \cos \theta)^{\ell N} \sin \theta d\theta d\phi
+  \frac{\ell N'}{2^{\ell N} 4\pi} \int_{{S}^2}  {\hskip -5pt} [h(\Omega) - h(\Omega') ] (1+ \cos \theta)^{\ell N} d \Omega.  \label{INN}\end{align}
A direct computation leads to 
\begin{align}\frac{\ell N' h(\Omega')}{2^{\ell N} 4\pi} \int_{{S}^2} \hskip -5 pt (1+ \cos \theta)^{\ell N }\sin \theta 
d\theta d\phi = h(\Omega') \frac{\ell(N+1)}{2^{\ell N+1}}\frac{2^{\ell N+1}}{\ell N+1}  \to  h(\Omega'),   \label{limlimlim}\end{align}
as $N\to \infty$.
To conclude the proof, we need to show that 
\begin{equation} \frac{\ell N'}{2^{\ell N} 4\pi} \int_{{S}^2}  {\hskip -5pt} [h(\Omega) - h(\Omega') ] (1+ \cos \theta)^{\ell N} d \Omega  \to 0  \quad \mbox{for $N\to \infty$.}\end{equation}
Actually, it is sufficient to establish that
\begin{align} \label{intdef}
\frac{\ell N'}{2^{\ell N} 4\pi} \int_{A}  {\hskip -5pt} |h(\Omega) - h(\Omega')| |1+ \cos \theta|^{\ell N} d \Omega  \to 0  \quad \mbox{for $N\to \infty$,}\end{align}
where $A \subset {S}^2$ is an open neighborhood of $\Omega'$, in particular the one appearing in the hypothesis where $f$ is $C^1$.
In fact, on ${S}^2 \setminus A$ we have   $\left|\frac{1+ \cos \theta}{2}\right| \leq K<1$  for some $K \in (0,1)$  so that
$\ln K <0$ and
\begin{align}  \frac{\ell N'}{4\pi}   \left|\frac{1+ \cos \theta}{2}\right|^{\ell N}  \leq \frac{\ell(N+1)}{4\pi} e^{\ell N \ln K} \to 0 \quad \mbox{for $N\to \infty$.}\label{usefullater}\end{align} 
Therefore,
\begin{align} \label{estim2}
\lim_{N\raw\infty} 
\frac{\ell N'}{2^{\ell N} 4\pi} \int_{{S}^2 \setminus A}  {\hskip -5pt} |h(\Omega) - h(\Omega') | |1+ \cos \theta|^{\ell N} d \Omega
\leq 2 \|h\|_\infty  \ell N'e^{\ell N \ln K}
= 0 .\end{align}
Restricting the inital $set A$  if necessary, let us equip $A$ with a local chart (of the differentiable structure induced from $\mathbb{R}^3$) obtained by the canonical projection onto the $x,y$ plane (we use this chart because the chart of the coordinates 
$\theta, \phi$ is singular at $\Omega'$, here coinciding with the north pole). It is not difficult to see that, in this coordinate patch where we can safely assume $\cos \theta >0$, we have 
\begin{equation} \int_{A}  {\hskip -5pt} |h(\Omega) - h(\Omega') |\cdot | 1+ \cos \theta|^{\ell N} d \Omega
=  \int_A   {\hskip -5pt} [h(x,y) - h(0,0) ] \frac{(1+ \sqrt{1-x^2-y^2})^{\ell N}}{\sqrt{1-x^2-y^2}} dx dy\end{equation}
where we exploited the fact that the induced measure from $\mathbb{R}^3$ is  $dxdy/\sqrt{1-x^2-y^2}$ in that coordinate patch.
Assuming $f$ of class $C^1$ in coordinates $x,y$ on $A$, if necessary redefine again $A$ as  a smaller open neighborhood of  $(0,0)$
whose closure (which is compact) is contained in the initial  $A$. 
Lagrange's theorem applied to the segment joining $(x,y)$ and $(0,0)$ then leads to the estimate
\begin{equation} |h(x,y) - h(0,0)| = \left|\frac{\partial h}{\partial x}|_{(x',y')} x + \frac{\partial h}{\partial y}|_{(x',y')} y\right| \leq L^{(A)}_f  r\end{equation}
where $(x',y')$ is a point in $A$  depending on $(x,y)$, and 
\begin{align}L^{(A)}_h = \sup_A \sqrt{\left|\frac{\partial h}{\partial x}\right|^2 + \left|\frac{\partial h}{\partial y}\right|^2} < \infty,\label{Cf}\end{align}
which exists because $f$ is $C^1$ on the compact set $\overline{A}$,
and where we adopted plane polar coordinates $x= r \cos \vartheta$, $y= r \sin \vartheta$ with $r=\sqrt{x^2+y^2}$. 
Collecting all results,  using $z= \cos \theta =\sqrt{1-x^2-y^2}=  \sqrt{1-r^2}$,   we have 
\begin{equation}\frac{\ell N'}{2^N 4\pi} \int_{A}  {\hskip -5pt} |h(\Omega) - h(\Omega') \|1+ \cos \theta|^{\ell N} d \Omega 
\leq  \frac{L^{(A)}_h \ell N'}{2^{\ell N} 4\pi} \int_{\{(r,\vartheta) \:|\: 0\leq r \leq 1\}}   {\hskip -5pt} \frac{(1+ \sqrt{1-r^2})^{\ell N}}{\sqrt{1-r^2}} r^2 dr d\vartheta .\end{equation}
Integrating with respect to $\vartheta$,  (\ref{intdef}) holds. This ends the proof of {\bf (a)}, provided
\begin{equation} J_N =\frac{\ell N'}{2^{\ell N+1}} \int_0^1   {\hskip -5pt} \frac{(1+ \sqrt{1-r^2})^{\ell N}}{\sqrt{1-r^2}} r^2 dr \to 0 \quad \mbox{for $N\to \infty$.}\end{equation}
Changing variable to $x= \sqrt{1-r^2}$ and next to $t= \frac{1+x}{2}$, we find
\begin{align}J_N &= 2\ell N' \int_{1/2}^1 t^{\ell N+1/2}\sqrt{1-t} dt \leq 2\ell N' \int_{0}^1 t^{\ell N+1/2}\sqrt{1-t} dt\nonumber \\
&=  2\ell (N+1) \frac{\Gamma(3/2) \Gamma(\ell N+3/2)}{\Gamma(\ell N+3)}\end{align}
Stirling's estimate then yields $|J_N| \leq  L/\sqrt{\ell N}$ for some constant $L>0$. With the previous discussion, this gives the key to (\ref{intdef}) and hence to Assumption {\bf (a)}, viz.
\begin{align}\left|\frac{\ell N'}{2^{\ell N} 4\pi} \int_{A}  {\hskip -5pt} [h(\Omega) - h(\Omega') ] (1+ \cos \theta)^{\ell N} d \Omega\right|\leq  L L^{(A)}_h / \sqrt{\ell N}.  \label{JNO}\end{align}

\textbf{Proof of (b)}.  From (\ref{INN}), the identity in (\ref{limlimlim}), (\ref{estim2}), and (\ref{JNO}) we have 
\begin{align}
& \left| h(\Omega') - \frac{\ell N'}{4\pi} \int_{{S}^2}  h(\Omega)|\langle \Omega, \Omega'\rangle_N|^{2\ell } d \Omega\right|\nonumber \\ & \leq  |h(\Omega')|\left|1-\frac{\ell N+\ell}{\ell N+1} \right| +  2 \|h\|_\infty  \ell N'e^{\ell N \ln K} + \ell^{-1/2}L L^{(A)}_h /\sqrt{N}
\nonumber \\ &
\leq   \|h\|_\infty  \frac{|1-\ell|}{\ell N+1} + \|h\|_\infty   2\ell N'e^{\ell N \ln K} + \ell^{-1/2} LL^{(A)}_h  /\sqrt{N},\end{align}
where $K \in (0,1)$ does not depend on $h$.
With a standard argument one proves that, for some constant $C^{(A)}\geq 0$ independent of $h$, the constant $L_h^{(A)}$ in (\ref{Cf}) satisfies
\begin{equation}L^{(A)}_h  \leq C^{(A)} \|dh\|^{(A)}_\infty,\end{equation}
where, if ${\bf g}_\Omega$ is the natural inner product on $T_\Omega^*{S}^2$ induced from $\mathbb{R}^3$,
\begin{equation}\|dh\|_\infty = \sup_{\Omega\in A}\sqrt{{\bf g}_\Omega(d\overline{h},dh)}.\end{equation}
Inequality (\ref{delta2}) is therefore true
defining $C^{(A)}_\ell= \ell^{-1/2}L C^{(A)}$, since \begin{equation}\|h\|_\infty \left(\frac{|1-\ell|}{\ell N+1} + 2\ell (N+1)e^{\ell N \ln K}\right) \leq B_\ell\|h\|_\infty  /\sqrt{N}.\end{equation} 
Recalling that $\ln K <0$, we finally obtain 
 \begin{equation}B_\ell= \sup_{N \in \mathbb{N}}\sqrt{N} \left(\frac{|1-\ell|}{\ell N+1} + 2\ell (N+1)e^{\ell N \ln K}\right)< \infty\end{equation}   Notice that, by construction $B_\ell$ and $C_\ell^{(A)}$ do not depend on $\Omega'$.
 $\hfill$ $\Box$
\medskip

\noindent {\bf Proof of Lemma \ref{lemmaA}}. 
 We prove the claim for a real-valued  $h$, the extension the the complex case being trivial.
In the rest of the proof we always assume  that $A$ is sufficiently small according to Remark \ref{remA}.(1), 
 keeping the requirement $A\ni \Omega_\pm$.
In particular, we suppose that $A = A_+\cup A_-$ where $A_+$ and $A_-$ are sufficiently small open neighborhoods of $\Omega_+$ and $\Omega_-$ respectively. 

We start the proof by  observing that, taking advantage of a finite  partition of unit, we can decompose $h = h_+ + h_-$ where $h_\pm$  are measurable,  bounded  and $C^1$ in $A$ and satisfy
$h_+=0$ in a neighborhood of $\Omega_-$, and  $h_-=0$ in a neighborhood of $\Omega_+$. If the claim is valid for each of these functions, 
by linearity it is also valid for $h$. Therefore, in the rest of the proof
we assume that $h$ also vanishes in a neighborhood of $\Omega_-$  in addition to satisfying the hypotheses in the statement of the lemma (the other case can be treated similarly).

As a second observation, we notice that (c) in Assumption \ref{assumption} and Remark \ref{remCU}.(a),
and  the proof of Proposition \ref{lemmadelta} with (\ref{usefullater}),
 immediately imply that 
\begin{align}
 \frac{N+1}{4\pi}\int_{{S}^2 \setminus A}|\langle \Psi^{(0)}_N, \Omega \rangle|^{2} d\Omega &\to 0;\nonumber \\
  \frac{N+1}{4\pi}\int_{{S}^2 \setminus A}|\langle \Omega_\pm, \Omega \rangle|^{2} d\Omega&\to 0,
  \end{align} 
  respectively, 
  for every open set $A$ containing
$\Omega_\pm$. In view of those remarks and using 
\begin{align}
\limsup_n (a_n+b_n)&= \limsup_n a_n + \limsup_n b_n;\\
\liminf_n (a_n+b_n)&= \liminf_n a_n + \liminf_n b_n,
  \end{align} 
  if either $\{a_n\}_{n\in \mathbb{N}}$ or  $\{b_n\}_{n\in \mathbb{N}}$ has a 
limit in $\mathbb{R}$, we can write
\begin{align}
& \limsup_N \int_{{S}^2}  \frac{N+1}{4\pi} \left(|\langle \Psi^{(0)}_N ,\Omega
\rangle_N|^{2} - \half  |\langle \Omega, \Omega_+ \rangle_N|^{2}-\half  |\langle \Omega, \Omega_- \rangle_N|^{2}\right)h(\Omega)d\Omega\nonumber \\
& =\limsup_N \int_{A}  \frac{N+1}{4\pi} \left(|\langle \Psi^{(0)}_N ,\Omega
\rangle_N|^{2} - \half  |\langle \Omega, \Omega_+ \rangle_N|^{2}-\half  |\langle \Omega, \Omega_- \rangle_N|^{2}\right)h(\Omega)d\Omega \nonumber \\ &=\limsup_N \int_{A_+}  \frac{N+1}{4\pi} \left(|\langle \Psi^{(0)}_N ,\Omega
\rangle_N|^{2} - \half  |\langle \Omega, \Omega_+ \rangle_N|^{2}\right)h(\Omega)d\Omega,
\end{align} 
since the limit of the integration  over ${S}^2\setminus A$ is zero, and in the last line we exploited the fact that 
$h$ vanishes around $\Omega_-$. We can now decompose
\begin{align} & \int_{A_+} \frac{N+1}{4\pi} \left(|\langle \Psi^{(0)}_N ,\Omega
\rangle_N|^{2} - \half  |\langle \Omega, \Omega_+ \rangle_N|^{2}\right)h(\Omega)d\Omega \nonumber \\ & = h(\Omega_+)\int_{A_+}  \frac{N+1}{4\pi} \left(|\langle \Psi^{(0)}_N ,\Omega
\rangle_N|^{2} - \half  |\langle \Omega, \Omega_+ \rangle_N|^{2}\right)d\Omega \nonumber \\ & + \int_{A_+}  \frac{N+1}{4\pi} \left(|\langle \Psi^{(0)}_N ,\Omega
\rangle_N|^{2} - \half  |\langle \Omega, \Omega_+ \rangle_N|^{2}\right)(h(\Omega)- h(\Omega_+))d\Omega.
\end{align}
Taking advantage of  (\ref{invPSI})  and of the identity 
\begin{equation}
|\langle \Omega_{\pi-\theta, -\phi}, \Omega_\pm \rangle_N| =  |\langle \Omega_{\theta,\phi}, \Omega_\mp \rangle_N| 
\end{equation}
arising from   (\ref{ZeD24}),
and choosing  $A_-$ as the image of $A_+$ under the symmetry $$\theta \to \pi-\theta,\:\:  \phi \to -\phi$$ that swaps $\Omega_+$ and $\Omega_-$, the first integral on the right-hand side 
can be rewritten as
\begin{equation} \half h(\Omega_+)\int_{A_+\cup A_-}  \frac{N+1}{4\pi} \left(|\langle \Psi^{(0)}_N ,\Omega
\rangle_N|^{2} - \half  |\langle \Omega, \Omega_+ \rangle_N|^{2} - \half  |\langle \Omega, \Omega_- \rangle_N|^{2} \right)d\Omega.\end{equation}
Since  $A_+\cup A_- =A$, the limit for $N\to \infty$ of the integral above vanishes because it is the difference of the limit of the analogous integral extended to  the whole ${S}^2$,
 which vanishes 
due to the assumption 
(a), and the analogous limit when integrating over ${S}^2\setminus A$,  which vanishes as well, as already observed.
Hence 
\begin{align}
&\limsup_N \int_{{S}^2} \frac{N+1}{4\pi}  \left(|\langle \Psi^{(0)}_N ,\Omega
\rangle_N|^{2} - \half  |\langle \Omega, \Omega_+ \rangle_N|^{2}-\half  |\langle \Omega, \Omega_- \rangle_N|^{2}\right)h(\Omega)d\Omega \nonumber \\ & = \limsup_N\int_{A_+}  \frac{N+1}{4\pi} \left(|\langle \Psi^{(0)}_N ,\Omega
\rangle_N|^{2} - \half  |\langle \Omega, \Omega_+ \rangle_N|^{2}\right)(h(\Omega)- h(\Omega_+))d\Omega
\nonumber \\ & \leq \limsup_N\left|\int_{A_+}  \frac{N+1}{4\pi} \left(|\langle \Psi^{(0)}_N ,\Omega
\rangle_N|^{2} - \half  |\langle \Omega, \Omega_+ \rangle_N|^{2}\right)(h(\Omega)- h(\Omega_+))d\Omega\right|
\nonumber \\ & \leq  \limsup_N  \int_{A_+}  \frac{N+1}{4\pi} \left||\langle \Psi^{(0)}_N ,\Omega
\rangle_N|^{2} - \half  |\langle \Omega, \Omega_+ \rangle_N|^{2}\right| |h(\Omega)- h(\Omega_+)| d\Omega \leq G \epsilon,
\end{align}
where we exploited assumption (b) and the continuity of $h$ at $\Omega_+$, choosing the open set $A_+\ni \Omega_+$ such that $|h(\Omega)- h(\Omega_+)| <\epsilon$
 is guaranteed if $\Omega \in A_+$.
In summary,
\begin{equation}\overline{I}= \limsup_N \int_{{S}^2} \frac{N+1}{4\pi}  \left(|\langle \Psi^{(0)}_N ,\Omega
\rangle_N|^{2} - \half  |\langle \Omega, \Omega_+ \rangle_N|^{2}-\half  |\langle \Omega, \Omega_- \rangle_N|^{2}\right)h(\Omega)d\Omega \leq G\epsilon.\end{equation}
This entire reasoning can be repeated changing the sign in the integrand from schratch, i.e., referring to 
\begin{equation}\limsup_N \int_{{S}^2}  \frac{N+1}{4\pi} \left(\half  |\langle \Omega, \Omega_+ \rangle_N|^{2}+
\half  |\langle \Omega, \Omega_- \rangle_N|^{2}- |\langle \Psi^{(0)}_N ,\Omega
\rangle_N|^{2}\right)h(\Omega)d\Omega,\end{equation}
finding
\begin{equation}\limsup_N \int_{{S}^2} -\frac{N+1}{4\pi}  \left(|\langle \Psi^{(0)}_N ,\Omega
\rangle_N|^{2} - \half  |\langle \Omega, \Omega_+ \rangle_N|^{2}-\half  |\langle \Omega, \Omega_- \rangle_N|^{2}\right)h(\Omega)d\Omega \leq G\epsilon.\end{equation}
Since $\limsup_n(-a_n)= -\liminf_n a_n$, we conclude that
\begin{equation}-G\epsilon \leq \underline{I} = \liminf_N  \int_{{S}^2} \frac{N+1}{4\pi}  \left(|\langle \Psi^{(0)}_N ,\Omega
\rangle_N|^{2} - \half  |\langle \Omega, \Omega_+ \rangle_N|^{2}-\half  |\langle \Omega, \Omega_- \rangle_N|^{2}\right)h(\Omega)d\Omega.\end{equation}
In summary, 
\begin{equation}-G\epsilon \leq \underline{I} \leq \overline{I} \leq G\epsilon \quad \mbox{for every $\epsilon>0$}\end{equation}
  and where $G\geq 0$ is given.
Therefore,
\begin{equation}\lim_{N\to \infty} \int_{{S}^2} \frac{N+1}{4\pi}  \left(|\langle \Psi^{(0)}_N ,\Omega
\rangle_N|^{2} - \half  |\langle \Omega, \Omega_+ \rangle_N|^{2}-\half  |\langle \Omega, \Omega_- \rangle_N|^{2}\right)h(\Omega)d\Omega=0.\end{equation}
Using  Proposition \ref{lemmadelta},
we conclude that
\begin{align}
& \lim_{N\to \infty} \frac{(N+1)}{4\pi}
\int_{{S}^2}  \hskip -6 pt h(\Omega)
 |\langle \Psi^{(0)}_N , \Omega\rangle_N|^{2} d \Omega \nonumber \\ &  = \lim_{N\to \infty} \frac{(N+1)}{8\pi}
\int_{{S}^2}  \hskip -6 pt h(\Omega)
 |\langle \Omega_+ ,\Omega\rangle_N|^{2} d \Omega
+  \lim_{N\to \infty} \frac{(N+1)}{8\pi}
\int_{{S}^2}  \hskip -6 pt h(\Omega)
 |\langle \Omega_- ,\Omega\rangle_N|^{2} d \Omega \nonumber \\ & = \half  h(\Omega_+) + \half h(\Omega_-), \end{align}
 ending the proof.
 $\hfill$ $\Box$
\newpage

\noindent {\bf Proof of Lemma \ref{lemma23}}.  First of all, notice that the absolute value in the left-hand side of (\ref{tpb})
 can be rearranged into a more useful form:
\begin{align} &\int_{{S}^2}   \frac{N+1}{4\pi}\langle \Psi^{(0)}_N, \Omega\rangle_N h(\Omega) \langle \Omega,\Omega'\rangle_{N-M}
d\Omega - 
\langle \Psi^{(0)}_N, \Omega'\rangle_N h(\Omega') \nonumber  \\
&= \frac{N+1}{4\pi}\int_{{S}^2} d\Omega \left(
 \langle \Psi^{(0)}_N, \Omega\rangle_N \langle \Omega,\Omega'\rangle_{N-M}h(\Omega) -\langle \Psi^{(0)}_N, \Omega\rangle_N \langle \Omega,\Omega'\rangle_{N}h(\Omega')\right),
\end{align}
where we exploited (\ref{overcomplete}) in the second summand of the first line. We intend to prove the claim with this rearranged form.
Let us start by establishing the claim in the simplest case $f=1$, defining 
\begin{equation} I_N = \frac{N'}{4\pi}\int_{{S}^2} d\Omega \left(
 \langle \Psi^{(0)}_N, \Omega\rangle_N \langle \Omega,\Omega'\rangle_{N-M} -\langle \Psi^{(0)}_N, \Omega\rangle_N \langle \Omega,\Omega'\rangle_{N}\right),\end{equation}
where $N'=N+1$. The
Cauchy-Schwartz' inequality implies
\begin{align}  \label{NiN} |I_N| \leq \sqrt{\frac{N'}{4\pi} \int |\langle \Psi^{(0)}_N, \Omega\rangle_N |^2 d\Omega}
\sqrt{\frac{N'}{4\pi} \int |\langle \Omega,\Omega'\rangle_{N-M}  - \langle \Omega,\Omega'\rangle_{N} |^2d\Omega}.\end{align}
Here, eq.\ (\ref{important}) gives rise to 
\begin{equation}\langle \Omega,\Omega'\rangle_L = \left( \cos(\theta/2)\cos(\theta'/2)+ e^{i(\phi-\phi')}
\sin(\theta/2)\sin(\theta'/2)\right)^L = \langle \Omega,\Omega'\rangle_1^L, \end{equation}
so that 
\begin{align}|\langle \Omega,\Omega'\rangle_{N-M}  - \langle \Omega,\Omega'\rangle_{N} |^2&= |\langle \Omega,\Omega'\rangle_{N-M} |^2 |1-\langle \Omega,\Omega'\rangle_{M} |^2\nonumber \\
&= |\langle \Omega,\Omega'\rangle_{N-M} |^2 |1-\langle \Omega,\Omega'\rangle_{1} ^M|^2.\end{align}
Inserting this result in (\ref{NiN}),  we find
\begin{align}  |I_N|& \leq \sqrt{\frac{N'}{N'-M}}\sqrt{\frac{N'}{4\pi} \int |\langle \Psi^{(0)}_N, \Omega\rangle_N ,^2 d\Omega}\nonumber \\
&\times 
\sqrt{\frac{N'-M}{4\pi} \int|\langle \Omega,\Omega'\rangle_{N-M} |^2 |1-\langle \Omega,\Omega'\rangle_{1} ^M|^2d\Omega}.\label{NIN3}\end{align}
From a direct computation, we see that the map
${S}^2 \times {S}^2 \ni (\Omega,\Omega') \mapsto \langle\Omega ,\Omega'\rangle_1$ is nothing but the restriction to the unit sphere ${S}^2$  of  the map
\begin{equation}\mathbb{R}^3\times  \mathbb{R}^3\ni ( x,y,z, x',y',z')\mapsto \frac{(1+z+z'+zz'+ xx'+yy'+ixy'+ix'y)}{2\sqrt{(1+z)(1+z')}},\end{equation}
where $(x,y,z)$ and $(x',y',z')$ are the Cartesian coordinates of $\Omega$ and $\Omega'$ respectively.  From that, it is straightforward to establish  that, for $\Omega' \neq S$,  the function  \begin{equation}{S}^2 \ni \Omega\mapsto h_{\Omega'}(\Omega) = |1-\langle \Omega,\Omega'\rangle_{M} |^2 = 1+ |\langle \Omega,\Omega'\rangle_{1}|^{2M} -2 Re \langle \Omega,\Omega'\rangle^M_{1}\end{equation} vanishes for $\Omega=\Omega'$, and  is measurable and bounded.
Referring to the  atlas on ${S}^2$ consisting of the $6$ local charts given by  the canonical projections onto the $3$ coordinate $2$-planes, it is finally obvious that $h_{\Omega'}$ is  everywhere  smooth 
with respect to the differentiable structure induced from $\mathbb{R}^3$,  except for $\Omega=S$ (where $z=-1$).
We may therefore  apply (\ref{delta2}) to the special case $h(\Omega) = h_{\Omega'}(\Omega)$ --
which satisfies $h(\Omega')=0$ --  in (\ref{NIN3}). Exploiting also (\ref{lemmaAid}) with $g=1$ to handle the large-$N$ behavior of the first integral on the right-hand side
of (\ref{NIN3}),  which is bounded by some constant $H\geq 0$ when $N$ increases, we conclude that, if $N>M$,
\begin{align}\label{disfin1} |I_N| \leq \frac{K^{(A)}}{(N-M)^{1/4}},
\end{align}
for the  constant \begin{equation}K^{(A)} =H \sqrt{C\sup_{\Omega,\Omega' \in A} |h_{\Omega'}(\Omega)|+ D^{(A)}\sup_{\Omega,\Omega' \in A} \sqrt{{\bf g}_\Omega(d_\Omega h_{\Omega'}(\Omega), d_\Omega h_{\Omega'}(\Omega)})} .\end{equation}
Notice that with these definitions, $C$ and $D^{(A)}$ do not depend on the choice of the  function used here (viz.\ $h_{\Omega'}$),  whereas $D^{(A)}$ only depends on $A$, which is the same for all  possible choices of $\Omega' \in A$. Hence, no dependence on $\Omega'$ takes place.

Let us now turn attention to the general case  where now $h$ is a generic bounded measurable function that is $C^1(A)$, defining
\begin{equation} J_N = \frac{N'}{4\pi}\int_{{S}^2} d\Omega \left(
 \langle \Psi^{(0)}_N, \Omega\rangle_N \langle \Omega,\Omega'\rangle_{N-M}h(\Omega) -\langle \Psi^{(0)}_N, \Omega\rangle_N \langle \Omega,\Omega'\rangle_{N}h(\Omega')\right).\end{equation}
Inserting a vanishing term 
\begin{equation}0=  \frac{N'}{4\pi}\int_{{S}^2} d\Omega \left(
 \langle \Psi^{(0)}_N, \Omega\rangle_N \langle \Omega,\Omega'\rangle_{N-M}h(\Omega') -\langle \Psi^{(0)}_N, \Omega\rangle_N \langle \Omega,\Omega'\rangle_{N}h(\Omega')\right)\end{equation}
between the two summands on the right-hand side,
the triangle  inequality, the fact that $h(\Omega')$ is constant with respect to $\Omega$, and the definition of $I_N$ yield
\begin{equation}|J_N|  \leq \|h\|_\infty |I_N| + 
 \frac{N'}{4\pi}\int_{{S}^2} d\Omega 
 |\langle \Psi^{(0)}_N, \Omega\rangle_N|\cdot | \langle \Omega,\Omega'\rangle_{N-M}\|h(\Omega) -h(\Omega')|.\end{equation}
Applying the Cauchy-Schwartz inequality, we end up with
\begin{equation}|J_N|  \leq \|h\|_\infty |I_N| + 
 \sqrt{\frac{N'}{4\pi}\int_{{S}^2} 
 |\langle \Psi^{(0)}_N, \Omega\rangle_N|^2}d\Omega 
 \sqrt{\frac{N'}{4\pi}\int_{{S}^2}  | \langle \Omega,\Omega'\rangle_{N-M}|^2|h(\Omega) -h(\Omega')|^2 d\Omega}.\end{equation}
As in the previous case, in particular taking advantage of  (\ref{delta2}) to estimate the last integral and using (\ref{disfin1}) and noticing that $1/(N-M) > 1/N$, we end up with
\begin{equation}|J_N| \leq \frac{K^{(A)}\|h\|_\infty+\sqrt{ C \|h\|^2_\infty + D^{(A)} \|dF\|^{(A)}_\infty}}{(N-M)^{1/4}},\end{equation}
for  some constants $K^{(A)}, C,D^{(A)} \geq 0$, generally depending on $M$, but  independent of $\Omega'$, $h$ and $F$, where $F(\Omega) = |h(\Omega)-h(\Omega')|^2$ that is $C^1$ where $h$ is $C^1$.
$\hfill$ $\Box$

\section{Numerical evidence}\label{numerical}
 This appendix provides numerical evidence for equations \eqref{firstlim} - \eqref{firstlim2}. 
\subsection{Dicke components of  $\Psi^{(0)}_N$}\label{Dick} 
For any $N\in\mathbb{N}$,  the ground state eigenvector $\Psi_N^{(0)}$ lives in the symmetric subspace  $\mathrm{Sym}^N(\mathbb{C}^2)\subset\bigotimes_{n=1}^N\mathbb{C}^2$.
This non-obvious fact arises from the uniqueness of the 
CW-ground state vector (up to phases and normalization), which is ultimately a consequence of the Perron--Frobenius Theorem, and the fact that $h^{CW}_{1/N}$ is invariant under the natural action of permutation group of $N$ elements \cite{IL}, \cite[\S5.3]{vandeVen},  \cite{VGRL18}.

 In order to do computations with $\Psi_N^{(0)}$, it therefore suffices to represent this vector in an $(N+1)$-dimensional basis 
for $\mathrm{Sym}^N(\mathbb{C}^2)$. This is a big numerical advantage: diagonalizing 
a $(N+1)$-dimensional matrix is much more efficient for a computer rather than diagonalizing a $2^N$-dimensional matrix.
The Dicke  basis \er{Dickebasis} we  already introduced for this subspace therefore allows the expansion
\begin{align}\label{DPSI0}
    \Psi_N^{(0)}=\sum_{k=0}^Nc_N(k)|k,N-k\rangle,
\end{align}
where the coefficients $c_N(k)$ depend on $N$ and, again  from the {\em Perron-Frobenius  theorem}, the usual arbitrary phase affecting $ \Psi_N^{(0)}$ can be chosen in 
order that
 \begin{align} c_N(k) >0,\quad k=0,1,\ldots, N.\label{POS}
\end{align}
Both analytic asymptotics \cite{IL} and numerical computations \cite{vandeVen} of the coefficients $c_N(k)$ are known, but no analytic expression has been found so far. To compute the expression
$|\langle \Psi^{(0)}_N, \Omega_{\theta,\phi}\rangle|^{2l}$
popping up in equations \eqref{firstlim} (for $l=1/2$) and \eqref{firstlim2} (for $l=1$)
we  use eqs.\ (\ref{DPSI0}) and
(\ref{important}) for $\Psi^{(0)}_N$ and $|\Omega_{\theta,\phi}\rangle_N$ in terms of the Dicke basis. This way, the relevant inner product will be computed again in terms of the $N+1$ numerically favorable Dicke states, instead of $2^N$ basis vectors for $\bigotimes_{n=1}^N\mathbb{C}^2$.

Let us first focus on  the $\mathbb{Z}_2$-action $\zeta^{(1/N)}$ on $M_2(\mathbb{C})^N$, see text after \er{asigma1}. This automorphism is unitarily implemented by the following unitary operator:
\begin{align}
U_N  &= \underbrace{\sigma_1\otimes \cdots \otimes \sigma_1}_{N\: times}\in M_2(\mathbb{C})^N, \label{UN}\\
 \zeta^{(1/N)}(a) &= U_Na  U_N^{-1}
, \label{ZETA28}
\end{align} where $a\in M_2(\mathbb{C})^N$.
 Since $U_1=\sigma_1$, which swaps $|\!\uparrow\rangle$ and $|\!\downarrow\rangle$, we clearly have
\begin{equation}
U_N |k,N-k \rangle = |N-k, k \rangle.\label{ZeD}
\end{equation}
Passing to the coherent spin state basis, this gives
\begin{align}
U_N|\Omega_{\theta,\phi} \rangle_N &= e^{-iN\phi} |\Omega_{\pi-\theta,-\phi} \rangle_N;\label{ZeD2}\\
U_N  |\Omega_{\pm} \rangle_N &= |\Omega_{\mp} \rangle_N .\label{ZeD24}
\end{align}
As we already saw,
the  (algebraic) CW-ground state $\omega_{1/N}^{(0)}$ (\ref{ALST})  is invariant under the automorphism 
(\ref{ZETA28}). The  unit vector  $\Psi^{(0)}_N$ of $\omega_{1/N}^{(0)}$ must  therefore satisfy 
\begin{equation}
U_N\Psi^{(0)}_N = \pm  \Psi^{(0)}_N, \label{pm1}
\end{equation}
since $U_N^2=I$. By (\ref{ZeD}), for the components (\ref{DPSI0}), eq.\ \er{pm1}  can be rephrased as
\begin{equation}
 c_N(N-k)= \pm  c_N(k),
\end{equation}
 where the sign does not depend on $k$.  However,  because  $c_N(k)>0$ only the + sign can actually occur. Thus the $\mathbb{Z}_2$-invariance of the ground state is equivalent to
\begin{align}
c_N(k) = c_N(N-k),\quad k=0,1,\ldots, N,
\end{align}
and from (\ref{ZeD2}) we also have
\begin{align}
|\langle \Psi^{(0)}_N ,\Omega_{\theta,\phi}\rangle|^2 = |\langle \Psi^{(0)}_N,\Omega_{\pi-\theta,-\phi}\rangle|^2\label{invPSI}.
\end{align}
\subsection{Coefficients $c_N(k)$ for $N\geq 80$} We computed the components $c_N(k)$ of  $\Psi_N^{(0)}$  using {\sc matlab}. However, from  $N= 80$ onwards  our program was not able  to numerically distinguish  anymore between the lowest eigenvalue $\epsilon_0^{(N)}$ of $h^{CW}_{1/N}$ and its  first excited level $\epsilon_1^{(N)}> \epsilon_0^{(N)}$ in $\mathrm{Sym}^N(\mathbb{C}^2)$. As a consequence, within  this numerical approximation,  the $\epsilon_0^{(N)}$-eigenspace of $h^{CW}_{1/N}$ appears as a two-dimensional subspace $K^{(N)}$ of $\mathrm{Sym}^N(\mathbb{C}^2)$ and one needs to extract the actual ground state from the span of  the pair of apparent degenerate eigenvectors
$\Psi_N^{(0)\mbox{\footnotesize{\sc matlab}}}$ and  $\Psi_N^{(1))\mbox{\footnotesize{\sc matlab}}}$ of $h^{CW}_{1/N}$  with the  common eigenvalue 
$\epsilon_0^{(N)}$
computed by {\sc matlab}, which form an orthonormal basis of $K^{(N)}$.  This can indeed be done, because $K^{(N)}$  is invariant under the unitary representation $U^{(N)}$ (\ref{ZeD}) of the element -1 of $\mathbb{Z}_2$, which turns out to be {\em non-trivial} when restricted to that subspace. Hence
\begin{equation}
U_N|_{K^{(N)}} \neq I, 
\end{equation}
and since $U_N|_{K^{(N)}} $ is simultaneously unitary and selfadjoint, its spectrum consists only of two points $\pm1$. 
In other words, $K^{(N)}$ contains {\em exactly  one} (up to phases) unit vector 
$\Phi^{(N)}$ such that $U_N \Phi^{(N)} = \Phi^{(N)}$.  Since the true ground state of $h^{CW}_{1/N}$ satisfies the same condition and belongs to the same (approximate) subspace, we must have
\begin{equation}
 \Psi^{(0)}_N = \Phi^{(N)}.
\end{equation}
{\em Therefore,\footnote{Of course, with phases chosen such that the Perron--Frobenius condition \er{POS} holds.}
$\Psi^{(0)}_N $ is the unique  unit eigenvector 
 of $U_N$ with eigenvalue $1$.}
{\sc matlab} proposes a pair of orthonormal  vectors $\Psi_N^{(0)\mbox{\footnotesize{\sc matlab}}}$ and  $\Psi_N^{(1))\mbox{\footnotesize{\sc matlab}}}$, forming an orthonormal basis of $K^{(N)}$ which can be assumed to be of the form represented in the following picture, up to a change of the overall sign and the action of 
$U_N $ (which simply reflects the function around the vertical axis localized at $N/2$).

\begin{figure}[!htb]
\centering
   \includegraphics[width=11cm]{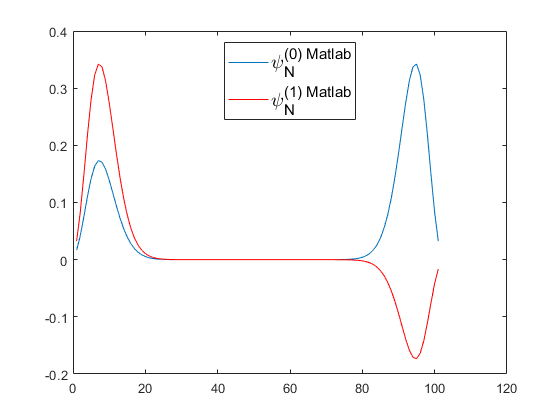}
 \caption{Plot of $\Psi_N^{(0)\mbox{\footnotesize{\sc matlab}}}$ (in blue) and $\Psi_N^{(1)\mbox{\footnotesize{\sc matlab}}}$ (in red) for $N=100, J=1, B=1/2$.}
    \label{bothplotsN=100}
\end{figure}
\noindent If $r_N$ denotes the ratio $r_N= H^{(N)}_L/H^{(N)}_R$, where $H^{(N)}_R\geq H^{(N)}_L$ is the height of the peak in the left part of the figure representing $\Psi_N^{(0)\mbox{\footnotesize{\sc matlab}}}$, and $H^{(N)}_L$ is defined analogously for the peak in the right part, it is not difficult to prove that the unique (up to phases) unit eigenvector 
$\Psi^{(0)}_N $ of $U_N$ with eigenvalue $1$ takes the form
\begin{align} \Psi^{(0)}_N  = \frac{1}{\sqrt{2}}\left(\frac{1+r_N}{\sqrt{1+ r^2_N}}  \Psi_N^{(0)\mbox{\footnotesize{\sc matlab}}}+ \frac{1-r_N}{\sqrt{1+ r^2_N}} \Psi_N^{(1)\mbox{\footnotesize{\sc matlab}}} \right)\label{ricettafine2}.\end{align} 
That is the desired ground state for $N\geq 80$.  Notice that, with $\Psi_N^{(0)\mbox{\footnotesize{\sc matlab}}}$ and $\Psi_N^{(1)\mbox{\footnotesize{\sc matlab}}}$ as computed by {\sc matlab}, the components 
$c_N(k)$ of $ \Psi^{(0)}_N$ also satisfy $c_N(k) \geq 0$ (instead of $c_N(k) > 0$ valid in the non-degenerate case).
\begin{remark}
{\em When $N< 80$, within our available computational precision {\sc matlab} \emph{is}  able to distinguish $\epsilon_0^{(N)}$ from $\epsilon_1^{(N)}$ and the computed vector $\Psi_N^{(0)\mbox{\footnotesize{\sc matlab}}}$
is such that $r_N=1$. Therefore, as expected,  (\ref{ricettafine2}) furnishes the  ground state 
\begin{align} \Psi^{(0)}_N  =  \Psi_N^{(0)\mbox{\footnotesize{\sc matlab}}}\label{ricettafine22}.\end{align} 
In the opposite direction, for $N>150$ we obtain $r_N =0$, so that  (\ref{ricettafine2})  reduces to
\begin{align} \Psi^{(0)}_N  = \frac{1}{\sqrt{2}}\left( \Psi_N^{(0)\mbox{\footnotesize{\sc matlab}}}+ \Psi_N^{(1)\mbox{\footnotesize{\sc matlab}}} \right)\label{ricettafine23}.\end{align}}
\hfill $\blacksquare$
\end{remark} 
\subsection{Numerical evidence for (a),(b),(c) in Assumption \ref{assumption}}
  We computed the integrals in  \eqref{firstlim} and  \eqref{decreasingsequence}
  for increasing values of $N$: see Table 1 and Table 2 below, respectively.
\begin{center}
\begin{tabular}
{p{2cm} p{7cm} }
 \multicolumn{2}{c }{Table $1$. Numerical values of the left-hand side on \eqref{firstlim} for increasing $N$.}  \\
 \hline\hline
 $N$ & Value of \eqref{firstlim}. \\
 \hline
 10  & 0.0060  \\
 20  & $4.0922 \cdot 10^{-4}$ \\
 30 &  $3.8941 \cdot 10^{-5}$\\
 60  & $-1.4394 \cdot 10^{-5}$\\
 90  & $-2.7404 \cdot 10^{-6}$  \\
 120  & $-4.2139 \cdot 10^{-7}$ \\
 150 & $-6.0988 \cdot 10^{-8}$  \\
 180 & $-8.6073 \cdot 10^{-9}$  \\
 \hline\hline
\end{tabular}
\end{center}
\begin{center}
\begin{tabular}
{p{2cm} p{5cm} p{5cm}}
 \multicolumn{3}{c }{Table $2$. Numerical values of the left-hand side on \eqref{decreasingsequence} for increasing $N$.}  \\
 \hline\hline
 $N$ &  Value of \eqref{decreasingsequence} for $l=1$ &  Value of \eqref{decreasingsequence} for $l=1/2$\\
 \hline
 10 & 0.2559 & 0.4185\\
 20 & 0.1065 & 0.2095\\
 30 & 0.0868 & 0.1860\\
 40 & 0.0765 & 0.1731\\
 50 & 0.0707 & 0.1649\\
 60 & 0.0666 & 0.1590\\
 70 & 0.0636 & 0.1547\\
 80 & 0.0614 & 0.1514\\
 90 & 0.0596 & 0.1488\\
 100 & 0.0582 & 0.1469\\
 110 & 0.0570 & 0.1452\\
 120 & 0.0561 & 0.1439\\
 130 & 0.0552 & 0.1427\\
 140 & 0.0546 & 0.1418\\
 150 & 0.0540 & 0.1409\\
 \hline\hline
\end{tabular}
\end{center}
From this table, it is clear that  for $l=1$ as well as $l=1/2$, eq.\ \eqref{decreasingsequence} is decreasing in $N$, and therefore uniformly bounded in $N$. In fact, from this table it appears that 
\begin{align}
&A_N=\int_{{S}^2}  \frac{N+1}{4\pi}|\langle \Psi^{(0)}_N ,\Omega
\rangle_N|^{2l}d\Omega\approx \nonumber \\ 
  &B_N=\int_{{S}^2} \bigg{(}
\frac{N+1}{4\sqrt{2}\pi} |\langle \Omega, \Omega_+ \rangle_N|^{2l} + \frac{N+1}{4\sqrt{2}\pi} |\langle \Omega, \Omega_- \rangle_N|^{2l}\bigg{)}d\Omega,
\end{align}
as $N$ becomes large. To be even more precise, we numerically computed the values of $A_N$ and $B_N$ for increasing values of $N$: see Table 3.
\newline
\newline
\begin{center}
\begin{tabular}
{p{2cm} p{3cm} p{3cm} p{3cm} p{3cm} }
 \multicolumn{5}{c }{Table $3$. $A_N$ and $B_N$ (as defined above) from \eqref{decreasingsequence} for increasing $N$.}  \\
 \hline\hline
 $N$ &  $A_N$ for $l=1/2$ & $B_N$ for $l=1/2$ & $A_N$ for $l=1$ & $B_N$ for $l=1$\\
 \hline
 10 & 2.3357 & 2.5471 & 0.9831 &0.9772 \\
 20 & 2.6489 & 2.6846 & 0.9950 & 0.9946 \\
 30 & 2.7285 & 2.7330 & 0.9983 & 0.9982\\
 40 & 2.7598 & 2.7574 & 0.9993 & 0.9993\\
 50 & 2.7759 & 2.7719 & 0.9997 & 0.9997\\
 60 & 2.7858 & 2.7816 & 0.9999 & 0.9999\\
 70 & 2.7926 & 2.7884 & 0.9999 & 0.9999\\
 80 & 2.7977 & 2.7935 & 1.0000 & 1.0000 \\
 90& 2.8015 & 2.7974 & 1.0000 & 1.0000\\
 100& 2.8046 & 2.8005 & 1.0000 & 1.0000\\
 110& 2.8071 & 2.8031  & 1.0000 & 1.0000\\
 120 & 2.8092 & 2.8052  & 1.0000 & 1.0000\\
 130& 2.8109 & 2.8070 & 1.0000 & 1.0000\\
 140& 2.8124 & 2.8085  & 1.0000 & 1.0000\\
 \hline\hline
\end{tabular}
\end{center}
This clearly suggests that for $l=1/2$ both integrals converge to $2\sqrt{2}\approx 2.828$.
Therefore, since the integral in  \eqref{decreasingsequence}  is bounded  by $A+B$,
 there is strong numerical evidence that   \eqref{decreasingsequence} is valid for some constant $G$, for example given by  the sum of $A_N$ and $B_N$, i.e., $G=4\sqrt{2}$. A similar result holds for the case $l=1$.
\smallskip

\noindent Furthermore, the validity of part (c) in \eqref{assumption} 
has been checked by comparing the graphs of the function \begin{equation}{S}^2 \ni \Omega \mapsto \frac{N+1}{4\pi}|\langle \Psi^{(0)}_N ,\Omega
\rangle_N|^{2\ell}\end{equation} with the graphs of the function \begin{equation}{S}^2 \ni \Omega \mapsto \frac{N+1}{4\pi 2^\ell} |\langle \Omega_+, \Omega \rangle_N|^{2\ell}+\frac{N+1}{4\pi 2^\ell} |\langle \Omega_-, \Omega \rangle_N|^{2\ell},\end{equation} since the latter satisfies (c) ((b) Remark \ref{remCU}) and the graph of the former becomes more and more indistinguishable from the graph of the latter as $N$ increases.
We display various plots of the graphs of both functions for two typical different values of $N$.  In order to make a clear comparison we avoid  single $3d$ plots, but instead plot two $2d$ plots, one as a function of $\theta$ for fixed $\phi=0$, and the other as a function of $\phi$ for fixed $\theta=\pi/6$. \footnote{Note that due to symmetry we also could have chosen the point $\theta=5\pi/6$. We indeed checked this numerically, but omitted the plots.} 
These pairs of $2d$ plots (for $l=1/2$ and $l=1$) are depicted in the next pages for $N=30$ and $N=250$, and as always $J=1, B=1/2$.

\newpage
\begin{figure}[!htb]
\centering
\includegraphics[width=9cm]{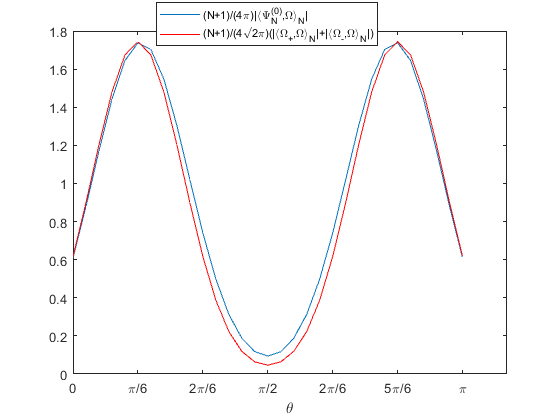}
 \label{newplot1}
\end{figure}
Figure 4: Plot for  $N=30$ of the functions, in blue and in red, respectively, 
 \begin{align*}
 \theta &\mapsto \frac{N+1}{4\pi}|\langle \Psi^{(0)}_N, \Omega_{\theta,0}\rangle|;\\
 \theta &\mapsto \frac{N+1}{4\sqrt{2}\pi}|\langle \Omega_{+}, \Omega_{\theta,0}\rangle|+\frac{N+1}{4\sqrt{2}\pi}|\langle \Omega_{-}, \Omega_{\theta,0}\rangle|
 \end{align*} 

\begin{figure}[!htb]
\centering
\includegraphics[width=9cm]{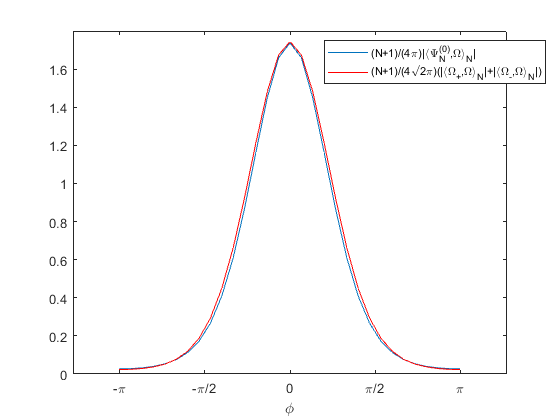}
 \label{newplot2}
\end{figure}
Figure 5: Plot for  $N=30$ of the functions, in blue and in red, respectively, 
 \begin{align*}
\phi & \mapsto \frac{N+1}{4\pi}|\langle \Psi^{(0)}_N, \Omega_{\pi/6,\phi}\rangle|;\\
\phi & \mapsto \frac{N+1}{4\sqrt{2}\pi}|\langle \Omega_{+}, \Omega_{\pi/6,\phi}\rangle|+\frac{N+1}{4\sqrt{2}\pi}|\langle \Omega_{-}, \Omega_{\pi/6,\phi}\rangle|
 \end{align*} 
 \newpage
 
  \begin{figure}[!htb]
\centering
\includegraphics[width=9cm]{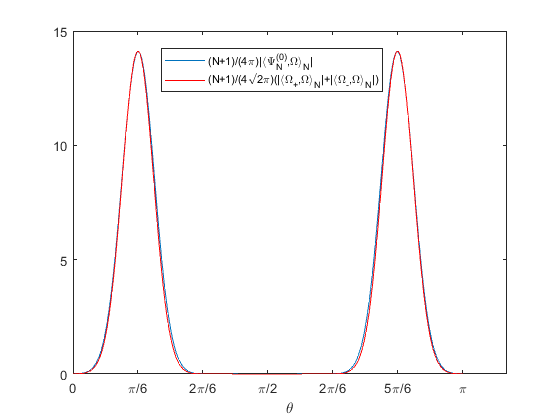}
 \label{newplot5}
\end{figure}
Figure 6: Plot for  $N=250$ of the functions, in blue and in red, respectively, 
 \begin{align*}
 \theta &\mapsto \frac{N+1}{4\pi}|\langle \Psi^{(0)}_N, \Omega_{\theta,0}\rangle|;\\
 \theta &\mapsto \frac{N+1}{4\sqrt{2}\pi}|\langle \Omega_{+}, \Omega_{\theta,0}\rangle|+\frac{N+1}{4\sqrt{2}\pi}|\langle \Omega_{-}, \Omega_{\theta,0}\rangle|
 \end{align*} 

 \begin{figure}[!htb]
\centering
\includegraphics[width=9cm]{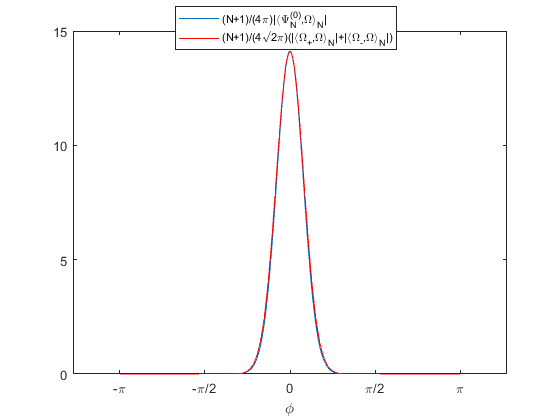}
 \label{newplot6}
\end{figure}
Figure 7: Plot for  $N=250$ of the functions, in blue and in red, respectively, 
 \begin{align*}
\phi & \mapsto \frac{N+1}{4\pi}|\langle \Psi^{(0)}_N, \Omega_{\pi/6,\phi}\rangle|;\\
\phi & \mapsto \frac{N+1}{4\sqrt{2}\pi}|\langle \Omega_{+}, \Omega_{\pi/6,\phi}\rangle|+\frac{N+1}{4\sqrt{2}\pi}|\langle \Omega_{-}, \Omega_{\pi/6,\phi}\rangle|
 \end{align*} 
 \newpage
  \begin{figure}[!htb]
\centering
\includegraphics[width=9cm]{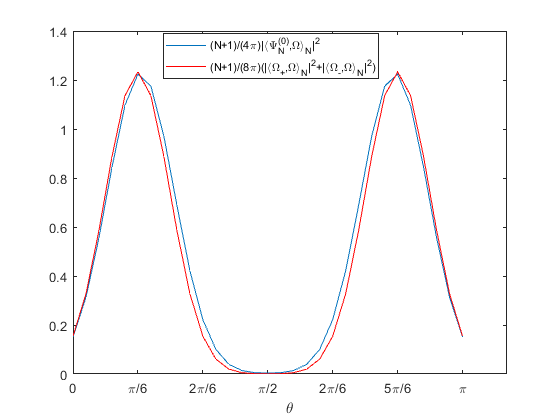}
 \label{newplot7}
\end{figure}
Figure 8: Plot for  $N=30$ of the functions, in blue and in red, respectively, 
 \begin{align*}
 \theta &\mapsto \frac{N+1}{4\pi}|\langle \Psi^{(0)}_N, \Omega_{\theta,0}\rangle|^2;\\
 \theta &\mapsto \frac{N+1}{8\pi}|\langle \Omega_{+}, \Omega_{\theta,0}\rangle|^2+\frac{N+1}{8\pi}|\langle \Omega_{-}, \Omega_{\theta,0}\rangle|^2
 \end{align*} 

\begin{figure}[!htb]
\centering
\includegraphics[width=9cm]{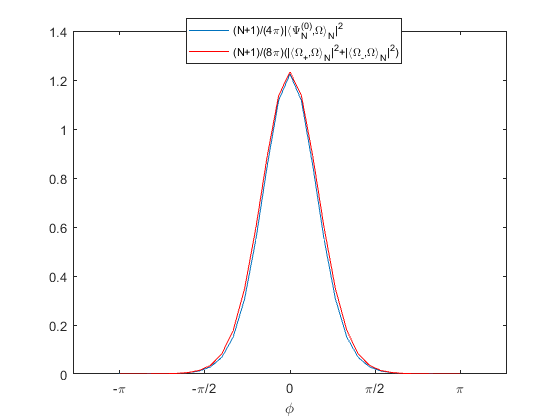}
 \label{newplot8}
\end{figure}
Figure 9: Plot for  $N=30$ of the functions, in blue and in red, respectively, 
 \begin{align*}
\phi & \mapsto \frac{N+1}{4\pi}|\langle \Psi^{(0)}_N, \Omega_{\pi/6,\phi}\rangle|^2;\\
\phi & \mapsto \frac{N+1}{8\pi}|\langle \Omega_{+}, \Omega_{\pi/6,\phi}\rangle|^2+\frac{N+1}{8\pi}|\langle 
\Omega_{-}, \Omega_{\pi/6,\phi}\rangle|^2
 \end{align*} 
 \newpage
  
 \begin{figure}[!htb]
\centering
\includegraphics[width=9cm]{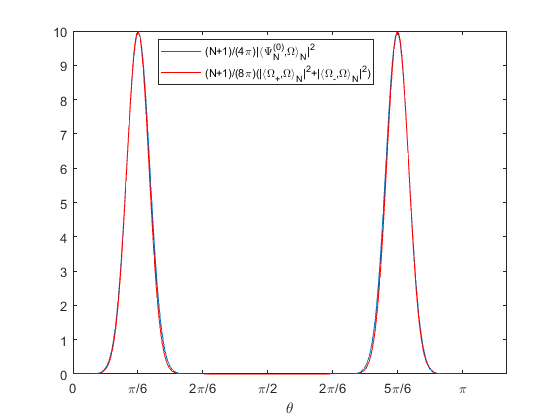}
 \label{newplot11}
\end{figure}
Figure 10: Plot for  $N=250$ of the functions, in blue and in red, respectively, 
 \begin{align*}
 \theta &\mapsto \frac{N+1}{4\pi}|\langle \Psi^{(0)}_N, \Omega_{\theta,0}\rangle|^2;\\
 \theta &\mapsto \frac{N+1}{8\pi}|\langle \Omega_{+}, \Omega_{\theta,0}\rangle|^2+\frac{N+1}{8\pi}|\langle \Omega_{-}, \Omega_{\theta,0}\rangle|^2
 \end{align*} 

\begin{figure}[!htb]
\centering
\includegraphics[width=9cm]{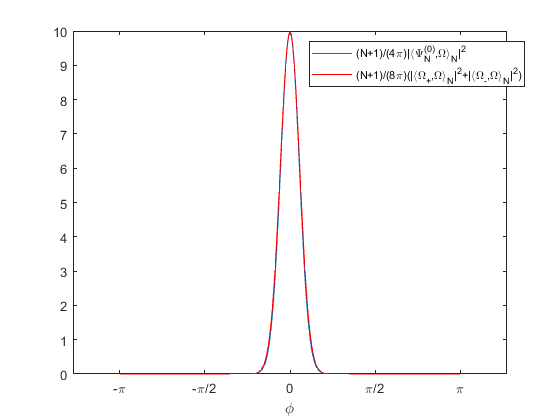}
 \label{newplot12}
\end{figure}
Figure 11: Plot for  $N=250$ of the functions, in blue and in red, respectively, 
 \begin{align*}
\phi & \mapsto \frac{N+1}{4\pi}|\langle \Psi^{(0)}_N, \Omega_{\pi/6,\phi}\rangle|^2;\\
\phi & \mapsto \frac{N+1}{8\pi}|\langle \Omega_{+}, \Omega_{\pi/6,\phi}\rangle|^2+\frac{N+1}{8\pi}|\langle 
\Omega_{-}, \Omega_{\pi/6,\phi}\rangle|^2
 \end{align*} 
\newpage
\noindent
Finally, we give another numerical fact  corroborating  \eqref{assumption}, namely that
the full width at half maximum (fwhm) of the function 
\begin{equation}
N\mapsto \frac{N+1}{4\pi}|\langle \Psi^{(0)}_N ,\Omega \label{last}
\rangle_N|^{2\ell} 
\end{equation}
vanishes as $N\to\infty$. 
We discretized $\theta$ and $\phi$ uniformly in $N$ points on $(0,\pi/6)$ and $(-\pi, \pi)$ respectively, so that \er{last}  becomes a $2d$ array of $N^2$ points. We then computed the number of points $a(N,\pi/6)$ at half height of the array $\frac{N+1}{4\pi}|\langle \Psi^{(0)}_N ,\Omega
\rangle_N|^{2\ell}$ at fixed $\pi/6$, but varying the discrete values of $\phi$. Then we repeated this step but now varying $\theta$ at fixed $\phi=0$. Similarly as before, we now define $b(N,0)$ to be the number of points at half maximum for $\phi=0$. This basically means that we count the number of points in a rectangle at half maximum of the total array.
It is clear that the area of the rectangle spanned by $a(N,\pi/6)$ and $b(N,0)$ includes all points of the function at half maximum. Some of the values are given in the graph below:\footnote{We display this for the case $\ell=1$, but  numerically checked that the same holds for $\ell=1/2$.} 
\begin{figure}[!htb]
\centering
   \includegraphics[width=9cm]{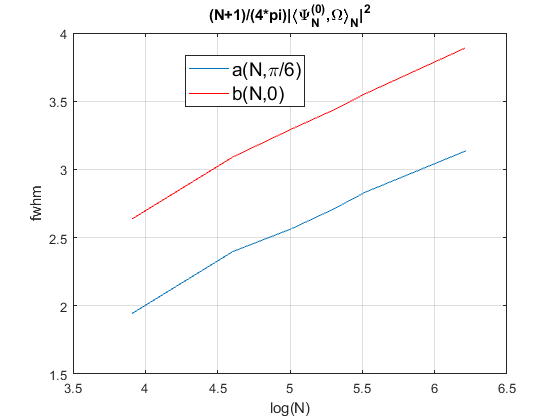}
 \caption*{Figure 12: Full width at half maximum  for the function \er{last}, for $N=50,150,200,250,500$ on a log scale . The red line corresponds to $\theta\in(0, \pi/2)$ and $\phi=0$, whilst the blue line corresponds to $\theta=\pi/6$ and $\phi\in(-\pi, \pi)$.}
    \label{fwhm}
\end{figure}

 \noindent It may be clear that the slope of both lines is about $0.5$, which means that the fwhm goes like $\sqrt{N}$. It is  also clear that $b(N,0)$ seems to be translated with respect to $a(N,\pi/6)$ be a factor $2$. We conclude that the number of points in the rectangle is approximately given by $\sqrt{N}\cdot\sqrt{N}\cdot 2=2N=O(N)$.  Using the above discretization, we then have about $\sqrt{N}$ steps of $\pi/2N$ each, and about $2\sqrt{N}$ steps of $2\pi/N$ so that in particular the spanned rectangle has a width of $2\pi^2/N=O(1/N)$. This means that the fwhm of the function \er{last} indeed vanishes as $N\to\infty$. 
 
\end{document}